\documentclass[aps,pra,eps2pdf,twocolumn,superscriptaddress]{revtex4-2}

\usepackage[utf8]{inputenc}
\usepackage{graphicx}
\graphicspath{{../}}
\pdfoutput=1

\usepackage{dsfont}
\usepackage{amsmath}
\usepackage{amssymb,bm}
\usepackage{amsthm}
\usepackage{algorithm}
\usepackage{algorithmicx}
\usepackage{float}
\makeatletter
\let\newfloat\newfloat@ltx
\makeatother

\usepackage[table,xcdraw]{xcolor}

\usepackage[]{algpseudocode}

\usepackage{mathtools}
\usepackage{url}
\usepackage{microtype}
\usepackage{qcircuit}
\usepackage{array, makecell}
\usepackage{boldline}
\usepackage{enumitem}

\usepackage{tikz}
\usetikzlibrary{decorations.pathreplacing}
\usetikzlibrary{positioning}
\usetikzlibrary{shapes,patterns}
\usetikzlibrary{calc,backgrounds}

\usepackage{color,dsfont}
\usepackage{ragged2e}
\usepackage[colorlinks=true, hyperindex, breaklinks, linkcolor=blue, urlcolor=blue, citecolor=blue]{hyperref}
\usepackage[normalem]{ulem}
\usepackage[capitalise]{cleveref}
\usepackage{mathrsfs}
\usepackage[caption=false]{subfig}
\usepackage{float}
\usepackage{verbatim}
\usepackage{latexsym}
\usepackage{setspace}
\usepackage{amsfonts}
\usepackage{stmaryrd}
\usepackage{xcolor}
\usepackage{enumitem}
\usepackage{lipsum}
\usepackage{multirow}
\AtBeginDocument{\tabcolsep=6pt}
\usepackage{booktabs}

\usepackage{graphicx} 
\usepackage{amsfonts}
\usepackage{amssymb}
\usepackage{mathrsfs}
\usepackage{theorem}
\usepackage{amsmath}
\usepackage{times}
\usepackage{color}
\usepackage{soul}

\usepackage{ifpdf}
\ifpdf
\usepackage{epstopdf}   
\fi
\usepackage{cleveref}

\DeclareMathOperator{\poly}{poly}

\definecolor{nicepurple}{RGB}{138,115,221}
\definecolor{KiannaPurple}{RGB}{172,7,249}

\definecolor{XRed}{RGB}{0,0,0}
\definecolor{XGreen}{RGB}{0,0,0}
\definecolor{XBlue}{RGB}{0,0,0}
\definecolor{XOrange}{RGB}{0,0,0}
\definecolor{XYellow}{RGB}{0,0,0}
\definecolor{XPurple}{RGB}{0,0,0}
\definecolor{XTurquoise}{RGB}{0,0,0}
\definecolor{XPink}{RGB}{0,0,0}

\definecolor{DarkBlue}{RGB}{20,120,150}
\definecolor{DarkRed}{RGB}{155,0,0}
\definecolor{DarkGreen}{RGB}{0,145,0}
\definecolor{VeryDarkGreen}{RGB}{0,130,0}
\definecolor{DarkYellow}{RGB}{156, 132, 44}
\definecolor{DarkPurple}{RGB}{108,25,96}

\newcommand{\ket}[1]{\ensuremath{\vert#1\rangle}}

\newcommand{\LP}{\mathcal{L}^{(MC,P)*}}
\newcommand{\LPRB}{\LP_{RB}}
\newcommand{\LPRG}{\LP_{RG}}
\newcommand{\LPBG}{\LP_{BG}}
\newcommand{\eE}{\mathcal{E}}

\DeclareMathOperator{\Sp}{Sp}

\newcommand{\F}{\mathbb{F}}
\DeclareMathOperator{\CNOT}{CNOT}

\def\R{\mathbb{R}}

\renewcommand{\vec}[1]{\ensuremath{\mathbf{#1}}}

\newcommand{\wt}{\ensuremath{\mathrm{wt}}}

\begin{document}

\title{Decoding Merged Color-Surface Codes and Finding Fault-Tolerant Clifford Circuits Using Solvers for Satisfiability Modulo Theories}
\author{Noah Shutty}
\affiliation{AWS Center for Quantum Computing, Pasadena, CA 91125, USA}
\affiliation{Stanford Institute for Theoretical Physics, Stanford University, Stanford, CA 94305, USA}
\author{Christopher Chamberland}
\affiliation{AWS Center for Quantum Computing, Pasadena, CA 91125, USA}
\affiliation{
IQIM, California Institute of Technology, Pasadena, CA 91125, USA}

\begin{abstract}

Universal fault-tolerant quantum computers will require the use of efficient protocols to implement
encoded operations necessary in the execution of algorithms. In this work, we show how solvers for satisfiability modulo theories (SMT solvers) can be used to automate the construction of Clifford circuits with certain fault-tolerance properties and
we apply our techniques to a fault-tolerant magic-state-preparation protocol. Part of the protocol requires
converting magic states encoded in the color code to magic states encoded in the surface code. Since the
teleportation step involves decoding a color code merged with a surface code, we develop a decoding
algorithm that is applicable to such codes.

\end{abstract}

\maketitle

\section{Introduction}
\label{sec:Intro}

Many problems in quantum computing require the construction of Clifford circuits with some desired properties. For instance, in topological quantum error correction, multi-qubit gates used to measure the stabilizers of the code must be implemented in a particular order to prevent small errors from propagating to large errors which reduce the effective distance of the code \cite{Yoder2017surfacecodetwist,Litinski2018latticesurgery,chamberland2020triangular,PrabhuReichardtv1}. In many cases, fault-tolerant circuits for syndrome extraction require the  use of extra ancilla qubits, known as {\it flag qubits}, whose role is to detect or prevent the propagation of errors arising from a small number of faults to large data qubit errors \cite{SteaneV1,KnillEC,AliferisV1,CRv1,CRv2,CB18,TCL20,ReichardtColorCode,CZYHC20,chamberland2020triangular,CR21,InkV1,InkV2,PrabhuReichardtv1}. For instance, in Refs.~\cite{chamberland2019fault,chamberland2020very}, it was shown that flag qubits could be used to fault-tolerantly prepare high-fidelity magic states without the use of top-down magic state distillation protocols, by which we refer to protocols which use encoded gates to implement all the Clifford operations of the distillation circuits. As such, the Clifford circuits are typically not fault-tolerant. We refer to protocols which prepare magic states using fault-tolerant Clifford circuits as bottom-up protocols. In many cases, fault-tolerant circuits used in bottom-up protocols are constructed from either first principles or brute-force numerical methods. 

In this paper, we show in \cref{sec:SMTsolvers} how to formulate the desired properties (or constraints) of Clifford circuits in a format compatible with satisfiability modulo theories (SMT). These problems can then be solved using SMT solvers such as Z3 \cite{de2008z3}. In \cref{sec:FaultTolerantHtype}, we apply these techniques to construct fault-tolerant circuits for preparing $\ket{H}$-type magic states encoded in the color code \cite{Bombin1,Bombin2,KubicaBevs1,LatticeColorCode}, where the physical qubits are constrained to live on a two-dimensional (2D) lattice interacting via nearest neighbors with low degree connectivity. Such constructions have the potential to be suitable for many quantum computing hardware architectures currently under development.  

Currently, the leading approach to protect logical information afflicted by noise during a quantum computation is to use a two-dimensional topological quantum error-correcting code \cite{kitaev2003fault}, such as the surface code \cite{bravyi1998quantum,dennis2002topological} or the color code. Such codes are then combined with magic state distillation and lattice surgery to perform universal fault-tolerant quantum computation. In particular, the surface code has several advantages over the color code \cite{fowler2012surface}.
For instance, the surface code has a much higher noise threshold than the color code and can achieve desired logical failure rates using fewer qubits for physical error rates expected in near term architectures. Variations of the surface code, such as the XZZX code \cite{ataides2021xzzx}, may provide some advantages over the surface code in settings where the underlying noise model exhibits some bias. However, the surface code still provides lower overhead costs to achieve a given logical error rate for most studied realistic noise models \cite{chamberland2020building}. Since the methods of \cref{sec:FaultTolerantHtype} are used to prepare magic states encoded in the color code, in \cref{sec:DecodingMerge} we show how magic states encoded in the color code can be converted to magic states encoded in the surface code. The schemes involve a teleportation protocol that is implemented using lattice surgery methods. In particular, using gauge fixing to perform an $X \otimes X$ logical Pauli measurement, the color codes and surface code are merged into one code. However, known decoders for surface codes and color codes are not suitable for correcting errors of the merged code. As such, we conclude this paper by presenting a decoding algorithm that can be used to decode the merged code and is hopefully of value for successfully converting states encoded in the color code to states encoded in the surface code.

\section{Quantum Circuit Design as an SMT Decision Problem}
\label{sec:SMTsolvers}
Quantum computers will require high-level quantum algorithms to be compiled to low-level gate implementations that are efficient, fault-tolerant, and compatible with the hardware constraints.
This compilation to a physically-implementable circuit is
a topic of intense scientific research, with significant effort invested in reducing the gate count and depth required to implement algorithms.

Many of the core primitives in a quantum computation are, or can be viewed as, the implementation of Clifford circuits \footnote{For instance, in the bottom-up magic state preparation protocol of \cite{chamberland2020very,chamberland2019fault}, a non-Clifford circuit involving controlled-Hadamard gates is viewed as a Clifford circuit suitably conjugated by $T$ gates.}. Unlike a general $n$-qubit unitary operation, which is specified by an exponential number of real values, an arbitrary $n$-qubit Clifford circuit can be specified by just $O(n^2)$ bits. In addition, these ``simple" circuits can be efficiently simulated using a classical computer.
Despite their mathematical simplicity, the compilation of Clifford circuits is sometimes performed by a skilled researcher, possibly aided by software that verifies the circuit has the desired computational and fault-tolerance properties.
This approach is time-consuming, unpredictable, and may not be as flexible as desired.

In this paper, we document an alternative approach to hand-designing Clifford circuits in which the constraints on a quantum circuit are formulated as an {\it SMT decision problem} which we define in \cref{sec:SMTDecisionProblems}. This problem can then be solved by an off-the-shelf SMT solver such as Z3 \cite{de2008z3}.
Despite SMT decision problems having exponential or worse time complexity for hard instances, ``automated reasoning" software libraries such as Z3 have been heavily optimized and refined through decades of research and are are now widely applied in formal software verification and electronic design automation, among many other domains. They can scale to solve problems containing thousands of variables in diverse domains through careful tuning of problem encoding and solver techniques \cite{barrett2018satisfiability,backes2019reachability,de2011satisfiability}.
In \cref{sec:notationClifford} we show how arbitrary computations from the Clifford group can be represented by bit-matrices, and how to solve for circuits implementing these operations using a limited gate set. We then explain how faults in the circuit can be symbolically propagated through to the end of the circuit, allowing constraints to be added to the SMT problem. In \cref{sec:faultPropagation,sec:flagConstraints}, we show how such constraints can be used to design circuits with guaranteed fault-tolerance properties.
In \cref{susec:Iterative}, we describe techniques for constructing the SMT problems iteratively, which enable more scalable solutions that in turn can be used to solve more difficult circuit design problems of practical interest.

\subsection{Notation and Definitions}\label{sec:notationClifford}

\subsubsection{SMT Decision Problems}\label{sec:SMTDecisionProblems}
Boolean formulas consists of expressions such as the following:
\begin{equation}\label{eq:BooleanFormulaExample}
F = (X_1\lor X_2)\land (X_2 \lor X_3) \land (\neg X_2\lor \neg X_1).
\end{equation}
These expressions involve some Boolean (i.e., binary) variables (the $X_i$ terms in \cref{eq:BooleanFormulaExample}) along with logical operators such as $\land,\lor,\oplus,$ and $\neg$.
A Boolean formula such as $F$ is {\it satisfiable} if there is a way to assign $0$ or $1$ to each of the $X_i$ such that $F$ evaluates to $1$. We call such an assignment of bits to the $X_i$ a {\it satisfying assignment}.
\cref{eq:BooleanFormulaExample} has the satisfying assignment $X_1 = X_3 = 0, X_2 = 1$.

{\it Satisfiability modulo theories} \cite{barrett2018satisfiability} extend the notion of a Boolean formula to an {\bf SMT formula} such as the following:
\begin{align}
    \label{eq:SMTformulaExample}
    &F_{\text{SMT}} = \nonumber \\
    &\left(X_1 + (X_1\oplus X_2)\land \left(X_1+X_2 +(X_3\neq 0) + X_4 \leq 1\right)\right) \leq 3.
\end{align}
These SMT formulas support variables and clauses over larger non-Boolean domains, such as the integers. They also support operators such as integer arithmetic ($+,-,\times,\div$) and comparison ($=,\neq, \leq, \geq$) along with the Boolean operators above. The {\it type} of an SMT expression is determined by the topmost operator in the parse tree; for example the formula in \cref{eq:SMTformulaExample} will evaluate to a Boolean due to the comparison operator.

An {\bf SMT decision problem} is an SMT formula such as $F_{\text{SMT}}$ which is of Boolean type (i.e., evaluates to a Boolean $\in \{0,1\}$). An {\it SMT solver}, such as Z3 \cite{de2008z3}, is a software program that uses heuristic strategies to find either a satisfying assignment of values to all of the variables $X_i$, such that $F_{\text{SMT}}(\{X_i\})$ evaluates to 1, or a formal proof that no such assignment exists.

SMT solvers exhibit good performance for a wide range of problems from program verification to network engineering \cite{barrett2018satisfiability,de2011satisfiability,backes2019reachability}. This performance improves each year as measured by competitions \cite{WCDHNR19,balyo2017sat}.
SMT solvers have only recently been applied to quantum circuit synthesis, gate scheduling, and qubit routing \cite{tan2020optimal,ding2019exact,murali2019formal,murali2019noise}. This work uses the bit-matrix representation of Clifford operations to efficiently encode whole-circuit design problems subject to fault-tolerance constraints into SMT decision problems. As we will explain in the subsequent sections, this key technique enables synthesis of large fault-tolerant circuits from scratch to implement nontrivial Clifford operations while maintaining compatibility with 2D hardware.

\subsubsection{Clifford Group}
The Clifford group on $n$ qubits $G$ (modulo a global phase $U(1)$) is isomorphic to the binary symplectic group $\Sp(2n, \mathbb{F}_2)$ whose elements may be considered matrices in $\F_2^{2n\times 2n}$ preserving the symplectic inner product. In the context of quantum computation, these matrices can be thought of as acting on a {\bf bit-vector representation} $\vec{x}\in \F_2^{2n}$ of a Pauli group stabilizer $\prod_{i=1}^{n}{X_i^{x_i}Z_i^{x_{n+i}}}$ of a quantum state, modulo the unimportant global phase.
For example, the $\CNOT$ gate acts on the basis $\{X_1, X_2, Z_1, Z_2\}$ of the vector space over $\F_2$ of the Pauli group (modulo phase) on two qubits, as follows:
\begin{align}
\Qcircuit @C=1em @R=.7em {
& \gate{X} & \ctrl{1} & \qw \\
& \qw      & \targ    & \qw
} \,\, &= \,\, \Qcircuit @C=1em @R=.7em {
& \qw & \ctrl{1} & \gate{X} \\
& \qw & \targ    & \gate{X}
} \label{eq:pauliX1PropagationCNOT} \\
\Qcircuit @C=1em @R=.7em {
& \qw & \ctrl{1} & \qw \\
& \gate{X}      & \targ    & \qw
} \,\, &= \,\, \Qcircuit @C=1em @R=.7em {
& \qw & \ctrl{1} & \qw \\
& \qw & \targ    & \gate{X}
} \label{eq:pauliX2PropagationCNOT} \\
\Qcircuit @C=1em @R=.7em {
& \gate{Z} & \ctrl{1} & \qw \\
& \qw      & \targ    & \qw
} \,\, &= \,\, \Qcircuit @C=1em @R=.7em {
& \qw & \ctrl{1} & \gate{Z} \\
& \qw & \targ    & \qw
} \label{eq:pauliZ1PropagationCNOT} \\
\Qcircuit @C=1em @R=.7em {
& \qw & \ctrl{1} & \qw \\
& \gate{Z}      & \targ    & \qw
} \,\, &= \,\, \Qcircuit @C=1em @R=.7em {
& \qw & \ctrl{1} & \gate{Z} \\
& \qw & \targ    & \gate{Z}
} \label{eq:pauliZ2PropagationCNOT}
\end{align}

Therefore, the binary matrix representing the $\text{CNOT}$ gate is
\begin{align}
\overline{\CNOT} = \left(\begin{array}{rrrr}
1 & 0 & 0 & 0 \\
1 & 1 & 0 & 0 \\
0 & 0 & 1 & 1 \\
0 & 0 & 0 & 1
\end{array}\right).
\end{align}
We call this matrix the {\bf bit-matrix representation} of the CNOT gate and denote it by $\overline{\CNOT}$.
Given a Pauli operator with bit-vector $\vec{e}$ acting on the input state to a Clifford circuit $C$, we can propagate the operator through the circuit as $\vec{e}' = \overline{C} \vec{e}$. That is, we left-multiply the bit-vector by the bit-matrix representation of the Clifford operation implemented by the circuit.

It will be helpful to define the {\bf reduced bit-matrices} $\overline{\CNOT}|_X$and $\overline{\CNOT}|_Z$ which are
$$\overline{\CNOT}|_X = \left(\begin{array}{rr}
1 & 0 \\
1 & 1
\end{array}\right) = \overline{\CNOT}|_Z^T.$$
These matrices characterize the action of the Clifford operation, when one Pauli type ($X$ or $Z$) is disregarded.

\subsubsection{Product-Sum Relation}\label{sssection:ProductSumRelation}
Given a Clifford circuit $C$ consisting of a series of $g$ gates with bit-matrices $G_1, \ldots, G_g$, we now describe how we can find the bit-matrix representation $\overline{C}$ of the entire circuit $C$.
Naively, we could multiply all of the gates in time order, finding $C =\prod_{i=1}^{g}{\overline{G}_i}$. This requires $g$ matrix multiplications, where there are $g$ gates in the circuit.
In our approach to designing quantum circuits with SMT solvers, the bit-matrices of the gates (i.e. the $G_i$) will be {\bf symbolic matrices}, meaning that their entries may be SMT formulas of some abstract variables rather than Boolean constants $\in \{0, 1\}$.
It turns out that when the bit matrices $G_1,\ldots, G_g$ are symbolic, the naive approach of multiplying all gates in time order results in unwieldy formulas with an extremely large {\it formula size}, as we will explain below. So, we do not use this naive approach, and will later explain the {\bf product-sum relation} which allows us to keep the formula size small when we design large quantum circuits.

Informally, we can define the formula size of a formula $F$ as:
\begin{multline}
    \text{size}(F) = \#\{\text{variable occurrences in }F\} +\\
    \#\{\text{logical connectives in }F\}.
\end{multline}
The exact definition of formula size is not as important as the empirical fact that formulas with larger size require more memory and processing time to manipulate when constructing and solving the SMT decision problem with an SMT solver.
It is therefore of fundamental importance to keep the formula size as small as possible to be able to design large quantum circuits.
Note that we can multiply and add symbolic bit matrices modulo 2 using the SMT operators $\land, \oplus$. For example, we can multiply and add symbolic $2\times 2$ matrices as follows:
\begin{multline}
\left(\begin{array}{rr}
{\color{XRed}x^{(0)}_{00}} & {\color{XGreen}x^{(0)}_{01}}\\
{\color{XBlue}x^{(0)}_{10}} & {\color{XOrange}x^{(0)}_{11}}
\end{array}\right)\left(\begin{array}{rr}
{\color{XTurquoise}x^{(1)}_{00}} & {\color{XYellow}x^{(1)}_{01}}\\
{\color{XPink}x^{(1)}_{10}} & {\color{XPurple}x^{(1)}_{11}}
\end{array}\right) = \\
\left(\begin{array}{rr}
({\color{XRed}x^{(0)}_{00}}\land {\color{XTurquoise}x^{(1)}_{00}}\oplus {\color{XGreen}x^{(0)}_{01}} \land {\color{XPink}x^{(1)}_{10}}) & ({\color{XRed}x^{(0)}_{00}}\land {\color{XYellow}x^{(1)}_{01}}\oplus {\color{XGreen}x^{(0)}_{01}} \land {\color{XPurple}x^{(1)}_{11}})\\
({\color{XBlue}x^{(0)}_{10}} \land {\color{XTurquoise}x^{(1)}_{00}} \oplus {\color{XOrange}x^{(0)}_{11}}\land {\color{XPink}x^{(1)}_{10}}) & ({\color{XBlue}x^{(0)}_{10}}\land {\color{XYellow}x^{(1)}_{01}} \oplus {\color{XOrange}x^{(0)}_{11}}\land {\color{XPurple}x^{(1)}_{11}}) 
\end{array}\right) \\
\left(\begin{array}{rr}
{\color{XRed}x^{(0)}_{00}} & {\color{XGreen}x^{(0)}_{01}}\\
{\color{XBlue}x^{(0)}_{10}} & {\color{XOrange}x^{(0)}_{11}}
\end{array}\right)\oplus\left(\begin{array}{rr}
{\color{XTurquoise}x^{(1)}_{00}} & {\color{XYellow}x^{(1)}_{01}}\\
{\color{XPink}x^{(1)}_{10}} & {\color{XPurple}x^{(1)}_{11}}
\end{array}\right) = \\
\left(\begin{array}{rr}
{\color{XRed}x^{(0)}_{00}} \oplus {\color{XTurquoise}x^{(1)}_{00}} & {\color{XGreen}x^{(0)}_{01}} + {\color{XYellow}x^{(1)}_{01}}\\
{\color{XBlue}x^{(0)}_{10}} \oplus {\color{XPink}x^{(1)}_{10}} & {\color{XOrange}x^{(0)}_{11}} \oplus {\color{XPurple}x^{(1)}_{11}}
\end{array}\right)
\end{multline}
The $2\times 2$ symbolic matrices on the left side above  have entries with formula size $1$. Their product on the right has entries with size $7$, while their sum has entries with size $3$.
As illustrated by this example, addition of symbolic matrices results in less of a formula size increase than multiplication.

We will now introduce a technique which allows us to construct the bit-matrix for $C$ with only $N$ matrix multiplications, where $N$ is the number of time steps of $C$. The technique works by replacing many of the symbolic matrix multiplications with additions.
Specifically, we will add instead of multiplying certain matrices corresponding to gates that act simultaneously on disjoint sets of qubits. As we have just explained above, symbolic matrix addition incurs a smaller size increase than multiplication. This technique therefore decreases the resulting formula sizes and improves solver performance, since $N \ll g$.

Given a bit-matrix representation $\overline{G} \in \F_2^{2n\times 2n}$ of a Clifford gate $G$ acting on $n$ qubits, which acts trivially on the $\ell$th qubit, it can be shown that 
\begin{align}
\overline{G}_{ij} = \delta_{ij} \forall i,j \in \left(\{\ell, \ell+n\}\times [2n]\right) \cup \left([2n]\times \{\ell, \ell+n\}\right),
\end{align}
where $[n]:=\{1,\ldots,n\}$.
In other words, the matrix $\overline{G}$ must leave invariant all possible Pauli operators on the $\ell^\text{th}$ qubit.
We now define the following notation for a bit matrix $\overline{G}$:
\begin{align}
\Delta \overline{G} := \overline{G}\oplus I_{2n\times 2n}.
\end{align}
From the previous observation we can see that the matrix $\Delta \overline{G}$ is supported only in the combinatorial rectangle with rows and columns indexed in the set $S$ of qubits supporting the gate corresponding to $G$.
Therefore the product of two gates $G_1G_2$  simplifies when acting on disjoint supports:
\begin{align}\label{eq:disjointSupportDeltaProductZero}
\overline{G}_1 \overline{G}_2 &= (I \oplus (\overline{G}_1 \oplus I)) (I \oplus (\overline{G}_2 \oplus I)) \nonumber \\
&= I \oplus (\overline{G}_1 \oplus I) \oplus (\overline{G}_2 \oplus I) \oplus \underbrace{(\overline{G}_1 \oplus I)(\overline{G}_2 \oplus I)}_{=0} \nonumber \\
&= I \oplus \Delta \overline{G}_1 \oplus \Delta \overline{G}_2.
\end{align}
In general, for $m$ simultaneous Clifford gates $G_1, \ldots, G_m$ acting on pairwise disjoint sets of qubits, we can compute the composite bit-matrix of all these gates as
\begin{align}
\prod_{i=1}^{m}\overline{G}_i = I \oplus \bigoplus_{i=1}^{m}\Delta \overline{G}_i.
\label{eq:productSumRelation}
\end{align}
We refer to \cref{eq:productSumRelation} as the {\it product-sum relation}.

We now give a small example of the product-sum relation for the circuit on 3 qubits shown in \cref{eq:circuitExampleProductSum}.
\begin{equation}\label{eq:circuitExampleProductSum}
\Qcircuit @C=1em @R=1em {
\lstick{} & \ctrl{1} & \qw \\
\lstick{} & \targ    & \qw \\
\lstick{} & \gate{H} & \qw
}
\end{equation}
We refer to the $\CNOT$ gate on qubits 1 and 2 as $\CNOT_{1,2}$ and the $H$ (Hadamard) gate on qubit 3 as $H_3$.
The bit-matrices for these gates are as follows:
\begin{align}
\overline{\CNOT_{1,2}} &= \left(\begin{array}{rrrrrr}
1 & 0 & 0 & 0 & 0 & 0 \\
1 & 1 & 0 & 0 & 0 & 0 \\
0 & 0 & 1 & 0 & 0 & 0 \\
0 & 0 & 0 & 1 & 1 & 0 \\
0 & 0 & 0 & 0 & 1 & 0 \\
0 & 0 & 0 & 0 & 0 & 1
\end{array}\right)
\overline{H_3}&= \left(\begin{array}{rrrrrr}
1 & 0 & 0 & 0 & 0 & 0 \\
0 & 1 & 0 & 0 & 0 & 0 \\
0 & 0 & 0 & 0 & 0 & 1 \\
0 & 0 & 0 & 1 & 0 & 0 \\
0 & 0 & 0 & 0 & 1 & 0 \\
0 & 0 & 1 & 0 & 0 & 0
\end{array}\right).
\label{eq:bitMatricesSimpleExampleProductSum}
\end{align}
These follow from the elementary propagation rules in \cref{eq:pauliX1PropagationCNOT,eq:pauliX2PropagationCNOT,eq:pauliZ1PropagationCNOT,eq:pauliZ2PropagationCNOT} as well as the relations $HX = ZH$ and $XH = HZ$. From these we may easily check that
\begin{multline*}
(\overline{H_3} \oplus I)(\overline{\CNOT_{1,2}} \oplus I)\\
= \left(\begin{array}{rrrrrr}
0 & 0 & 0 & 0 & 0 & 0 \\
0 & 0 & 0 & 0 & 0 & 0 \\
0 & 0 & 1 & 0 & 0 & 1 \\
0 & 0 & 0 & 0 & 0 & 0 \\
0 & 0 & 0 & 0 & 0 & 0 \\
0 & 0 & 1 & 0 & 0 & 1
\end{array}\right)
\left(\begin{array}{rrrrrr}
0 & 0 & 0 & 0 & 0 & 0 \\
1 & 0 & 0 & 0 & 0 & 0 \\
0 & 0 & 0 & 0 & 0 & 0 \\
0 & 0 & 0 & 0 & 1 & 0 \\
0 & 0 & 0 & 0 & 0 & 0 \\
0 & 0 & 0 & 0 & 0 & 0
\end{array}\right)\\
=0_{6\times 6},
\end{multline*}
as required in \cref{eq:disjointSupportDeltaProductZero}. One can also verify that adding $I \oplus \Delta \overline{\CNOT_{1,2}} \oplus \Delta \overline{H_3} = \overline{\CNOT_{1,2}} \overline{H_3}$ as per the product-sum relation given in \cref{eq:productSumRelation}.

The product-sum relation is used in Section~\ref{sec:symbolicBitMatrix} to reduce the number of costly symbolic bit-matrix multiplications in the construction of the SMT decision problem from scaling with $g$ (the number of gates) to scaling with $N$ (the number of timesteps). For shallow quantum circuits on many qubits, $N \ll g$, resulting in substantial savings on formula size.
The product of the $N$ symbolic matrices of dimensions $2n\times 2n$ can then be computed as a symbolic matrix whose entries have formula size $O(Nn^3)$.

\subsection{Solving for Clifford Circuits}
\label{sec:solving}
The reader may notice many familiar notions, which in this section are re-formalized as SMT formulas to enable the precise characterization of the Clifford circuit design problem as an SMT decision problem.
\subsubsection{Gate-Time Encoding}
To encode a Clifford circuit design problem as an SMT decision problem (as defined in \cref{sec:SMTDecisionProblems}), we require a format for encoding an arbitrary circuit supported by the hardware in terms of some Boolean variables $\{X_i\}$.
This encoding should be efficient and easy to implement in the SMT solver software.
We use the {\bf gate-time encoding} of a circuit, which we define as follows.
Suppose our quantum computer has $n$ qubits and that it supports $w$ distinct fundamental gate operations $\{G_1, \ldots, G_w\}$.
Finally, suppose that we wish to encode a circuit with at most $N$ time steps.
Then we encode the circuit by $wN$ symbolic Boolean variables indexed as $X_{ij}$ where $i\in [w]$ and $j\in [N]$.
By convention, the gate $G_i$ is applied at time step $j$ if and only if $X_{ij} = 1$.
The Boolean values $X_{ij}$ then specify an arbitrary-depth $N$ circuit $C_X$ consisting of $G_i$ gates.

For example, consider the layout of three qubits labeled $1,2,3$ on the planar graph below. We can imagine that this corresponds to a physical device on a 2D surface, where the qubits have nearest-neighbor interactions shown by the graph edges. More specifically, for any pair of qubits connected by an edge, we can implement any $\CNOT$ gate on that pair. Suppose also that we can implement any single qubit Pauli $X, Y, Z$, Hadamard $H$, or phase $S$ gate. Then we would have $w = 6 + 3 + 3 + 3 + 3 + 3 = 21$ distinct gates $G_1, \ldots, G_w$, and our gate set would be as follows:
\begin{center}
\begin{tikzpicture}
\begin{scope}
    \node[draw,circle] (a) at (0,0) {$1$};
    \node[draw,circle] (b) at (1,0) {$2$};
    \node[draw,circle] (c) at (0.5,.78) {$3$};
    \draw[-] (a) -- (b);
    \draw[-] (a) -- (c);
    \draw[-] (b) -- (c);
\end{scope}
\begin{scope}[xshift=120,yshift=10]
\node (Gs) at (0,0) {$\begin{aligned}
    \{G_1, \ldots, G_w\} &= \\
    \{ & \CNOT_{1,3}, \CNOT_{3,1},\\
    & \CNOT_{1,2}, \CNOT_{2,1},\\
    & \CNOT_{2,3}, \CNOT_{3,2},\\
    & X_1, X_2, X_3, Z_1, Z_2, Z_3,\\
    & Y_1, Y_2, Y_3, H_1, H_2, H_3,\\
    & S_1, S_2, S_3\}
    \end{aligned}$};
\end{scope}
\end{tikzpicture}
\end{center}
The product-sum relation then gives a symbolic expression for the bit-matrix of the Clifford operation $\overline{C}$ performed by the circuit $C$ in terms of Boolean variables $X_{ij}$ and the bit-matrices for the individual gates $\overline{G_i}$. This expression is given by
\begin{align}\label{eq:productSumFormulaCircuit}
    \overline{C} = \prod_{j=1}^{N}{\left (I \oplus \bigoplus_{i=1}^{w}{X_{ij} \Delta\overline{G_i}} \right )}.
\end{align}

At high level, our technique to construct an SMT decision problem is to construct a symbolic bit-matrix $P$ which is simply the bit-matrix $\overline{C}$ of the circuit $C$, but where the entries $\overline{C}_{ij}$ are now formulas involving the variables $X_{kl}$ that determine the circuit $C$, as well as some auxiliary variables $Y_{i}$ and the constants $0, 1$.
The auxiliary variables are used to avoid exponentially compounding increases in formula size when multiplying the $N$ timestep matrices, and will be explained in \cref{sec:symbolicBitMatrix}.

Supposing we have some Clifford operation $O$ whose bit-matrix is $\overline{O}$,
we can enforce that the circuit implements the desired Clifford operation simply by requiring that 
\begin{equation}
\label{eq:CequalsO}
    \overline{C}=\overline{O}.
\end{equation} In practice, it is easy to compute the bit-matrix $\overline{O}$ as long as a the operation $O$ is well-defined -- we can for example use a long, unoptimized circuit which implements $O$ non-fault-tolerantly to compute the bit-matrix $\overline{O}$. We give more details on how $\overline{C}$ is constructed in Section~\ref{sec:symbolicBitMatrix}.

Additional requirements on the circuit can be added by conjuncting this core equality with any number of additional constraints.
We begin in Section~\ref{sec:gateExclusion} by describing how some simple requirements related to the validity of the gate scheduling can be added to the SMT decision problem.
We explain how similar techniques are used to enforce geometric locality and enable joint co-design of hardware layout with low-level gate scheduling.
In Section~\ref{sec:faultPropagation} we explain how faults can be symbolically propagated through the symbolic bit-matrix and show an example application of this technique for fault-tolerant surface code syndrome extraction.
We generalize this technique in Section~\ref{sec:flagConstraints} to design $v$-flag fault tolerant circuits to implement desired clifford operations.

\subsubsection{Symbolic Bit-Matrix Construction}\label{sec:symbolicBitMatrix}
For each $i\in [w]$ we find the bit-matrix representation $\overline{G_i}\in \F^{2n\times 2n}$ of the $i$th gate. We then use the product-sum relation to express the symbolic bit-matrix for the circuit specified by $X_{ij}$, using \cref{eq:productSumFormulaCircuit}.

It will be helpful for Section~\ref{sec:faultPropagation} to define more generally a sequence of symbolic bit-matrices for the partial circuits consisting of the last $N-k$ steps of the circuit $C$:
\begin{equation}\label{eq:partialBitMatrices}
    \overline{C}^{(k)} := \prod_{j=k+1}^{N}{\left(I \oplus \bigoplus_{i=1}^{w}{X_{ij} \Delta\overline{G_i}}\right)}.
\end{equation}
Clearly, we have that $\overline{C} = \overline{C}^{(0)}$ and, furthermore, all of the matrix multiplications over $\F_2$ can be implemented with any SMT solver using the fundamental operations of multiplication (i.e., logical AND) and addition mod 2 (i.e., exclusive-or XOR) on the formulas constituting the symbolic matrix, as explained in \cref{sssection:ProductSumRelation}.

\begin{table}
\begin{center}
\begin{tabular}{|l|l|l|}
\hline
Arguments & And(arguments) & FastAnd(arguments) \\ \hline
$X, 0, 1, 1$       & `$X\land 0\land 1\land 1$'      & 0                           \\
$X, Y, 1, 1$       & `$X\land Y \land 1 \land 1$'    & $X\land Y$                  \\
$1, 1, 1, 1$       & `$1\land 1 \land 1 \land 1$'    & 1                           \\
\hline
\end{tabular}
\caption{Partial simplification by pooling constants in high-arity functions, here the $\text{FastAnd}$ function. In the table, $X$ and $Y$ are two Boolean variables.}\label{table:fastAndXorNot}
\end{center}
\end{table}

We will next explain two optimizations that we find dramatically reduce the formula size of the symbolic bit-matrices $\overline{C}^{(k)}$.  

First, we explain the use of auxiliary variables in symbolic matrix multiplication.
Suppose we are given three symbolic matrices $X,Y,Z$ whose dimensions are all $n\times n$ and whose entries all have formula size $1$, and we must compute their symbolic product $XYZ$.
Using the multiplication of two symbolic matrices as a primitive, we could compute their symbolic product matrix as follows (naive approach):
\begin{enumerate}
    \item Compute the symbolic matrix $YZ$, whose entries have formula size $\Theta(n)$.
    \item Compute the symbolic matrix $X(YZ)$, whose entries have formula size $\Theta(n^2)$.
\end{enumerate}
More generally, the formula size of the entries of the product of $N$ matrices using the naive approach is $\Theta(n^{N-1})$.
Rather than compute $XYZ$ directly, we can introduce the auxiliary symbolic matrix $W = (w_{ij})$ and proceed as follows (optimized approach):
\begin{enumerate}
    \item Compute the symbolic matrix $YZ$, whose entries have formula size $\Theta(n)$.
    \item Add the formula equivalent to $W=YZ$ to the SMT problem formula, increasing the formula size by $\Theta(n^3)$.
    \item Compute the symbolic matrix $XW$, whose entries have formula size $\Theta(n)$.
\end{enumerate}
More generally, the formula size of the entries of the product of $N$ matrices using the optimized approach remains $O(n)$, but we incur a cost of $O((N-2)n^3)$ in the problem size. However this additional cost is negligible compared to the scaling of the naive approach above.

Second, we explain how incremental simplification is used to optimize formula size when manipulating formulas with lots of constants. 
SMT solvers such as Z3 \cite{de2008z3} generally operate in two stages. In the first stage, the formula is constructed to be solved according to the user's wishes. In the second stage, the solver applies a heuristic set of approaches to simplify the formula, derive lemmas, and eventually solve the formula or prove that it is unsatisfiable.
Importantly, the formula given by the user in the first stage is left ``as-is" during the first stage -- even simple formulas such as the following parity expression:
$$0 \oplus 0 \oplus 0 \oplus \ldots \oplus 0 \oplus 1$$
are not reduced or simplified (e.g. in the above example, to the literal value 1) in any way. This leads to a huge slowdown in constructing the SMT problem as manipulating these large symbolic formulas has a much larger memory and time overhead than directly manipulating single bits.
The SMT solver's simplification routines may be manually triggered on the partial expressions that build up a formula, but this brings its own associated slowdowns since in many cases the solver spends time performing nontrivial simplifications.
Moreover, we have observed that a pre-simplified formula can take longer to solve, as it is generally better to leave decisions about nontrivial simplifications, substitution, etc. to the finely tuned heuristics of the solver at solving time rather than to manually force potentially nontrivial simplifications during problem construction.

To avoid the worst of the slowdowns when manipulating large symbolic bit-matrices, we implemented variadic subroutines $\text{FastOr}, \text{FastAnd}, \text{FastXor}$ which pool all constant arguments and only call the symbolic SMT Boolean functions when needed. An example showing how this reduces the formula size is shown in \cref{table:fastAndXorNot}.
We found that this optimization leads to a $\times 100$ speedup in the construction of the matrices $P^{(k)}$ and is thus extremely important for scaling up to large circuits involving dozens of qubits and time steps.
Here is a small example. When evaluating one term in the inner sum in \cref{eq:productSumFormulaCircuit} we must symbolically compute the product
$X_{ij}\Delta \overline{G}_{i}$, where we recall that $X_{ij}$ is a Boolean variable and $\Delta \overline{G}_{i}$ is a bit-matrix of (literal) Boolean values.
As a toy example, imagine that the matrix is equal to 
\begin{align}
    \Delta \overline{G}_{i} =\left(\begin{array}{rr}
0 & 1 \\
1 & 0
\end{array}\right), 
\end{align}
then performing the multiplication using FastAnd will give the result
\begin{align}
    X_{i,j}\Delta \overline{G}_{i} = \left(\begin{array}{rr}
0 & X_{ij} \\
X_{ij} & 0
\end{array}\right),
\end{align}
whereas using Z3's logical operators directly will give
\begin{align}
    X_{i,j}\Delta \overline{G}_{i} =  \left(\begin{array}{rr}
`X_{ij}\land 0' & `X_{ij}\land 1' \\
`X_{ij} \land 1' & `X_{ij} \land 0'
\end{array}\right).
\end{align}
For large matrices with many 0 entries this simplification has a dramatic effect.

Another simplification we can make pertains to the use of auxiliary variables in bit matrix construction. 

\subsubsection{Constraints on Valid Circuits}\label{sec:gateExclusion}
Valid circuits must generally satisfy some {\bf gate exclusion relations}. We describe these in terms of a set of SMT assertions $\{f_i\}$ that are conjuncted into the constraint satisfaction problem as $\bigwedge\limits_{i} f_i$.
In a typical setup there can be at most one gate acting on each qubit at any one time step. We can capture this as, for a time step $t\in [N]$ and a qubit $q\in [n]$,
\begin{equation}\label{eq:gateExclusionRelations}
f^{(t,q)} = \left(\sum_{i\in [w], q \in \text{supp}(G_i)} X_{i t} \leq 1\right),
\end{equation}
in which the sum is over the integers (i.e., not modulo 2). We then obtain the combined valid circuit constraints as $$\text{IsValidCircuit}(\{X_{ij}\}) = \bigwedge\limits_{t\in [N], q\in [n]} f^{(t,q)}.$$

In some cases we may modify the gate exclusion relations~\eqref{eq:gateExclusionRelations}.
Here are several examples:
\begin{itemize}
\item We may choose to represent particular gates by products of those in our gate set $G_1, \ldots, G_w$. For example, we may represent the Pauli $Y$ gate on a single qubit as acting with both the $X$ and $Z$ gates simultaneously. This would decrease the number of distinct gates $w$, making the circuit representation more efficient.
\item We may limit the number of gates of a particular type that are applied, because e.g. the noise rate or overhead cost may be especially high for that gate type.
\item We may enforce that few or no idling locations occur, as idling qubits add additional fault locations to the circuit.
\item We may limit the number of distinct long-range gates between distant qubits, in case we can only manage to implement a few such gates and wish to use them sparingly.
\item We may limit the number of distinct qubits with which any one qubit $q\in [n]$ interacts, so as to minimize the degree of the connectivity graph.
That is, we may enforce a condition such as
 \begin{align}
 \left(\sum_{i\in [w], q\in\text{supp}(G_i)}{\underbrace{\bigvee_{t\in [N]}{X_{it}} }_{\text{gate }i\text{ is used at least once}}}\right) \leq d,
 \end{align}
 in which $d$ is the desired maximum degree.
\item We may bind the gates used in one circuit design problem with the gates used in a different problem -- that is, we may co-design multiple protocols simultaneously, with a global degree or other gate constraint enforced jointly across all the protocols. For example, suppose that we are designing two protocols labeled $1$ and $2$ which share a common set of qubits $[n]$ and possible gates indexed by $1, \ldots, w$. Suppose that we wish for the protocols to be implementable on the same hardware, which is subject to a maximum degree connectivity of $d$. Then we can enforce a condition such as
\begin{align}
 \left(\sum_{i\in [w], q\in\text{supp}(G_i)}{
 \bigvee_{t\in [N]}{X^{(1)}_{it}} \lor 
 \bigvee_{t\in [N]}{X^{(2)}_{it}} 
} \right) \leq d,
\label{eq:jointDegreeConstraints}
 \end{align}
 in which the $\{X^{(k)}_{ij}\}_{i,j}$ are the gate-time encoding variables for protocol $k$.
 
\end{itemize}

In all of these cases we can construct the appropriate gate exclusion relations $f_i$, using standard techniques to encode them as SMT decision formulas.

\subsubsection{Topological fault-tolerance using symbolic fault-propagation}\label{sec:faultPropagation}

\begin{figure*}
    \centering
    \begin{tikzpicture}
        \node at (0,0) {\includegraphics[height=10em]{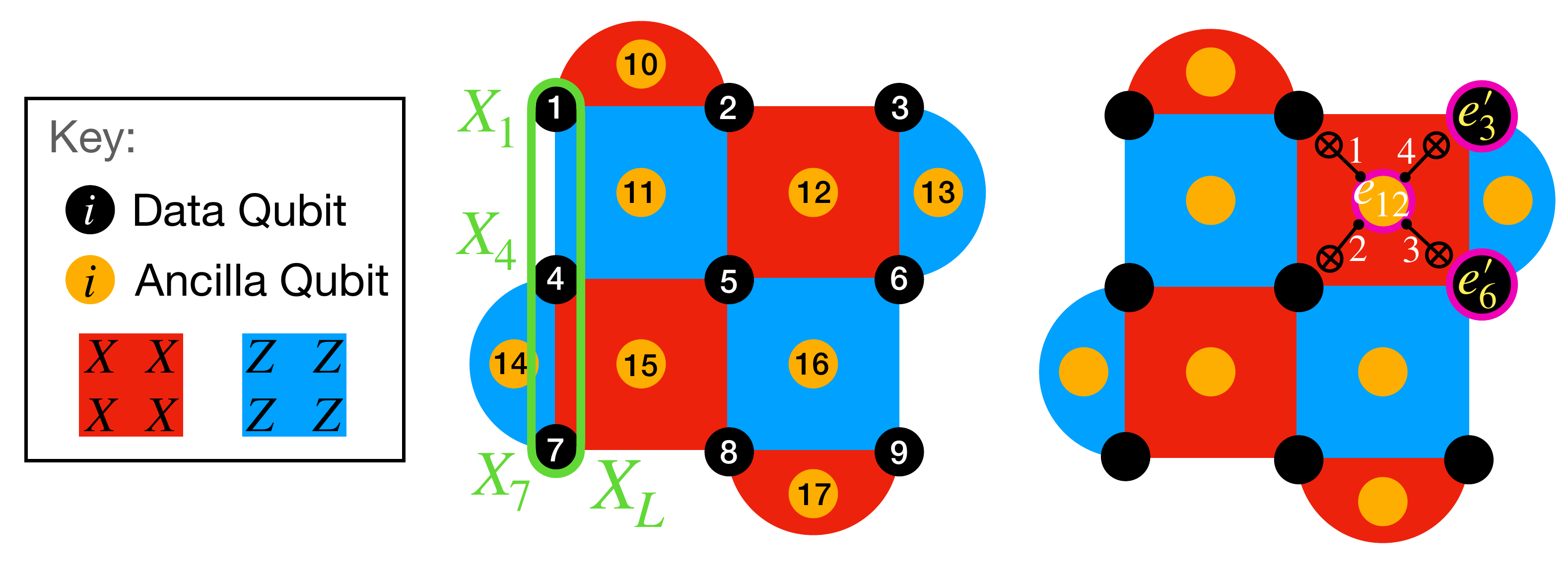}\includegraphics[height=10em]{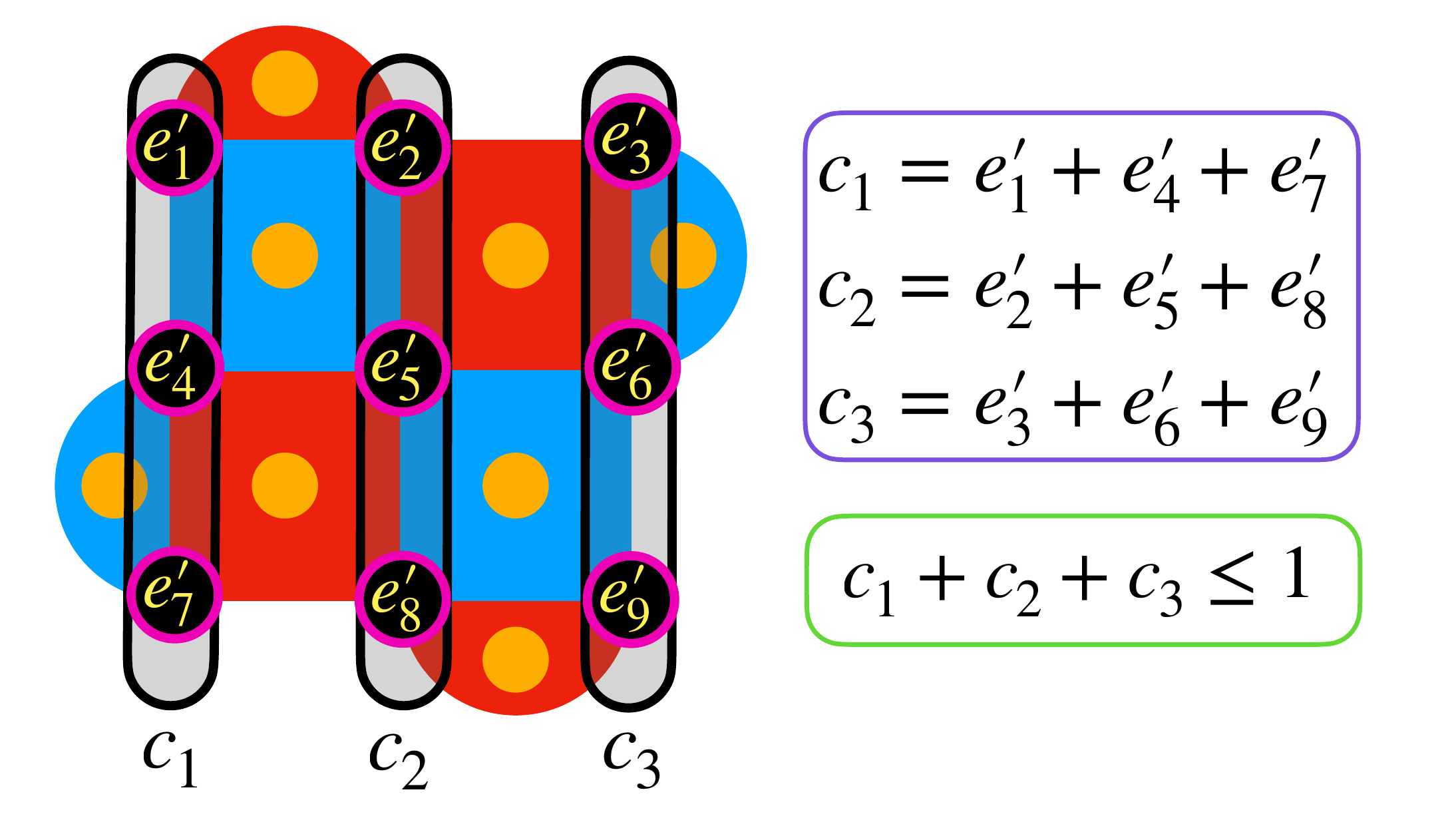}};
        \node at (-3.525,1.9) {(a)};
        \node at (0,     1.9) {(b)};
        \node at (3.525, 1.9) {(c)};
    \end{tikzpicture}
    \caption{ Illustration of how fault propagation constraints can be used to find a fault-tolerant gate scheduling for the surface code such that at most $(d-1)/2$ faults does not result in a logical error. (a) $3\times 3$ surface code with a vertical logical $X_L$ operator. (b) For a particular syndrome extraction circuit with the $\CNOT$ gates and time steps shown for one of the $X$-type stabilizers, a single fault resulting in an $X$ error $\vec{e}$ on qubit $12$ after the second time step propagates to a weight-two $X$ error on the data qubits 3 and 6 that is parallel to the logical $X_L$ operator. (c) Symbolically propagating the error arising from a fault at that spacetime location as $\vec{e}' = \overline{C}^{(2)}\vec{e}$, we can impose the fault propagation constraints (right) to ensure that the fault does not propagate parallel to a logical operator. By imposing this constraint for each possible fault location for $X$-type errors, we ensure that no single faults in the syndrome extraction circuit propagate parallel to a $X_L$ logical operator. Similar constraints can be written to prevent $Z$-type errors from propagating parallel to a $Z_L$ logical operator.}
    \label{fig:propagatePerpendicular}
\end{figure*}

We now introduce {\it symbolic fault propagation} and show how it can be applied to design fault-tolerant syndrome extraction circuits for topological codes such as the surface code.
Since arbitrary noise operators can be expressed as linear combinations of Pauli operators, we can model the noise as a distribution on Pauli operators at each spacetime location (i.e., idle, gate, state-preparation and measurement) in a given circuit.
Specifically, we describe a Pauli noise operator (ignoring global phases) as a column vector $\vec{e} \in \F_2^{2n}$ in which the first $n$ rows correspond to $X$ errors on the $n$ qubits, and the second $n$ rows correspond to $Z$ errors.

Suppose that a noise process occurs at time step $k$ resulting in some initial Pauli noise operator vector $\vec{e}$. We {\bf symbolically propagate} the error arising from the fault through the remainder of the circuit by computing the symbolic vector $\vec{e}' = \overline{C^{(k)}} \vec{e}$ (where the partial circuit bit-matrices are defined in \cref{eq:partialBitMatrices}).
We then impose constraints as needed on the final propagated error to ensure that the scheme is fault-tolerant up to the full code distance.
The combined fault propagation constraint is then:
$$\bigwedge_{\text{single fault }\vec{e}} \neg\text{NotFaultTolerant}(\vec{e}'),$$
in which $\text{NotFaultTolerant}$ 
is a Boolean formula $\text{NotFaultTolerant}(e_1', \ldots, e_{2n}')$ on $2n$ variables which decides whether the propagated error at the single fault location invalidates the desired fault-tolerance properties of the circuits. For example, the gate scheduling chosen for the syndrome extraction circuit of a topological code must satisfy the property that errors propagate perpendicularly to the appropriate logical operator. An example is provided in \cref{fig:propagatePerpendicular}.

In a noise model where two-qubit gates are afflicted by two-qubit Pauli errors, we may wish to enforce these constraints only for faults occurring on those two-qubit gates whose associated Boolean variables $X_{ij}$ are set to 1. In this case, we can amend the combined constraints as:
\begin{align}
\bigwedge_{\substack{\text{single fault }\vec{e} \\ \text{after timestep }k}} \left(\neg\text{NotFaultTolerant}(\vec{e}')\right)\lor\text{IsInvalidFault(X, \vec{e}, k)},
\label{eq:NotFaultTolIsInvalid}
\end{align}
in which $\text{IsInvalidFault}(X, \vec{e}, k)$ is a Boolean formula which decides whether the error $\vec{e}$ could have occurred at that spacetime location.

\subsubsection{Flag Fault-Tolerance}
\label{sec:flagConstraints}

Another interesting application of symbolic fault propagation is in the construction of $v$-flag circuits used to fault-tolerantly measure a given operator (for instance, a stabilizer of an error correcting code).
Following the definition introduced in \cite{CB18}, a $v$-flag circuit for measuring a stabilizer $g_i$ has the property where for any set of $t \le v$ faults resulting in the error $E$ such that $\text{min}(\text{wt}(E), \text{wt}(Eg_i)) > t$, at least one the flag qubits are measured nontrivially.
Here $\text{wt}(P)$ corresponds to the weight of an operator $P$.
More details can be found in Ref.~\cite{CB18}. In what follows, we say that a circuit \textbf{flagged} if at least one of the flag qubits is measured nontrivially.
For each qubit $i$ which is to be measured, we let $P_i\in \{X, Y, Z\}$ be the Pauli measurement basis.
As the $P_i$-basis measurement of qubit $i$ must deterministically give a trivial $+1$ measurement outcome in the absence of any faults, the initial stabilizers of the input state must be mapped by the desired Clifford
operation C to a set of stabilizer generators, which generate a stabilizer group that includes the stabilizer $P_i$ on the final state.

To symbolically verify whether a propagated error resulting from $\leq t$ faults causes the circuit to flag, it suffices to obtain the symbolic propagated error and to verify whether any flag qubit $i$ gives a nontrivial $P_i$-basis measurement outcome.
For instance, if qubit $i$ is measured in the $P_i=X$ basis, we verify that there is a $Z$ or $Y$ error on that qubit in the symbolic propagated error $\mathbf{e}'$.

We may allow the SMT solver to decide which qubits are to play the role of flag qubits.
In this case, to each possible flag qubit $i$ we add a Boolean variable $\text{IsFlag}_{i}$, where $\text{IsFlag}_i$ will be set to 1 if and only if qubit $i$ is a flag qubit in the final protocol.
Similarly we may allow the SMT solver to choose the measurement basis $P_i$ for each measured qubit $i$ by adding Boolean variables $\text{MeasuredInX}_i$ and $\text{MeasuredInZ}_i$, with the convention that the $Y$ basis is chosen if both are set to $1$, and with an added constraint that $\text{MeasuredInX}_i\lor\text{MeasuredInZ}_i=1$. For convenience, we may still refer to this encoded Pauli variable as $P_i$.
As we have just mentioned, the $P_i$ basis Pauli measurement outcome must deterministically give $+1$ when there are no faults, so the desired Clifford operation $C$ must be compatible with the choice of $P_i$ and $\text{IsFlag}_i$.
The solution that we use is to make $C$ itself depend on the setting of these variables, so that the circuit does not flag when there are no faults.

For each possible fault in the circuit we associate a tuple $(k, \vec{e}, \vec{e}', S)$ where $k\in [N]$ is the time step such that the fault occurs, $\vec{e}\in \F_2^{2n}$ is the vector representation of the Pauli noise operator resulting from the fault, $\vec{e}' = \overline{C^{(k)}}\vec{e}$ and $S$ is the stabilizer in whose measurement circuit the fault occurred.

We define several functions returning SMT formulas, as follows.
The \text{MinWt} function returns a formula which evaluates to the minimum integer weight of the propagated error resulting from a given set of $t$ faults when multiplied by the $\ell$ distinct stabilizers $\{Q_1, \ldots, Q_\ell\} = \{S_i : i \in [t]\}$ in whose measurement circuits the $t$ faults occurred\footnote{We make this distinction since multiple faults can occure within the same stabilizer measurement circuit.}:
\begin{align}\label{eq:minWeightFormula}
    \text{MinWt}&\left(\left\{\left(k_i, \vec{e}_i, \vec{e}_i', \overline{S}_i\right) : i \in [t]\right\}\right) \nonumber \\
    &= \min_{x\in \{0,1\}^{\ell}}{ \wt\left[ \left(\prod_{j=1}^{\ell}{\overline{S}_j^{x_j}}\right) \sum_{i=1}^{t}{\vec{e}_i'} \right]}.
\end{align}
Note that the integer $\min$ function can be implemented as an SMT formula using the comparison operators and the If-Then-Else operator, which are both supported \cite{de2008z3}.
The $\text{NontrivialOutcome}(i, P_i, \vec{e}')$ function returns a formula that evaluates to 1 if the $P_i$-basis measurement outcome of qubit $i$ gives a nontrivial -1 outcome in the presence of the error $\vec{e}'$. As the operation $C$ is guaranteed to give a +1 measurement outcome when there are no faults, this is easily computed as
\begin{align}
\label{eq:nontrivialOutcomeDef}
&\text{NontrivialOutcome}(i, \text{MeasuredInX}_i, \text{MeasuredInZ}_i, \vec{e}') = \nonumber \\
&(e_i'\land \text{MeasuredInZ}_i) \oplus (e_{n+i}' \land \text{MeasuredInX}_i ).
\end{align}
The IsFlagged function returns a formula which evaluates to a Boolean 1 value if and only if there is a flag qubit which gives a nontrivial measurement outcome
\begin{align}
    &\text{IsFlagged}\left(\vec{e}', \left\{\text{IsFlag}_1, \ldots, \text{IsFlag}_n\right\}\right)= \nonumber \\
    &\bigvee_{i=1}^{n}{\text{IsFlag}_i \land \text{NontrivialOutcome}(i, P_i, \vec{e}')},
\end{align}
where for brevity we have abbreviated the variables $\text{MeasuredInX}_i, \text{MeasuredInZ}_i$ as simply $P_i$.

The $\text{IsValidFaultSet}$ function returns a formula which evaluates to 1 if and only if the faults occur at a valid origin point:
\begin{align}\label{eq:isvalidfaultset}
\text{IsValidFaultSet}\left(\left\{\left(k_i, \vec{e}_i, \vec{e}_i', \overline{S}_i\right) : i \in [t]\right\}\right) = \nonumber \\
 \bigwedge_{i=1}^{t} \bigvee_{\substack{j=1 \\ \text{supp}(G_j)\supseteq \text{supp} \vec{e}_i}}^{w}{X_{i,j}},
\end{align}
where the support of a gate $\text{supp}(G)$ is the set of qubits on which it acts, and the support of a noise operator with bit-vector $\vec{e}\in \F_2^{2n}$ is simply $\text{supp}(\vec{e}) = \{i \in [n] : e_i\lor e_{n+i} \}$.
Note that $\vec{e}$ is not a free variable as it is known at the time of SMT decision problem creation. Therefore the $\text{IsValidFaultSet}$ function returns a small SMT formula since the logical OR in \cref{eq:isvalidfaultset} is efficiently implemented by the program which constructs the SMT formula, rather than symbolically encoded in the formula itself.

The $\text{IsNotTFlagFaultTolerant}$ function returns a formula which evaluates to 1 if and only if the passed error violates the $t-$flag property of the circuit. That is,
\begin{align}\label{eq:isNotTFlagdef}
    \text{IsNotTFlagFaultTolerant}\left( \left\{\left(k_i, \vec{e}_i, \vec{e}_i', \overline{S}_i\right) : i \in [t]\right\} \right) = \nonumber \\
    (\text{MinWt} > t) \land (\neg \text{IsFlagged}) \land \text{IsValidFaultSet},
\end{align}
where we have suppressed all arguments except $\text{IsNotTFlagFaultTolerant}$ for brevity.

To design a $v$-flag circuit, we construct an SMT problem with fault tolerance constraints $\text{IsVFlag}$ which returns a formula evaluating to $1$ if and only if the circuit is $v$-flag.
\begin{align}
\label{eq:isVFlagConstraint}
    \text{IsVFlag}(\{X_{ij}\}, \{P_i\}, \{\text{IsFlag}_i\}) = \nonumber \\
    \bigwedge_{t\leq v}\bigwedge_{\left\{\left(k_i, \vec{e}_i, \vec{e}_i', \overline{S}_i\right) : i \in [t]\right\}} \neg \text{IsNotTFlagFaultTolerant},
\end{align}
where we have again omitted the arguments for clarity, and it is understood that the conjunction is over all possible sets of $t\leq v$ errors which occur at $t$ distinct fault locations.

For Calderbank-Shor-Steane (CSS) code syndrome measurement circuits built from just $\CNOT$ gates, we may only concern ourselves with the propagating error type (e.g., the $X$ type errors when measuring the $X$ stabilizers).
For this purpose we can consider just the restricted bit matrices $P^{(k)}|_X$, saving a factor of 4 on the size of the symbolic bit-matrices $\overline{C}^{(k)}$.

\begin{figure*}
\begin{center}
\begin{tikzpicture}

\begin{scope}[xshift=0]
    \node[scale=0.8] (interactionGraphTitle) at (0.5, 1.4) {Qubit Interaction Graph};
    \node[draw,circle] (q2) at (0,0) {$2$};
    \node[draw,circle] (q3) at (1,0) {$3$};
    \node[draw,circle] (q1) at (0.5,.78) {$1$};
    \draw[thick,-] (q1) -- (q2);
    \draw[thick,-] (q1) -- (q3);
\end{scope}
\begin{scope}[xshift=120,yshift=10]

\node[color=DarkBlue] at (-3.7,-1) {(a) Design Objectives};
\draw[color=DarkBlue] (-5.5,-1.35) -- (-5.5,1.5) -- (5.9,1.5) -- (5.9,-1.35) -- cycle;

\node[above, yshift=23.5, scale=0.8] at (Gs) {Fundamental gate operations};
\node (Gs) at (0,0) {$\begin{aligned} &\,\,w= 2, N = 2\\
&\begin{cases}
    G_1 = \CNOT_{1,2}\\
    G_2 =\CNOT_{1,3}
    \end{cases}\end{aligned}$};
\node (BitMatrices) at (0,-3.2) {$\begin{cases}
\Delta\overline{G_1}|_{X} = \left(\begin{array}{rrr}
0 & 0 & 0 \\
1 & 0 & 0 \\
0 & 0 & 0
\end{array}\right) \\
\Delta\overline{G_2}|_{X} = \left(\begin{array}{rrr}
0 & 0 & 0 \\
0 & 0 & 0 \\
1 & 0 & 0
\end{array}\right)
\end{cases}$};
\draw[->] (Gs) -- (BitMatrices);

\draw[color=DarkRed] (-5.5,-1.45) -- (5.9,-1.45) -- (5.9,-4.9) -- (-5.5,-4.9) -- cycle;
\node[color=DarkRed, text width=4.5cm] at (-3,-4.2) {(b) Reduced Bit Matrix Encoding};

\begin{scope}[xshift=110]
\node (desiredOp) {\Qcircuit @C=1em @R=.7em {
& \ctrl{2} & \qw \\
& \targ & \qw \\
& \targ & \qw \
}};
\node[left,xshift=-20] at (desiredOp) {$O =$};
\node[above,yshift=23,scale=0.8] (desiredOpTitle) at (desiredOp) {Desired quantum operation};
\node[right,xshift=-44,yshift=-80] (Oreduced) at (desiredOp) {$\overline{O}|_X = \left(\begin{array}{rrr}
1 & 0 & 0 \\
1 & 1 & 0 \\
1 & 0 & 1
\end{array}\right)$};
\draw[->] (desiredOp) -- (Oreduced);
\end{scope}

\begin{scope}[xshift=-100]
\begin{scope}[yshift=-170,xshift=70]
\node at (2.6,.25) {$\overline{C}_X = I \oplus \underbrace{\left[X_{21}\Delta\overline{G_1} + X_{22}\Delta\overline{G_2} \right]}_{t=2}
\underbrace{\left[X_{11}\Delta\overline{G_1} + X_{12}\Delta\overline{G_2} \right]}_{t=1}
$};

\draw[color=DarkGreen] (-4.45,.9) -- (6.95,.9) -- (6.95,-.6) -- (-4.45,-.6) -- cycle;
\node[color=DarkGreen, text width=4cm] at (-2.2,0) {(c) Symbolic Bit Matrix of Circuit};

\node at (2.8,-1.15) {$\text{IsValidCircuit} = (X_{11} + X_{12} \leq 1)\land (X_{21} + X_{22} \leq 1)$};

\node[color=DarkYellow, text width=4.33cm] at (-2.05,-1.5) {(d) Circuit Validity \\ Constraint};
\draw[color=DarkYellow] (-4.45,-.75) -- (6.95,-.75) -- (6.95,-2) -- (-4.45,-2) -- cycle;

\node (Fsmt) at (-1.4,-2.5) {$F_{\text{SMT}} = (\overline{C}|_X = \overline{O}|_X)\land \text{IsValidCircuit}$};

\node[align=left,text width=5cm,left,yshift=20] (smtProb) at (7.,-4.15) {SMT Decision Problem: \\
Find $X_{11}, X_{12}, X_{21}, X_{22}\in \{0,1\}$ \\
such that $F_{\text{SMT}} =1$.};

\draw[->] (Fsmt) -- (smtProb);

\begin{scope}[yshift=-100]
\draw[color=DarkPurple] (-4.45,1.4) -- (6.95,1.4) -- (6.95,-.6) -- (-4.45,-.6) -- cycle;
\end{scope}

\node[color=DarkPurple] at (-2.45,-3.6) {(e) Problem Construction};

\end{scope}
\end{scope}
\end{scope}
\end{tikzpicture}
\end{center}
\caption{An example of our framework for encoding a quantum circuit design as an SMT decision problem, applied to a simple set of circuit design objectives.
(a) We start with some hardware layout on $3$ qubits with which we can implement 2 fundamental quantum gates $\{G_1, G_2\}$. We also have a Clifford circuit describing a quantum operation $O$.
We wish to implement $O$ in $N=2$ timesteps.
(b) We first encode the available gates $G_1, G_2$ and the operation $O$ as bit matrices.
Since we only have $\CNOT$ gates, both off diagonal blocks of all involved bit-matrices are automatically $0$, so it is sufficient to consider just the reduced bit matrices $\overline{G_1}, \overline{G_2}, \overline{O}|_X$.
(c) We use four Boolean variables $X_{11}, X_{12}, X_{21}, X_{22}$ to encode the circuit where $X_{ti} = 1$ only if gate $i$ is applied at timestep $t$.
We evaluate the symbolic reduced bit matrix $\overline{C}|_X$.
(d) We prepare the expression $\text{IsValidCircuit}$ which evaluates to 1 only if each qubit is acted on by at most one gate at each timestep.
(e) We then write the SMT formula $F_{\text{SMT}}$ so that it evaluates to $1$ only if the $\{X_{ti}\}$ variables encode a physically implementable circuit which implements the Clifford operation $O$.
We use an off-the-shelf SMT solver like Z3 \cite{de2008z3} to find a solution, or a proof that no solutions exist.
}\label{fig:smallExampleSingleStabilizerNoFlags}
\end{figure*}
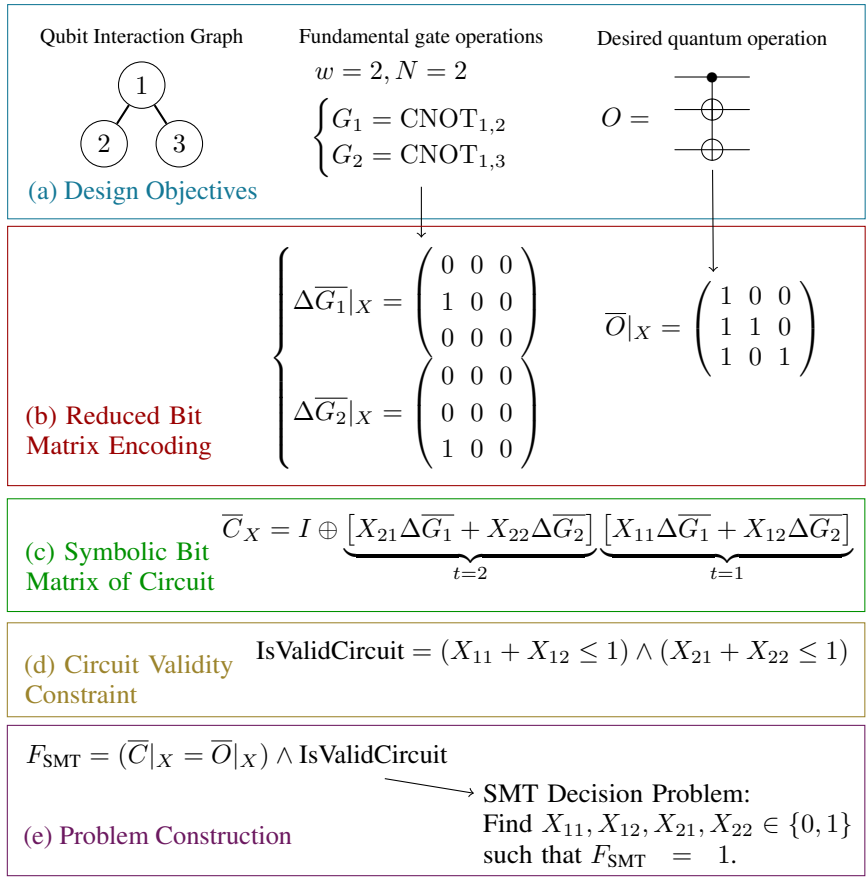

\cref{fig:smallExampleSingleStabilizerNoFlags} shows a simple example to illustrate the construction of the SMT formula to design a circuit to measure a two-qubit stabilizer with no flag qubits.

\subsection{Iterative Solving}
\label{susec:Iterative}
A circuit with depth $N$ on $n$ qubits with $w$ possible gates at each time step, has at most $N(w+q)$ distinct fault locations.
As explained in \cref{sec:flagConstraints} we can construct a SMT formula (\cref{eq:isVFlagConstraint}) which evaluates to 1 if and only if the found circuit is $v$-flag.
However, this formula will have size (number of clauses in the logical AND) that scales with the number of fault locations in the circuit.
Specifically, there are up to $\sum_{t=1}^{v}{N(w+q) \choose t}$ distinct fault combinations which give constraints in \cref{eq:isVFlagConstraint}.
This large number of possible fault combinations results in a large SMT problem which is difficult to construct and likely not possible to solve directly for circuits with thousands of fault locations.

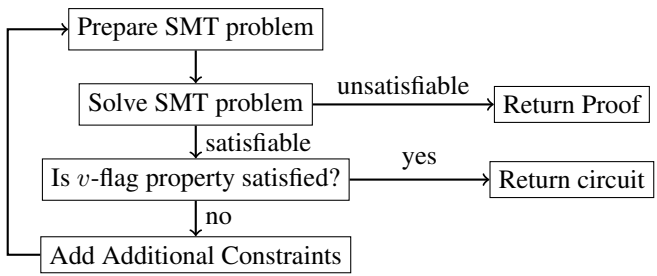
\begin{figure}
    \centering
    \begin{tikzpicture}
            \node[draw] (prepare) at (0,0) {Prepare SMT problem};
            \node[draw] (solve) at (0,-1) {Solve SMT problem};
            \node[draw] (unsat) at (5,-1) {Return Proof};
            \node[draw] (check) at (0,-2) {Is $v$-flag property satisfied?};
            \node[draw] (done) at (5,-2) {Return circuit};
            \node[draw] (add) at (0,-3) {Add Additional Constraints};

            \draw[->,thick] (prepare) -- (solve);
            \draw[->,thick] (solve) -- (check) node[midway,right] {satisfiable};
            \draw[->,thick] (solve) -- (unsat) node[midway,above] {unsatisfiable};
            \draw[->,thick] (check) -- (done) node[midway,above] {yes};
            \draw[->,thick] (check) -- (add) node[midway,right] {no};
            \draw[->,thick] (add) -- (-2.5,-3) -- (-2.5,0) -- (prepare);
    \end{tikzpicture}
    \caption{In iterative $v$-flag circuit solving, an initial SMT formula is constructed without flag constraints. After a solution is found, the conditions in \cref{eq:isVFlagConstraint} are checked. If any violated constraints are found, then some are added to the SMT formula and the problem is re-solved. The process terminates if either a $v$-flag circuit is found or a problem is proven to be unsatisfiable.}
    \label{fig:iterativeVFlagSolver}
\end{figure}

To circumvent this problem, we make use of an iterative approach as shown in Figure~\ref{fig:iterativeVFlagSolver}. We first construct the SMT problem instance $F_{\text{SMT}}$ as simply
\begin{equation}
\label{eq:noFlagConstraintsSMTproblem}
    F_{\text{SMT}} = \left(\overline{O} = \overline{C}\right),
\end{equation}
that is, without any fault-tolerance constraints.
We then check the $v$-flag property for all sets of $t\leq v$ errors occurring at $t$ distinct fault locations.
These sets of errors correspond precisely to the clauses in  \cref{eq:isVFlagConstraint}.
If any set $\{ (k_i, \vec{e}_i, \vec{e}_i', \overline{S}_i) : i \in [t]\}$ is found which violates the $v$-flag constraint, that is, such that
$$\text{IsNotTFlagFaultTolerant}\left(\{ (k_i, \vec{e}_i, \vec{e}_i', \overline{S}_i) : i \in [t]\}\right) = 1,$$
then we let
\begin{align}
    &\text{AdditionalConstraint} =\nonumber \\ &\neg\text{IsNotTFlagFaultTolerant} \left(\{ (k_i, \vec{e}_i, \vec{e}_i', \overline{S}_i) : i \in [t]\}\right),
\end{align}
and we then update the formula as
$$F_{\text{SMT}} \leftarrow F_{\text{SMT}} \land \text{AdditionalConstraint}.$$
We then ask the SMT solver to re-solve $F_{\text{SMT}}$. We repeat this process until either the problem is shown undecidable, or no fault combinations are present in the circuit which violate the $v$-flag property, as shown in \cref{fig:iterativeVFlagSolver}.

We expect two reasons that iterative solving works better than specifying all the constraints at the beginning of the protocol. The first is that many of the constraints are redundant, in a formal sense or a statistical sense. For example, two faults which occur late in the circuit at far away positions (as measured by the qubit connectivity graph distance) cannot possibly both flag the same flag qubit -- therefore as long as each of these faults flag {\it independently}, then the combination of both faults will also flag. Therefore the constraint $\neg \text{IsNotTFlagFaultTolerant}$ for the fault set containing these two faults is redundant in a formal sense with the constraints for each of the faults on their own.
Other sets of $t$ faults may have intersecting {\it lightcones} such that the constraint for the fault set is not formally redundant, and yet the majority of solutions to the problem do not violate that fault set's $t$-flag constraint $(\neg \text{IsNotTFlagFaultTolerant})$.

Interestingly, our numerical simulations have shown that iterative solving cuts runtimes to assemble the SMT problem instances dramatically, and enables scaling the approach beyond $1$-flag to $v$-flag (with $v \ge 2$) which simply is not possible with the naive approach of constructing the entire SMT problem up front.

\section{Fault-tolerant $\ket{H}$-type magic state preparation using SMT solvers}
\label{sec:FaultTolerantHtype}

\begin{figure*}
    \centering

    \begin{tikzpicture}[scale=0.65]
    
        \node[scale=1.4] (inputstate) at (-6,0) {$\ket{H_L}_{\text{n.f.}}$};
        \node[left,xshift=-30] at (inputstate) {(a)};
        \node[draw,scale=1.4] (H1) at (-3,0) {$H_m^{(d)}$};
        \node[draw,scale=1.4] (ED1) at (0,0) {$ED^{(d)}$};
        \node[scale=1.4] (ldots) at (3,0) {$\ldots$};
        \node[draw,scale=1.4] (Hlast) at (6,0) {$H_m^{(d)}$};
        \node[draw,scale=1.4] (EDlast) at (9, 0) {$ED^{(d)}$};
        \node[scale=1.4] (outputstate) at (12,0) {$\ket{H_L}_{\text{f.}}$};

        \draw [decorate,decoration={brace,amplitude=10pt,raise=20}]
(H1) -- (EDlast) node [black,midway,yshift=40,scale=1.05] {
$(d-1)/2$ times};
        \draw [->,thick] (inputstate) -- (H1);
        \draw [->,thick] (H1) -- (ED1);
        \draw [->,thick] (ED1) -- (ldots);
        \draw [->,thick] (ldots) -- (Hlast);
        \draw [->,thick] (Hlast) -- (EDlast);
        \draw [->,thick] (EDlast) -- (outputstate);
    \end{tikzpicture}
    
    \vspace{0.5cm}
    
    \begin{tikzpicture}
    \node (legend) at (0,-5.5) {\includegraphics[width=2.5cm]{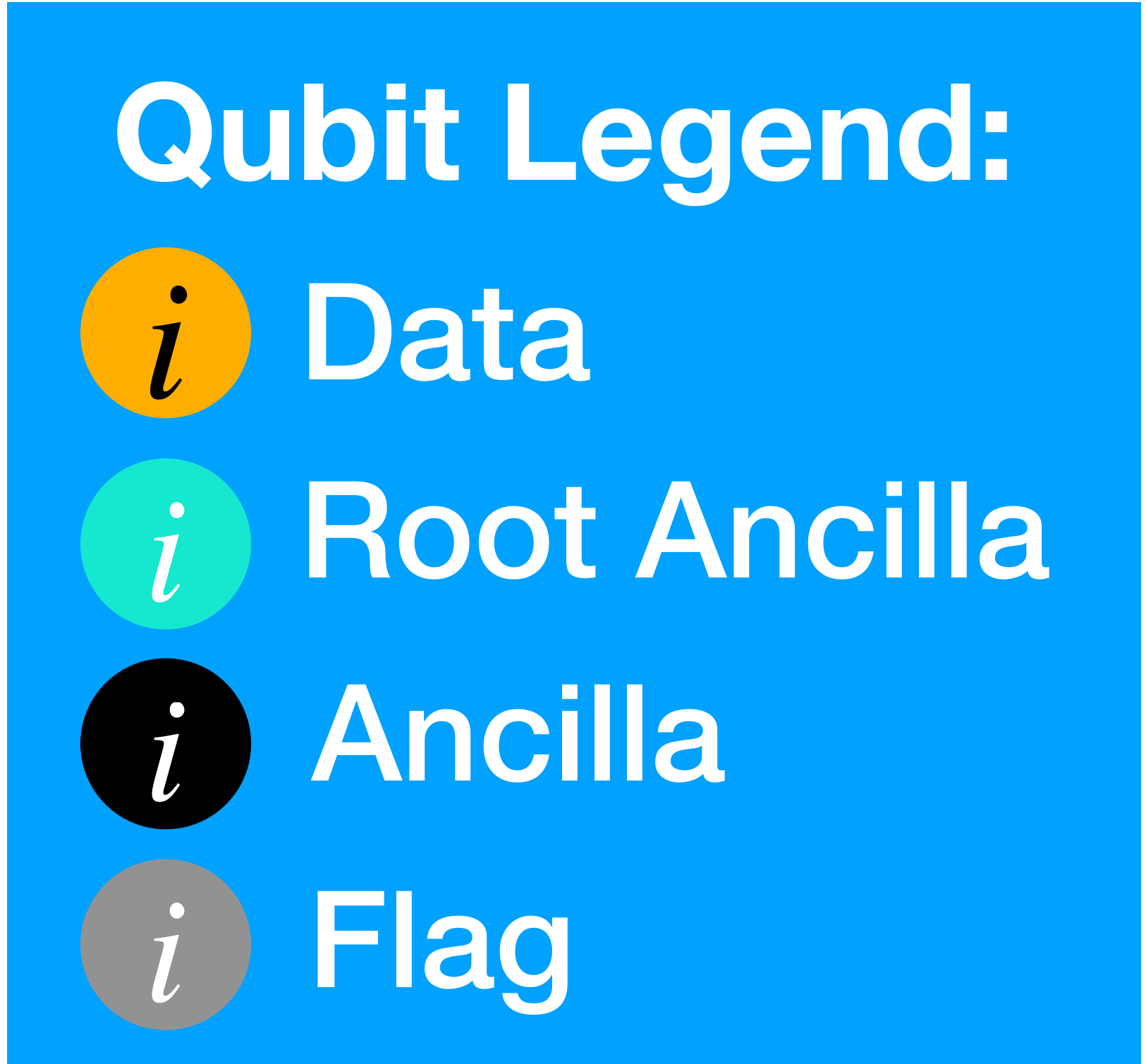}};
	\node [draw=red, thick] (GEO) at (0, 2) {\includegraphics[width=6cm]{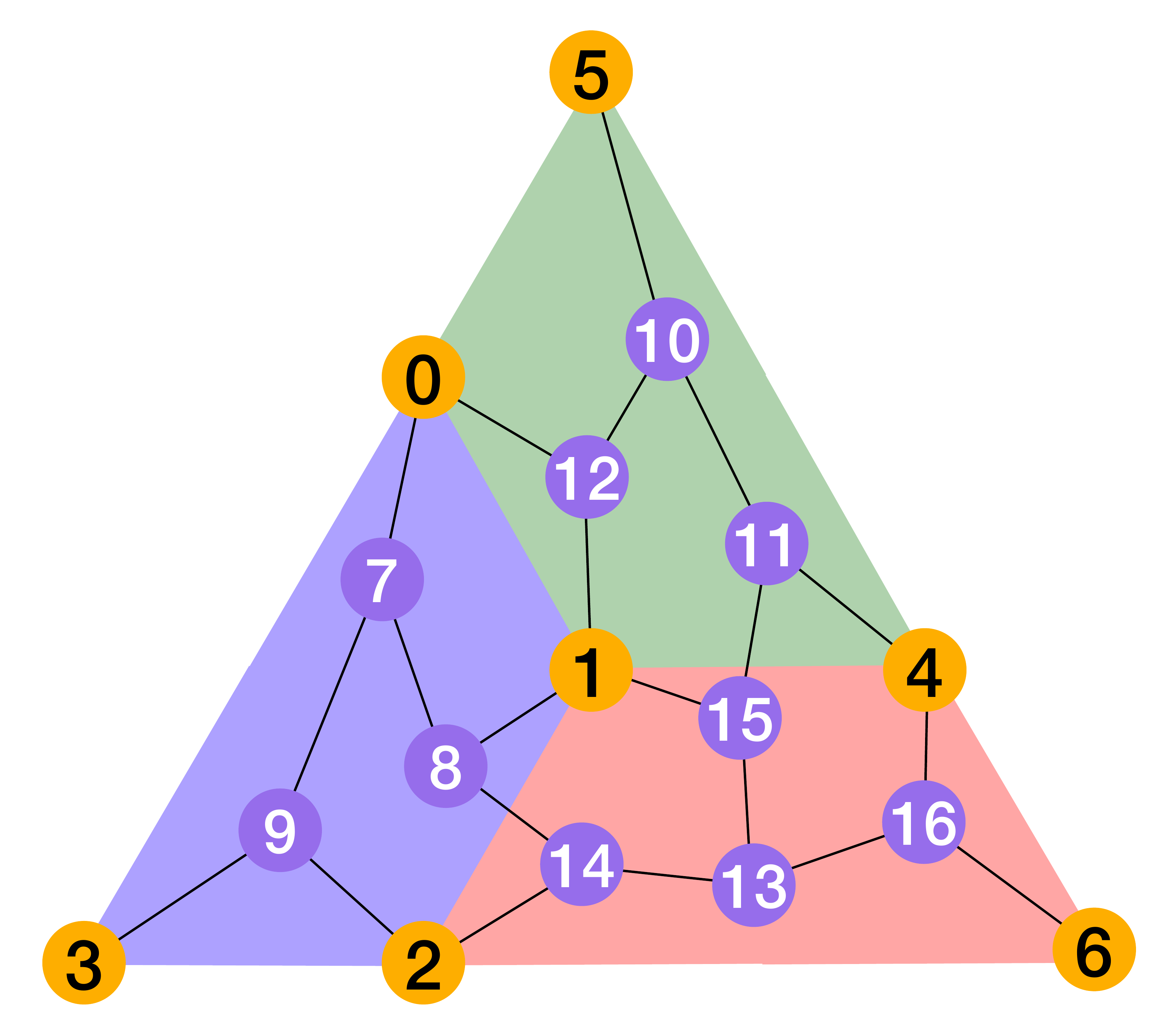}};
	\node[left,xshift=-181.5,yshift=50] at (GEO) {(b)};
	\node [yshift=85] at (GEO) {\bf Qubit Interaction Graph};
	\node[draw=nicepurple, thick] (gateset) at (0, -3) {\bf Gate Set $\{G_i\}$};
	\node[draw=blue, thick] (Z3) at (0, -4) {\bf SMT Problem};

	\node [draw=green, thick] (H) at (-5,-4) {\includegraphics[width=6cm]{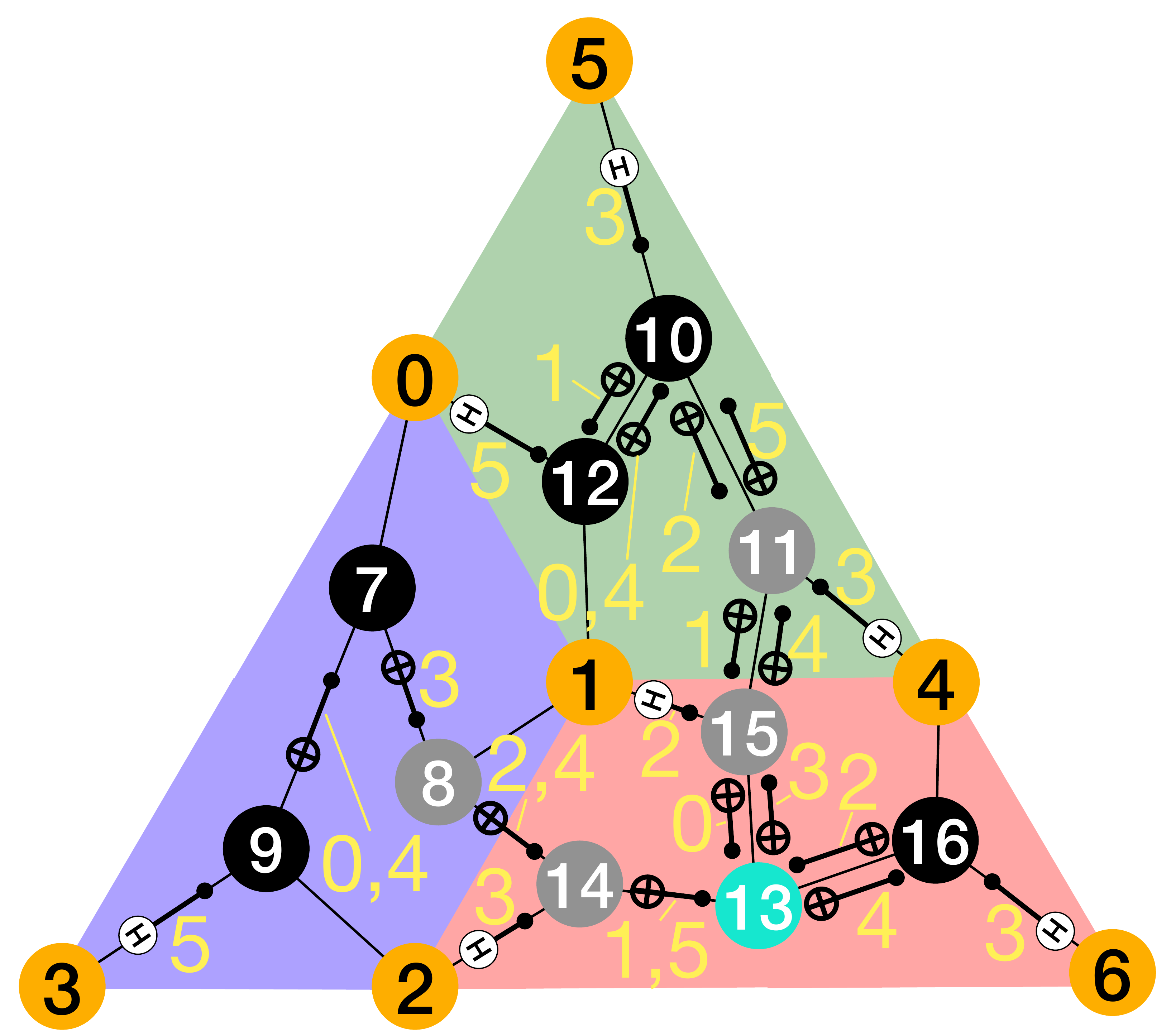}};
	\node [draw=green, thick] (S) at (5,-4) {\includegraphics[width=6cm]{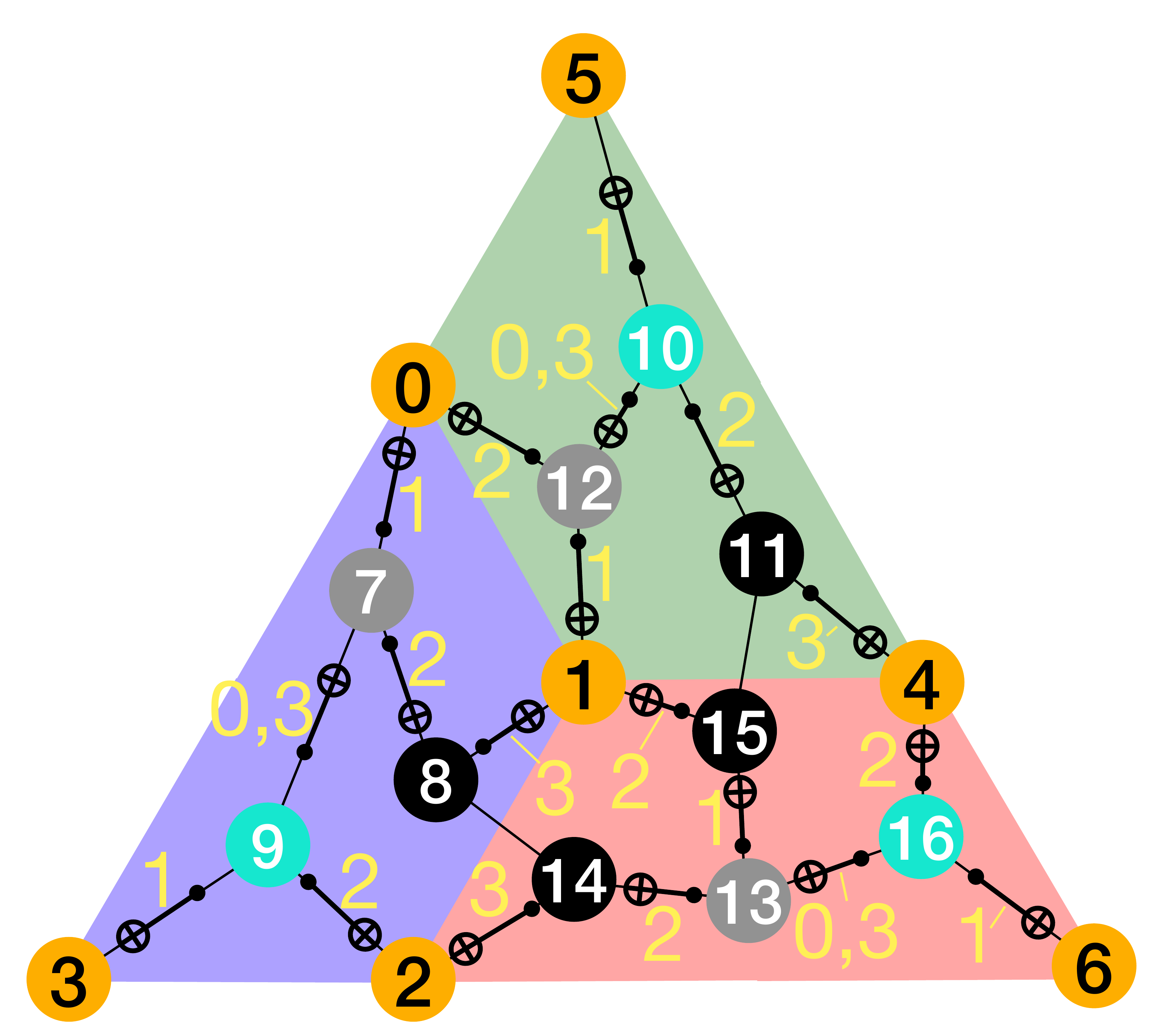}};

	\node[yshift=-90] at (H) {$H^{(d)}_m$};
	\node[yshift=-90] at (S) {$ED^{(d)}$};

	\draw[->,thick] (GEO) -- (gateset);
	\draw[->,thick] (gateset) -- (Z3);

	\draw[->,thick] (Z3) -- (H);
	\draw[->,thick] (Z3) -- (S);
    \end{tikzpicture}

    \caption{Co-designing fault-tolerant 1-flag protocols. In (a), the circuits $H_m^{(d)}$ are circuits for measuring the logical Hadamard $H_L = H^{\otimes n}$ of a distance $d$ color code. The circuits $ED^{(d)}$ correspond to one round of stabilizer measurements for a distance $d$ color code. Given a graph of qubit interactions and an abstract bit-matrix description of several desired quantum computations, the solver produces a protocol for each desired computation which is compatible with the interaction graph, satisfies joint degree and other gate constraints  (Section~\ref{sec:gateExclusion}), and is 1-flag (Section~\ref{sec:flagConstraints}). 
    \label{fig:geometryToProtocol}}
\end{figure*}
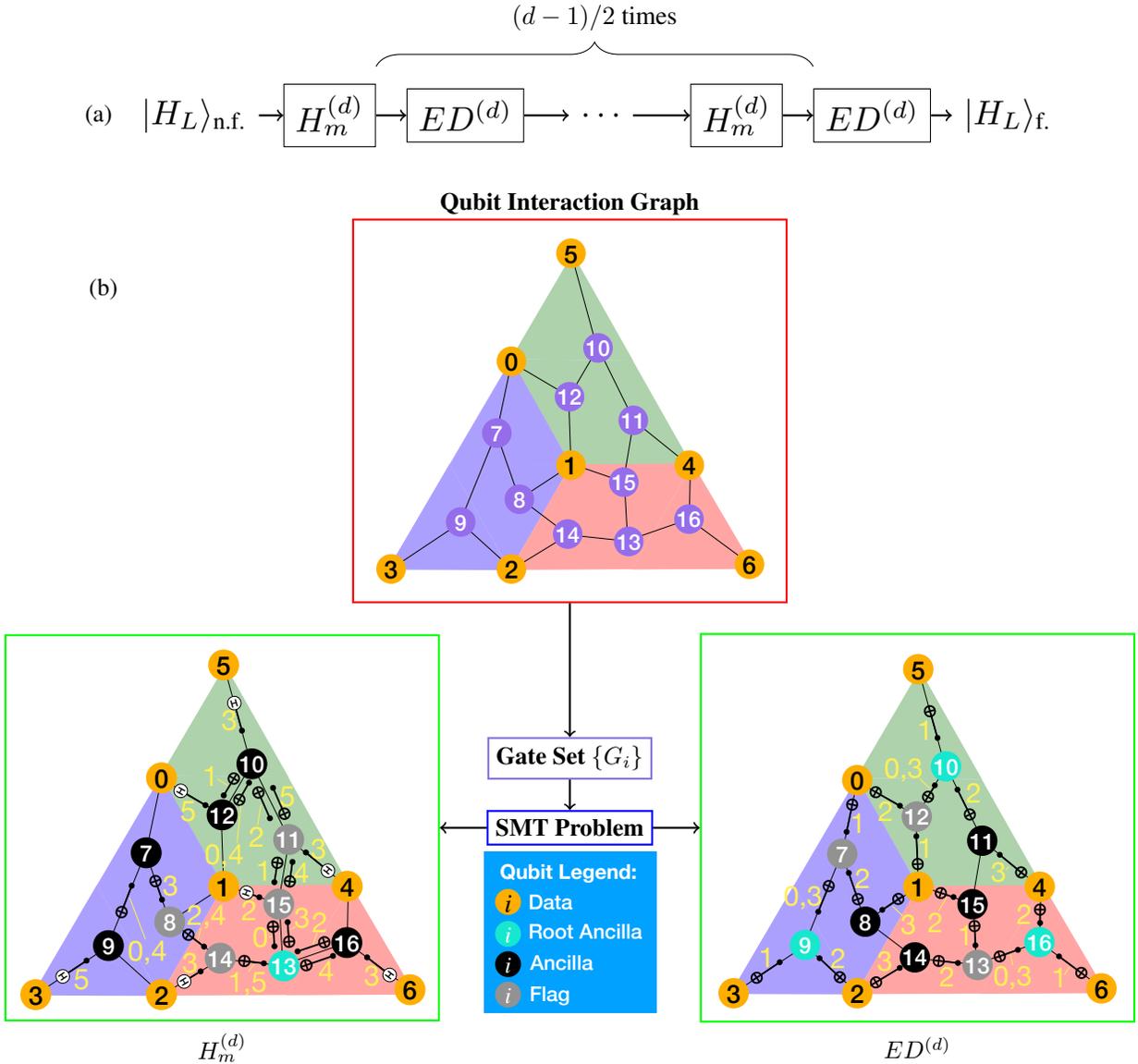

The leading approach to implementing quantum algorithms on a universal fault-tolerant quantum computer is to use magic state distillation (MSD) \cite{bravyi2005universal} in combination with lattice surgery techniques.
Alternative approaches to MSD for achieving universality, such as code-switching \cite{paetznick2013universal,anderson2014fault,Bombin2,bombin2016dimensional}, have been proposed.
Such alternative approaches are not always compatible with the 2D hardware constraints and have been shown to require larger resource overhead costs in their implementation \cite{litinski2019magic,chamberland2020building,beverland2021cost}.

$T$-type magic states (with $\ket{T} = (\ket{0} + e^{i \pi /4}\ket{1})/\sqrt{2}$) can be used as a resource state to fault-tolerantly implement logical $T$ gates. In Refs.\cite{fowler2012surface,gameofsurface}, $\ket{T}$ states  were prepared by first encoding several $\ket{T}$ states in distance $d=1$ surface codes using non-fault-tolerant methods and growing the codes to a final distance $d'\gg 1$. Afterwards, such states encoded in a distance $d'$ surface code were injected in a MSD protocol to distill them to a desired target logical failure rate determined by the size of the quantum algorithm being implemented.
Since the growing scheme is not fault-tolerant, a single fault (in the input $\ket{T}$ state) can result a logical error in the injected magic states prior to the implementation of the MSD protocol. Consequently, many rounds of MSD are required, where each operation is encoded in a large distance surface code, to generate high fidelity magic states such that they can be used in an algorithm.

An alternative proposal was put forth in \cite{chamberland2020very,chamberland2019fault}. In this approach, an encoded $\ket{H}$-type magic state (with $\ket{T} = e^{i \pi /8} H S^{\dagger}\ket{H}$ where $H$ and $S$ are Hadamard and phase gates) with code distance $d>1$ is directly prepared using a fault-tolerant protocol, meaning that any errors arising from at most $(d-1)/2$ faults cannot lead to a logical error. The protocols in \cite{chamberland2020very,chamberland2019fault} make use of the color code, which has the convenient property that the logical $\bar{H}$ gate is transversal (along with all other Clifford operations).
In this protocol, a physical ($d=1$) $\ket{H}$ state is grown using non-fault-tolerant methods to a $d>1$ encoded $\ket{\bar{H}}$ state. Subsequently, the grown state is injected in a bottom-up fault-tolerant magic state preparation protocol, where $(d-1)/2$ rounds of transversal logical Hadamard measurements and color code syndrome measurements are performed. If any of the syndrome or flag-qubit measurements during the state-preparation protocol are nontrivial, indicating the presence of at least one fault, the protocol is aborted and begins anew. An illustration for the sequence of such operations is shown in \cref{fig:geometryToProtocol}a. In \cite{chamberland2020very,chamberland2019fault}, the intermediate code distances $d\in \{3, 5, 7\}$ were analyzed, and despite additional space-time resource overhead costs due to rejection events, the scheme was shown to reduce overhead costs compared to previous MSD schemes in order to achieve a desired logical failure rate. The main reason for the overhead reduction is due to the fault tolerant nature of the preparation circuits, which can be implemented using physical Clifford operations. If the prepared magic states still require higher fidelities before being used in a quantum algorithm, such states can be injected in a MSD protocol requiring only a small number of distillation rounds

Although bottom-up protocols have shown promising results in providing low cost methods for preparing high fidelity magic states, one of the main challenges stems from finding flag-based circuits with the desired fault-tolerance properties. For the case of preparing $\ket{H}$-type magic states encoded in a distance $d$ color code, one requires $v$-flag circuits for measuring the logical Hadamard $H_L$ and stabilizers of the color code (the circuits $H^{(d)}_m$ and $ED^{(d)}$) in \cref{fig:geometryToProtocol}a. Further, for many hardware architectures, the qubits in such flag-based circuits are constrained to be laid out on a two-dimensional plane where the qubits can only interact with nearest neighbors. Further, the degree of the interactions must remain low \cite{CZYHC20}.

In \cref{sec:SMTsolvers} we showed how SMT solvers can be used to find Clifford circuits with a set of desired properties. One application of the techniques shown in \cref{sec:flagConstraints} was in finding $v$-flag circuits for fault-tolerantly measuring stabilizers of an error correcting code. In this section, we show how the protocols introduced in \cref{sec:flagConstraints,susec:Iterative} can be used to construct $v$-flag $H^{(d)}_m$ and $ED^{(d)}$ circuits with nearest-neighbor and low degree connectivity constraints imposed by many quantum hardware architectures \cite{CZYHC20,chamberland2020building}. 

\subsection{Qubit Interaction Graph}
Recall that we label the $n$ qubits by the integers 1 through $n$, and denote this set $[n]=\{1, \ldots, n\}$.
We now designate a subset $A\subset [n]$ of qubits that are to be prepared in some Pauli eigenstates and measured at the end of the circuit.
In our case, for designing $ED^{(d)}$ and $H_m^{(d)}$ circuits, $A$ contains all flag qubits, ancilla qubits, and root ancilla qubits.
The root ancilla qubit is distinguished from the other ancillas since, for $H^{(d)}_m$ and $X$-type stabilizer circuits in $ED^{(d)}$, it is initialized in a $+1$ $X$ eigenstate, whereas the other ancillas are initialized in $+1$ $Z$ eigenstates.
We refer to these flag, ancilla, and root ancilla qubits as the $A$-qubits.
The data qubits are then $[n]\setminus A$.
Since our target hardware is a 2D device where only nearest-neighbor qubits can interact, we create a planar graph $G_{\text{qubits}} = ([n], E)$ in which the vertices correspond to qubits and the edges are between qubits that support two-qubit gates.
This graph is called the {\bf qubit interaction graph}.
An example of such a qubit interaction graph is shown in Figure~\ref{fig:geometryToProtocol} (b).
In our graph we ensure that no two data qubits share an edge, as two-qubit gates are prohibited between data qubits.
This graph is designed by hand to use a low degree of connectivity between all qubits and still be connected, such that in the absence of faults it should be possible to implement the desired circuit.
We then assemble our set of gates.
Because we know the Clifford operations $ED^{(d)}$ and $H_m^{(d)}$\footnote{The $H_m^{(d)}$ operation is only Clifford when the data qubits wires are conjugated by the $T$ gates as shown in \cite{chamberland2020very}. This means $H_m^{(d)}$ is technically non-Clifford, but the $T$ gate conjugation on the data qubits does not affect the fault-tolerance properties, so we are free to design the conjugated circuit using our techniques and then apply the $T$ conjugation to the resulting circuit.} can be implemented using only preparation and measurement in the Pauli $X$ and $Z$ bases and $\CNOT$ gates, our gate set $\{G_1, \ldots, G_w\}$ consists only of $\CNOT$ gates. 
Specifically, for each edge $(u, v)\in E$ where $u, v\in A$, we add $\CNOT_{u,v}$ and $\CNOT_{v,u}$ to our gate set $\{G_i\}$.
For edges $(u, v)\in E$ where one of $u, v$ is a data qubit ($u\notin A$ or $v\notin A$), we only add a $\CNOT$ from the non-data qubit $\in A$ to the data qubit.

Note that controlled-Hadamard gates are equivalent to $\CNOT$ gates up to conjugation by a single-qubit non-Clifford gate. These single-qubit corrections do not propagate errors and can be placed at the beginning and end of the $H^{(d)}_m$ circuit.
As such, when deriving fault-tolerant $H^{(d)}_m$ circuits, it suffices to treat all controlled-Hadamard gates as $\CNOT$ gates. Lastly, note that if an error correction scheme was being developed rather than an error detection scheme, the type of data qubit errors would matter and, as such, the type of two-qubit gate used for measuring the logical Hadamard operator would need to be considered. We leave such considerations to future work on error-correction schemes.

\subsection{Constructing SMT Formula Constraints for $ED^{(d)}$ and $H_m^{(d)}$ Circuit Synthesis}
Once the gate set $\{G_i\}$ is specified, we feed it as input to the functions described in \cref{sec:solving}.
These functions construct SMT formulas which are then used to construct the entire SMT decision problem given below in \cref{eq:bigEquationAllConstraintsHEDdesign}.
The $A$-qubits which are shown in purple in \cref{fig:geometryToProtocol} must each be assigned one of three roles: ancilla, root ancilla, or flag qubit.
The parity of the measurement outcomes of the root ancilla as well as all other ancillas which are not flag qubits is used to obtain the measurement outcome for the operator being measured. Specifically, the measured operator, which is either a code stabilizer for $ED^{(d)}$ or the logical Hadamard for $H^{(d)}_m$, has 0 or more ancilla qubits and exactly one root ancilla qubit associated with it. The measurement outcome for this measured operator is then encoded as the product of the measurement outcomes across all these ancillas (including the root) that are measured and interact with the root ancilla. We allow the solver to choose which qubits have the role of flag, ancilla, or root ancilla qubit by creating Boolean variables $\{\text{IsFlag}_i : i\in [A]\}$ and $\{\text{IsRoot}_i : i\in [A]\}$.
We choose the following encoding:
For a qubit $i$, if $i$ is a root ancilla qubit, then we have $\text{IsRoot}_i=1, \text{IsFlag}_i = 0$.
If $i$ is a non-root ancilla qubit, then we have $\text{IsRoot}_i=0, \text{IsFlag}_i = 0$. 
f $i$ is a flag qubit, then we have $\text{IsRoot}_i=0, \text{IsFlag}_i = 1$.
We create a formula $\text{IsValidRoleAssignment}$
which evaluates to 1 if and only if there is exactly one root ancilla in each code stabilizer for the syndrome measurement circuits (or for the entire protocol in the case of the $H_m^{(d)}$ circuit), and that the ancillas, root ancillas, and flag qubits are all distinct.

For an $X$-type stabilizer measurement and logical $H_L$ measurement, we know that the root ancilla is prepared in a $+1$ eigenstate of the $X$ operator. The non-root ancillas and flag qubits are prepared in a $+1$ eigenstate of the $Z$ operator. The (root and non-root) ancillas are measured in the $X$ basis. The flag qubits are measured in the $Z$ basis.
For a $Z$-type stabilizer measurement, the $X$ and $Z$ bases are all swapped, that is, we replace $X$ with $Z$ and $Z$ with $X$ in the preceding description. As such, we do not need to produce a solution for $Z$-type stabilizer measurements. 
Here, as in \cref{sec:flagConstraints} we refer to the measurement bases of each qubit by $\{P_i\}$, with the understanding that this denotes a pair of symbolic Boolean values $(\text{IsMeasuredInXBasis}_i, \text{IsMeasuredInZBasis}_i)$ as explained in \cref{sec:flagConstraints}. In our case, for each qubit, the preparation and measurement bases of the qubit are the same and we use $P_i$ to refer to this one basis.

We then set a number of time steps $N^{(H)}$ and $N^{(ED)}$ for each circuit to complete. Then we declare the gate-time encoding variables $\{X_{ij}^{(H)} : i\in [w], j \in [N^{(H)}]\}$ and $\{X_{ij}^{(ED)} : i\in [w], j \in [N^{(ED)}]\}$ for each protocol; these variables are used in separate SMT decision problems (one for each of $ED^{(d)}, H_m^{(d)}$), but we refer to these just as $X_{ij}$ when we are speaking about a generic protocol of the two. 
We construct the gate exclusion relations constraint for each circuit as explained in \cref{sec:gateExclusion}. We label these SMT formulas by $\text{GateExclusion}^{(H)}, \text{GateExclusion}^{(ED)}$. Since the construction of these formulas proceeds analogously for both protocols, we refer to these formulas generically as $\text{GateExclusion}$.

Although the qubit interaction graph shown in \cref{fig:geometryToProtocol} already has maximum degree $3$ as desired, we can also start with a higher degree graph and enforce a global degree connectivity constraint with variables $X_{ij}^{(H)}$ and $X_{ij}^{(ED)}$ in \cref{eq:jointDegreeConstraints}, as explained in \cref{sec:gateExclusion}.

We then construct the bit-matrix $\overline{O}$ that describes the Clifford operation we would like to implement, as used in \cref{eq:CequalsO}. Since we allow the SMT solver to choose which qubits are flag, ancilla, and root ancilla qubits, we must use a symbolic matrix to represent $\overline{O}$ as explained previously in \cref{sec:flagConstraints}.
Recall that we have variables $P_i$ for the preparation and measurement basis.
For each qubit $i\in A$ we construct a bit-vector $S_i$ which describes the initial stabilizer of the input state acting on this qubit.
As this depends on $P_i$ we must make this a symbolic bit-vector so that it represents the appropriate initial Pauli stabilizer depending on the role of the qubit.
To understand how to construct the symbolic desired bit-matrix $\overline{O}$, let us consider its four quadrants: the upper left quadrant corresponding to $X$ to $X$ propagation, the lower right quadrant corresponding to $Z$ to $Z$ propagation, and the off-diagonal quadrants.
The off-diagonal quadrants are set to 0.
Recall that we call the upper left quadrant $\overline{O}|_{X}$ and the lower right quadrant $\overline{O}|_Z$ reduced bit-matrices.
Due to the symmetry of $\CNOT$ gate propagation for $X$ and $Z$ type Paulis (as shown in \cref{eq:pauliX1PropagationCNOT,eq:pauliX2PropagationCNOT,eq:pauliZ1PropagationCNOT,eq:pauliZ2PropagationCNOT}), since our circuits are only composed of $\CNOT$ gates we have that $\overline{O}|_X = \overline{O}|_Z^T$. By this symmetry, we must therefore only specify one of the two quadrants, so we specify how the $\overline{O}|_X$ quadrant is constructed.
Further, we do not have to fully specify $\overline{O}|_X$, as the value of $(\overline{O}|_X)_{j,i}$ does not matter for any $i$ which is a non-root ancilla or a flag qubit with $j\neq i$. The reason that $(\overline{O}|_X)_{j,i}$ does not matter for such $j$ and $i$ is the particular input state we have chosen. These off-diagonal values  $(\overline{O}|_X)_{j,i}$ for non-root columns $i$ can be set arbitrarily by right-multiplying $\overline{O}$ by $\overline{\CNOT_{i,j}}$. These $\CNOT_{i,j}$ gates would have no effect on the input state since the non-root qubits are initialized in the $\ket{0}$ state. In more formal terms, we only specify the symbolic bit-matrix $\overline{O}$ up to right-multiplication by an arbitrary Clifford operation which we know stabilizes the incoming state anyway.
Therefore we just set these rows of the symbolic bit-vector for this column equal to a wildcard value $*$.

The remaining columns of $\overline{O}|_X$ are associated with the root ancillas of all stabilizers and the data qubits.
These columns must be constrained exactly.
In particular, the initial $X$ stabilizer on the root ancilla must propagate to an $X$ type Pauli operator supported on all ancillas (including the root) in its stabilizer, along with the data qubits included in that stabilizer.
Therefore for each $i\in A$, if $\text{IsRoot}_i=1$, we must have the $i^{\text{th}}$ column of $\overline{O}_{XX}$ set equal to the vector supported on all the root and non-root ancillas and data qubits of this stabilizer. This vector can be constructed symbolically using the $\text{IsRoot}_i, \text{IsFlag}_i$ variables to determine whether it is supported on an $A$-qubit.
Furthermore, for all the data qubits $i\in [n]\setminus A$, there should be no propagation of $X$ errors from the data qubits onto the $A$-qubits, as the data qubits are always the {\it target} of $\CNOT$ gates.
Therefore for each $i\in [n]\setminus A$, the $i$th column of $\overline{O}_{XX}$ should be 0 except for a 1 in the $i$th row.

We have now explained how all of the entries of $\overline{O}$ would be determined, up to right-multiplication by arbitrary Clifford stabilizers of the incoming state.
We do not have a particular setting of the qubit roles as these are determined by $\{\text{IsFlag}_i\}$ and $\{\text{IsRoot}_i\}$. We handle this by replacing the constraint \cref{eq:CequalsO}, with a SMT formula we label $\text{HasDesiredEffect}$. This is easy to construct by going column-by-column through the restricted bit-matrix columns of $\overline{C}|_X$. For the $i$th column, we construct a symbolic bit-vector which is determined symbolically by $\text{IsRoot}_i$ and $\text{IsFlag}_i$ as discussed above using standard If-Then-Else support from the SMT solver. We then set the non-wildcard rows of this symbolic bit-vector equal to the corresponding rows of the symbolic bit-matrix of the entire circuit as computed by the product-sum formula (\cref{eq:productSumFormulaCircuit}).

Finally we use the techniques of \cref{sec:flagConstraints} to add $v$-flag constraints to the SMT problem. Recall that we can produce the SMT formula $\text{IsVFlag}$ in its entirety using the function shown in \cref{eq:isVFlagConstraint}, and providing the appropriate $\{X_{ij}\}$, $\{P_i\}$, and $\{\text{IsRoot}_i\}$ variables for the relevant protocol as inputs.
The final SMT decision problem for designing the circuit is then given by
\begin{align}
    \label{eq:bigEquationAllConstraintsHEDdesign}
    F_{\text{SMT}} = \text{IsValidRoleAssignment} \land\text{GateExclusion}\nonumber \\
     \land \text{HasDesiredEffect} \land \text{IsVFlag}
\end{align}
This formula $F_{\text{SMT}}$ is then fed into an off-the-shelf SMT solver such as Z3 \cite{de2008z3}.
For large problem instances we apply the iterative solving techniques of \cref{susec:Iterative} to speed up the symbolic construction of $F_{\text{SMT}}$ and the time taken by the solver to find a solution.

We remark that using the methods described above, the circuit $H_m^{(3)}$ in \cref{fig:geometryToProtocol} has low degree and is obtained systematically. In Fig. 7 (a) of Ref.~\cite{chamberland2020very}, a similar alternative circuit to $H_m^{(3)}$ was obtained ``by hand" and is more highly structured. However, this alternative circuit has much higher degree (see \cref{table:ProtocolComparison}). SMT solvers offer a systematic alternative to hand-design when finding optimal low degree circuits conforming to realistic hardware constraints.
Such optimal solutions might have less apparent structure than those that are hand-designed.
Fortunately, SMT solvers can automatically generate proofs of optimality with respect to protocol depth, degree, etc.
For example, we used Z3 to prove that the protocols shown in \cref{fig:geometryToProtocol} have the minimum possible number of timesteps for a degree 3 qubit interaction graph given the geometric and other problem constraints.

\begin{table}[]
\begin{tabular}{l|ll|}
\cline{2-3}
                                                      & \multicolumn{1}{l|}{\textbf{\begin{tabular}[c]{@{}l@{}}Fig. 7 of\\ Ref.~\cite{chamberland2020very}\end{tabular}}}               & \multicolumn{1}{l|}{\textbf{\begin{tabular}[c]{@{}l@{}}\cref{fig:geometryToProtocol} of\\ This Work\end{tabular}}}                                      \\ \hline
\multicolumn{1}{|l|}{\textbf{Device Requirements}}   & \multicolumn{1}{l|}{\cellcolor[HTML]{FFFFFF}}          &     \\ 
\multicolumn{1}{|l|}{Connectivity Degree}             & \multicolumn{1}{l|}{\cellcolor[HTML]{FFFFFF}{\color[HTML]{9A0000} \textbf{6}}} & \cellcolor[HTML]{FFFFFF}{\color[HTML]{036400} \textbf{3}}  \\ 
\multicolumn{1}{|l|}{Number of ancilla qubits}   & \multicolumn{1}{l|}{\cellcolor[HTML]{FFFFFF}{\color[HTML]{036400} \textbf{9}}} & \cellcolor[HTML]{FFFFFF}{\color[HTML]{9A0000} \textbf{10}} \\ 
\multicolumn{1}{|l|}{\textbf{$H^{(d)}_m$ Circuit}}  & \multicolumn{1}{l|}{\cellcolor[HTML]{FFFFFF}}       &                                                       \\ 
\multicolumn{1}{|l|}{Timesteps}                       & \multicolumn{1}{l|}{\cellcolor[HTML]{FFFFFF}\textbf{8}}                        & \cellcolor[HTML]{FFFFFF}\textbf{8}                         \\ 
\multicolumn{1}{|l|}{Degree used}                     & \multicolumn{1}{l|}{\cellcolor[HTML]{FFFFFF}{\color[HTML]{9A0000} \textbf{5}}} & \cellcolor[HTML]{FFFFFF}{\color[HTML]{036400} \textbf{3}}  \\ 
\multicolumn{1}{|l|}{\textbf{$ED^{(d)}$ Circuit}} & \multicolumn{1}{l|}{\cellcolor[HTML]{FFFFFF}}            &                                                                                   \\ 
\multicolumn{1}{|l|}{Timesteps}                       & \multicolumn{1}{l|}{\cellcolor[HTML]{FFFFFF}{\color[HTML]{9A0000} \textbf{7}}} & \cellcolor[HTML]{FFFFFF}{\color[HTML]{036400} \textbf{6}}  \\ 
\multicolumn{1}{|l|}{Degree used}                     & \multicolumn{1}{l|}{\cellcolor[HTML]{FFFFFF}\textbf{3}}                        & \cellcolor[HTML]{FFFFFF}\textbf{3}                         \\ \hline
\end{tabular}
\caption{A comparison of protocols from Fig. 7 of Ref.~\cite{chamberland2020very} with corresponding protocols in this work shown in \cref{fig:geometryToProtocol}}\label{table:ProtocolComparison}
\end{table}

\subsection{Complexity of finding $v$-flag circuits using SMT solvers}
\label{subsec:complexity}

In the preceding sections we explained several optimizations for the encoding of a fault tolerant quantum circuit design problem into an SMT decision formula $F_{\text{SMT}}$.
These optimizations reduce the formula size significantly. For example, as explained in \cref{sssection:ProductSumRelation}, the product-sum relation allows us to build formulas with size scaling in the number of time steps $N$, rather than the (much larger) total number of gates $g$.
For the design of $v$-flag syndrome extraction circuits of a distance $d$ code, these optimizations can ensure that
$$\text{size}\left(F_{\text{SMT}}\right) = O\left(\poly(d)^v\right).$$
These formulas therefore have polynomial size whenever $v$ is a constant.
The naive brute-force algorithm to solve $F_{\text{SMT}}$ has runtime $O(2^{\text{size}(F_{\text{SMT})}})$.
SMT solvers cannot provide a better runtime guarantee, so our worst-case complexity is still
$$O\left(2^{\poly(d)^v}\right),$$
which is doubly-exponential if $v=\Omega(d)$.
However, despite the doubly-exponential runtime in the worst case, we found that performance is ``good enough" for typical problem instances such as small code distances relevant for near-term quantum devices. Our suite of optimizations allowed us to solve for stabilizer and $H_m^{(d)}$ measurement circuits for color code distances up to $d=7$ with reasonable runtime.
As a benchmark, we ran our solver for a 90 qubit system of a distance 5 code to co-design the following three circuits:
\begin{enumerate}
    \item A 1-flag $H_m^{(d)}$ circuit, with a $\CNOT$ depth of 14.
    \item A 1-flag $ED^{(d)}$ circuit, with a $\CNOT$ depth of 6.
    \item An additional 1-flag syndrome extraction circuit for the merged surface-color code described in \cref{sec:DecodingMerge}, with a $\CNOT$ depth of 9. Such 1-flag circuits are required due to the weight-6 stabilizers along the boundary of the surface code and color code. 
\end{enumerate}

This design problem has 44 data qubits and 46 ancilla qubits and uses distance $d=5$ codes. In what follows, the numerics were obtained by running the solver on specialized z1d.metal Amazon Web Services (AWS) EC2 instances, which provide an all-core sustained clock frequency of up to 4.0 GHz.
Our solver constructs and solves $F_{\text{SMT}}$ to find all three circuits in 1 h 14 min with a degree constraint limiting each qubit to interacting with at most 3 nearest-neighbor qubits. This time reduces to just 58 minutes when the degree constraint is loosened to 4.

The heuristics used by the SMT solver (Z3 \cite{de2008z3}) are extremely effective compared to the brute-force strategy of guessing every possible circuit. In the $H_m^{(d)}$ subproblem described above, there are 141 possible $\CNOT$ gates at each timestep, for a total of 1974 possible $\CNOT$ gates in the circuit. The number of possible circuits is therefore at least $2^{1974}$, which is significantly larger than the number of atoms in the known universe. As such, the heuristics of the solver are hugely beneficial. One possible reason for the effectiveness of SMT solvers on our SMT formulas is that all of our variables are over finite (Boolean) domains and we do not use {\it quantifiers} such as ``$\exists$" and ``$\forall$" . The problem of deciding whether an SMT formula containing such features is satisfiable is undecidable in general. To illustrate the challenge presented by such formulas, consider the SMT formula $(a^{100} + b^{100} = c^{100}) \land(a>1 \land b>1 \land c>1)$ where the variables $a,b,c$ range over the integers. This formula is unsatisfiable, as implied by Fermat's last theorem \cite{wiles1995modular}. Nonetheless, it would be extremely surprising if contemporary SMT solvers could generate a proof of the unsatisfiability of this formula in any reasonable amount of time.

We add that it is not always necessary to scale the code distance $d$ to obtain a family of schemes which is ultimately fault tolerant.
For example, the error detection scheme for $H_m^{(d)}$ would have an unacceptably high rejection rate for code distances $d \ge 9$ due to the very large number of fault locations. Therefore in practical settings, $\ket{H}$-type magic states would be prepared using our method for distances $d \le 7$. Such states would then be teleported to states encoded in the surface code (see \cref{sec:DecodingMerge} below). Afterwards, a magic state encoded in the surface code would be grown to a larger code distance using gauge fixing and lattice surgery methods. Lastly, a top-down magic state distillation protocol would be used to further increase the fidelity of the state. Therefore, to limit the cost of the top-down protocols, it is crucial that the injected magic states have the highest possible fidelities, warranting the use of sophisticated methodologies such as SMT solvers.

\section{Decoding Merged Surface Code and Color Code}
\label{sec:DecodingMerge}

\subsection{Motivation}

In \cref{sec:FaultTolerantHtype} we provided tools to find circuits that can be used to fault-tolerantly prepare $\ket{H}$-type magic states encoded in the color code. However, a limitation of such protocols is that the final magic states are encoded in the color code rather than the surface code. As previously discussed, the surface code offers much better performance for quantum memory and computation implemented via lattice surgery. In order to ensure that the prepared magic states can be used in a competitive scheme for universal fault-tolerant quantum computation, in this section we consider a protocol for teleporting states encoded in the color code into states encoded in the surface code. The protocol requires the use of lattice surgery techniques \cite{FowlerGidneyLatticeSurgery,Litinski2018latticesurgery,gameofsurface,ChamberlandLatticeSurgery,ChamberlandLatticeSurgery2,PsiQLattice}, where gauge fixing results in a merged surface/color code, after-which the codes are split up again to terminate the two-qubit teleportation step. Our approach is an adaptation of the technique proposed in \cite{nautrup2017fault}, although there are several key differences and extensions. Firstly, we do not require additional data qubits for the code obtained by merging the surface code and color code (which we refer to as the merged code). Secondly, we explicitly construct a decoder for the the merged code (an aspect which was not addressed in \cite{nautrup2017fault}). 

In Ref.~\cite{vasmer2021morphing}, the authors describe a decoding algorithm for ``hybrid color-toric" (HCT) codes constructed by local modifications to 2D color codes that create localized ball-like toric code regions. These HCT codes and their associated decoding algorithm differ from the merged color-surface code that we consider in the work (see \cref{algorithm:decoding}). In the HCT codes in Ref.~\cite{vasmer2021morphing} there is a correspondence between the vertices of the syndrome lattice of the HCT code and the vertices of the ``original" unmodified color code lattice. This allows their local lift procedure (see \cite{vasmer2021morphing}, Appendix G) to be described in terms of the standard local lift of the color code introduced by \cite{kubica2019efficient}, as it would be applied to the vertices of the original unmodified color code. In our hybrid code, there is no such correspondence, so the HCT local lift is not well-defined and we must define the lifting procedure from scratch. Furthermore, the HCT decoder would need to be modified in order to be used for codes with boundaries such as our code; the authors of \cite{vasmer2021morphing} suggest but do not explicitly describe how this would be done.

\begin{figure}
\begin{center}
$$\Qcircuit @C=1em @R=1em {
\lstick{\ket{\psi}} & \multigate{2}{M_{XX}} &     \gate{M_{Z}}  &  \cctrl{2}       &            & \\
                          &      \pureghost{M_{XX}}                          &     \cctrl{1}        &                       &            & \\
\lstick{\ket{0}}     & \ghost{M_{XX}}           &     \gate{Z}         &   \gate{X}        &       \qw     & \rstick{\ket{\psi}}
}$$
\caption{Quantum circuit for teleporting a logical state $\ket{\psi}$ from the color code to the surface code.}\label{fig:teleport}
\end{center}
\end{figure}
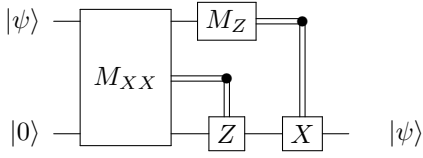

Figure~\ref{fig:teleport} depicts the teleportation circuit for transferring a state encoded in the color code to a state encoded in the surface code. To implement this circuit fault-tolerantly using the hardware constraints mentioned above, lattice surgery must be used to measure the two-qubit $X \otimes X$ operator. After measuring $X \otimes X$, repeated rounds of error correction must be performed on the merged code in order to prevent both timelike and spacelike errors from giving the wrong parity of the measurement outcome. As such, a decoder is required to process the stabilizer syndrome measurements of the merged code during the lattice surgery protocol. 

When performing gauge fixing to merge the color code and surface codes, the $X$-operator measurements at the boundaries of the two-codes anticommute with some $Z$ stabilizers of the original codes, causing them to merge into elongated stabilizers. The stabilizers of the unmerged codes, and the stabilizers of the merged code (including merging operators) are shown in Figure~\ref{fig:stabilizersMerge}. We also note that the theoretical framework of Ref.~\cite{LatticeColorCode} can be used to obtain the semitransparent domain walls (i.e. stabilizer generators between the surface code and color code) allowing one to perform lattice surgery between the color code and surface code.  

In Section~\ref{sec:notation}, we introduce some notation and provide some definitions. In Section~\ref{sec:decoder}, we provide the details of our decoding algorithm for the merged code.

\subsection{Notation and Definitions}\label{sec:notation}

\begin{figure*}
\begin{center}
    \begin{tikzpicture}            
            \node (xgraph) at (-2.6,3) {
                \includegraphics[width=0.4\linewidth]{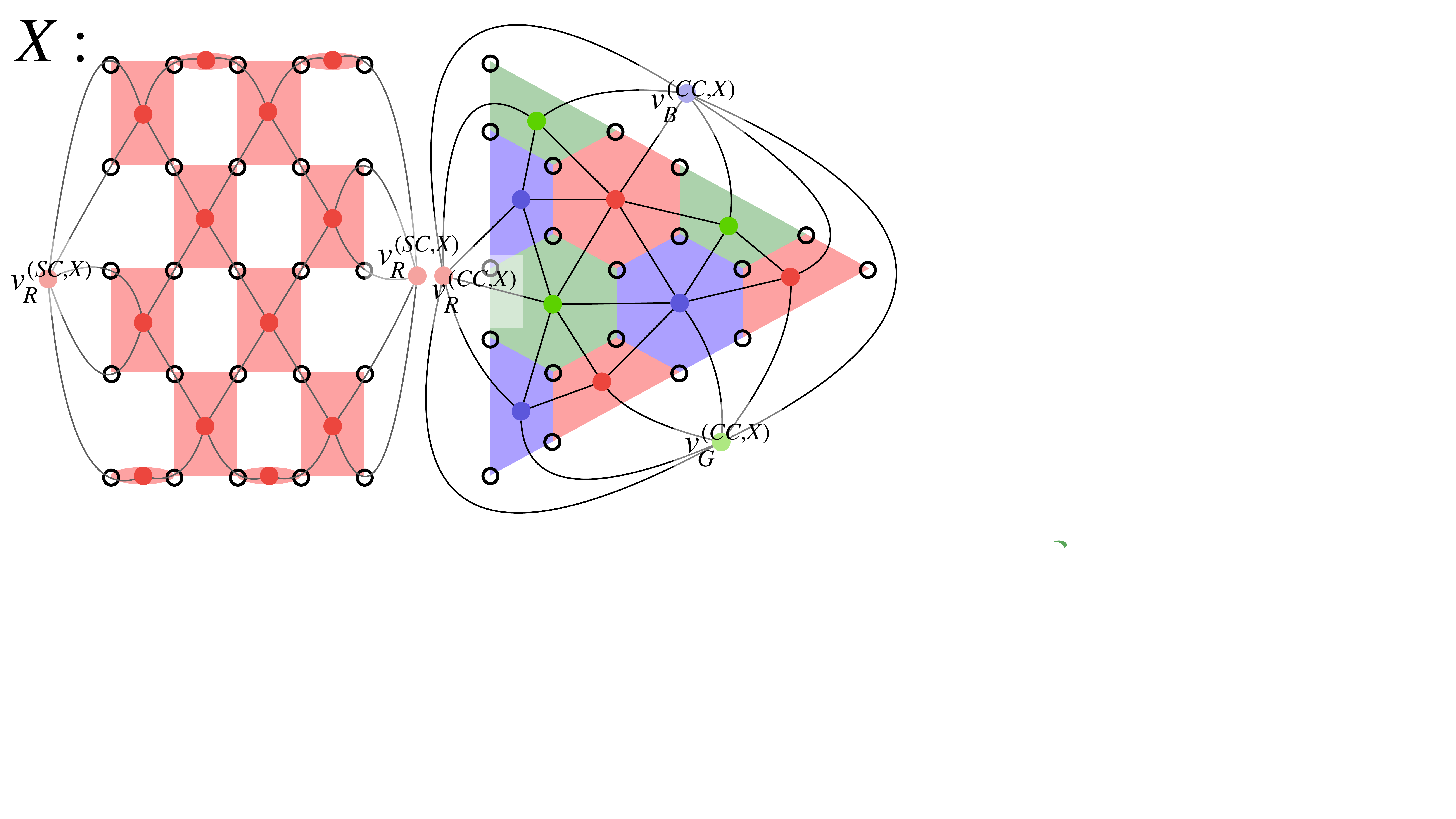}
                \includegraphics[width=0.4\linewidth]{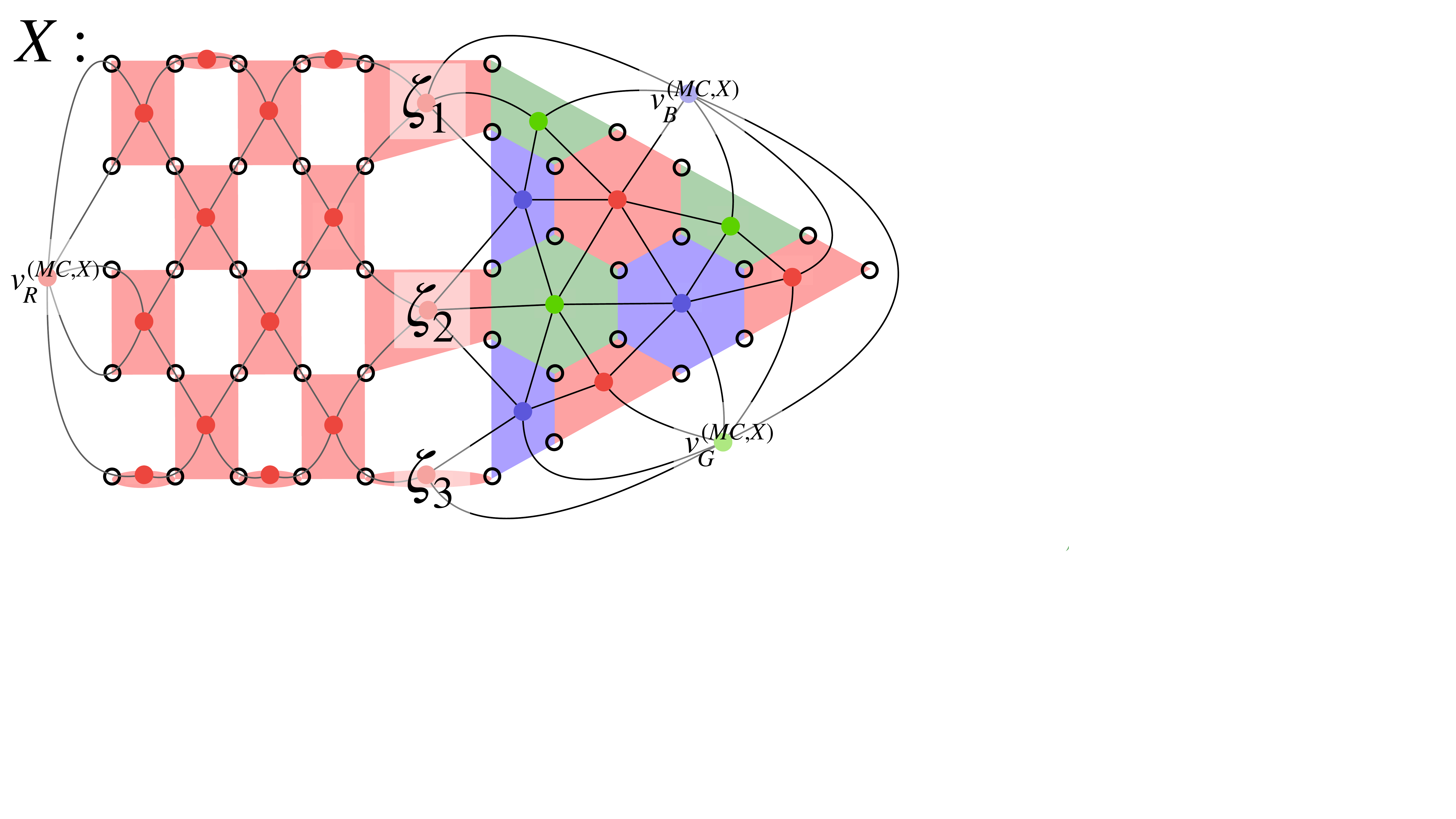}
            };
            \node (zgraph) at (-2.5,-2) {
                \includegraphics[width=0.4\linewidth]{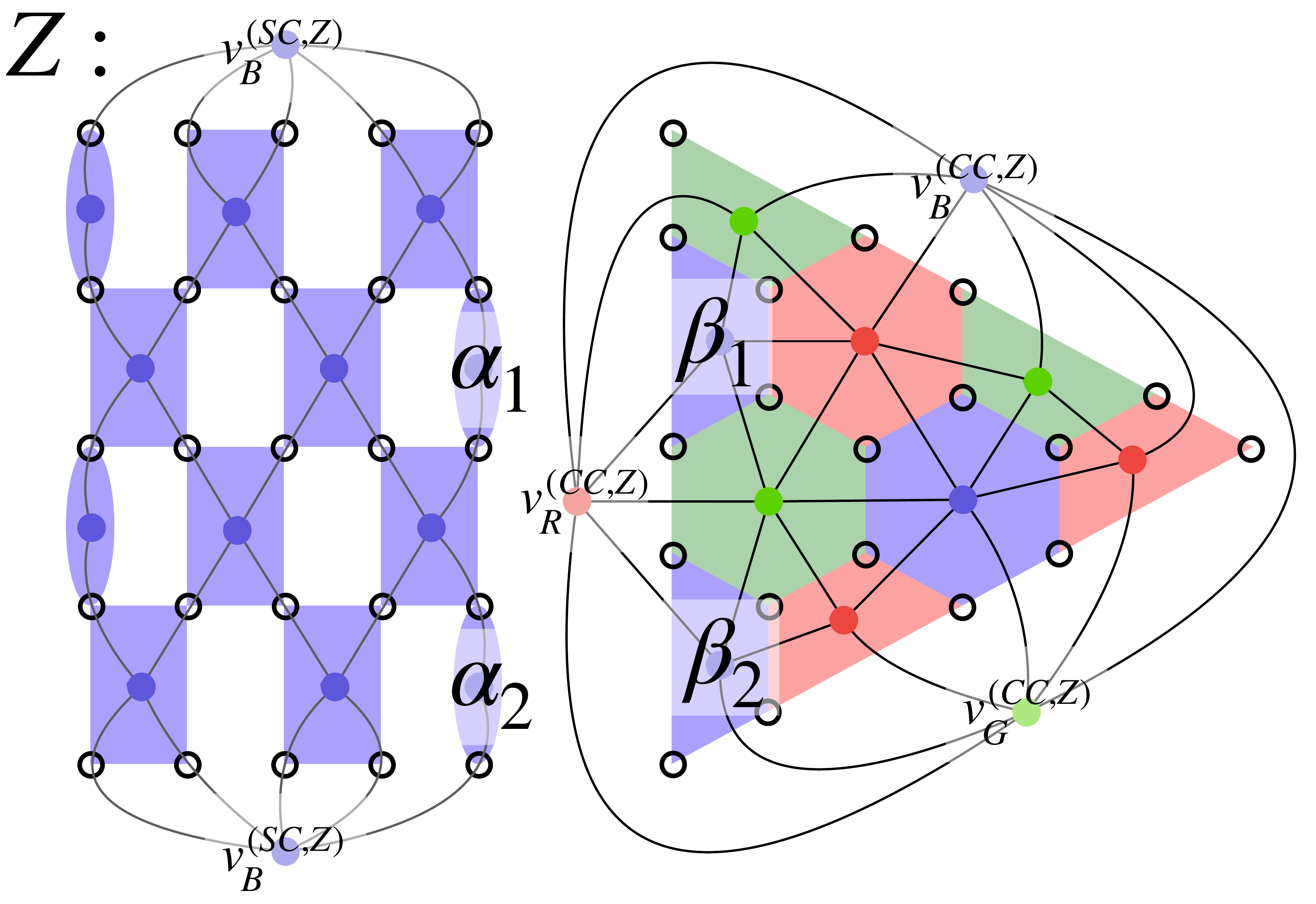}
                \includegraphics[width=0.4\linewidth]{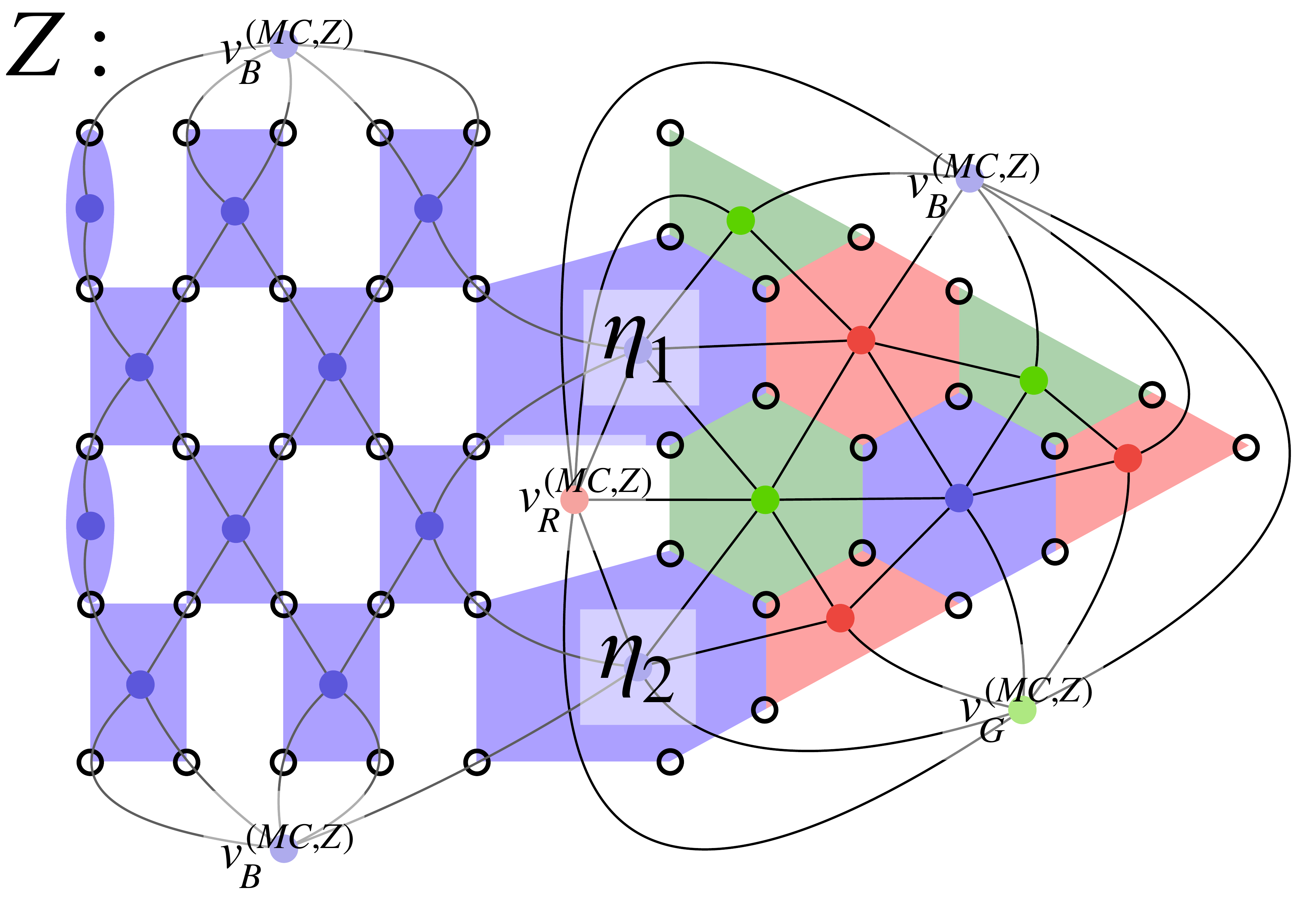}
            };
            \node[left] at (-6.2,0.5) {(a)};
            \node[left] at (-6.2,-4) {(c)};
            \node[left] at (0,0.5) {(b)};
            \node[left] at (0,-4) {(d)};

            \node at (-8.15,-4.5) {surface code};
            \node at (-4.5,-4.5) {color code};
            \node at (1.35,-4.5) {MC};
        \end{tikzpicture}
\end{center}
\caption{$X$ and $Z$ syndrome graphs $\LP$ of merged surface code and color code. Surface code qubits live on edges, color code qubits live on faces.
The restriction decoder \cite{kubica2019efficient,chamberland2020triangular} {\bf lifts} 1D paths produced by a toric code decoder to recover 2D face-qubits.}\label{fig:stabilizersMerge}
\end{figure*}
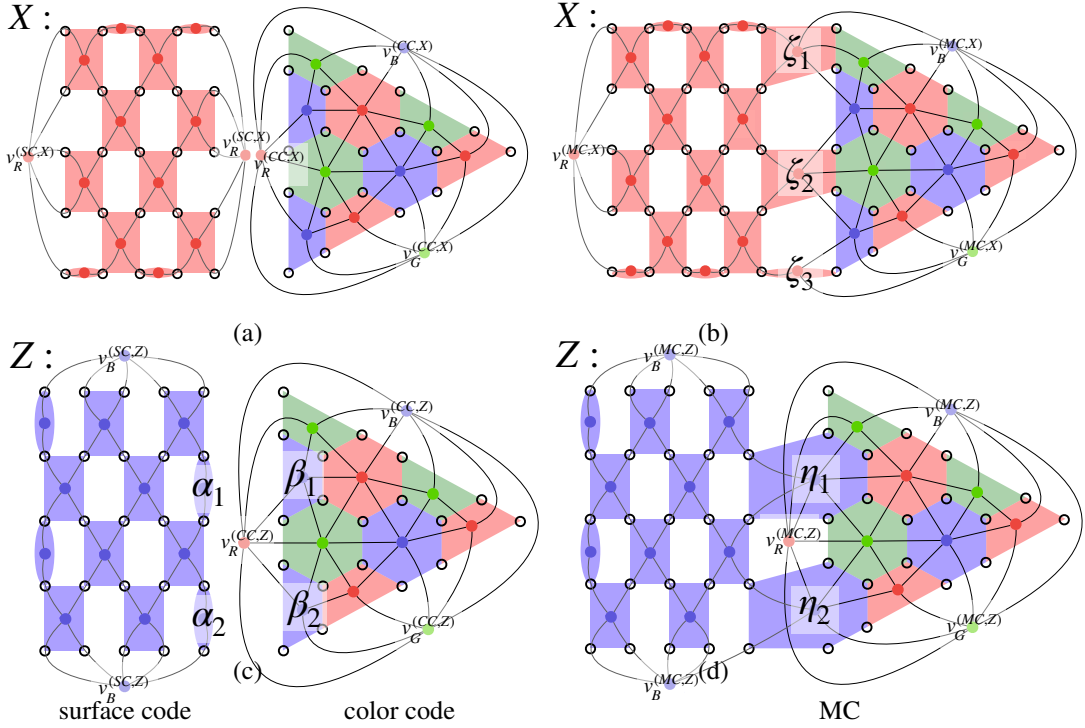

In what follows, we set $L \geq 3$ to be odd.

A triangular color code (CC) of distance $d=L$ is a CSS code with identical $X$ and $Z$ stabilizers where the data qubits can be placed at the vertices of a two-dimensional hexagonal lattice. We consider an orientation as shown in Figure~\ref{fig:stabilizersMerge} (a) and (c), right side (for distance 5). We have a 3-coloring of red (R), green (G), and blue (B) tiles corresponding to stabilizer generators of the color code.

The surface code (SC) of distance $d=L$ on a square lattice is a CSS code with differing $X$ and $Z$ stabilizer subgroups. The stabilizers for a $d=5$ surface code are shown on the left-hand side of Figure~\ref{fig:stabilizersMerge} (a) and (c). We assign colors red (R) and blue (B) to the $X$ and $Z$ stabilizers of the surface code, respectively.

For $P\in \{X, Z\}$, we define the CC (SC) $P$-syndrome graph $\mathcal{L}^{(CC,P)*}$ ($\mathcal{L}^{(SC,P)*}$), whose vertices are associated with stabilizer measurement outcomes of $P$-type stabilizers of the color (surface) code. In addition, we add virtual boundary vertices corresponding to virtual stabilizers, whose color is indicated by its subscript. For the color code, we add three $P$-type virtual stabilizers to $\mathcal{L}^{(CC,P)*}$: $$\left\{v_R^{(CC,P)}, v_G^{(CC,P)}, v_B^{(CC,P)}\right\};$$ 
these are illustrated as isolated dots in Figure~\ref{fig:stabilizersMerge}. The support of $v_{C}^{(CC,P)}$ is the set of qubits of the color code which are not contained in any (real) $C$-colored $P$-type stabilizers.

For the surface code, we add one virtual stabilizer to each of $\mathcal{L}^{(SC,X)*}$ and $\mathcal{L}^{(SC,Z)*}$: the $X$-type red virtual stabilizer $v_R^{(SC,X)}$ is added to $\mathcal{L}^{(SC,X)*}$ and the $Z$-type blue virtual stabilizer $v_B^{(SC,Z)}$ is added to $\mathcal{L}^{(SC,Z)*}$. The support of $v_R^{(SC,X)}$ is the set of all surface code qubits that are contained in exactly one (real) $X$-type stabilizer, while the support of $v_B^{(SC,Z)}$ is the set of all qubits which are contained in exactly one (real) $Z$-type stabilizer.
Finally, we add edges to the syndrome graphs $\mathcal{L}^{(P,CC)*}, \mathcal{L}^{(P,SC)*}$ between all pairs of stabilizers (including virtual stabilizers) sharing one or more common qubits.

We will now explain how gauge fixing is implemented to combine the SC and the CC into a single merged code.
Consider the orientation shown in Figure~\ref{fig:stabilizersMerge} for a distance 5 surface code and color code.
The rightmost weight-$2$ blue $Z$ stabilizers are labeled $\alpha_1, \ldots, \alpha_{(d-1)/2}$ from top to bottom.
The leftmost weight-$4$ blue $Z$ stabilizers are labeled $\beta_1, \ldots, \beta_{(d-1)/2}$ from top to bottom.
We also denote by $\zeta_1, \ldots, \zeta_{(d+1)/2}$ the red $X$-type operators which are labeled in Figure~\ref{fig:stabilizersMerge} (b). 

To produce a merged code from a surface code and color code, we gauge fix by measuring the $\zeta_i$ operators.
After performing the measurement, the $\zeta_i$ operators are now red $X$ stabilizers of the resulting merged code.
Clearly $\zeta_1$ anticommutes with both $\alpha_1$ and $\beta_1$ but commutes with $\alpha_1 \beta_1$. Therefore upon measuring $\zeta_1$, we no longer have  $\alpha_1$ and $\beta_1$ as stabilizers, but the operator $\alpha_1\beta_1$ remains a stabilizer.
We denote this new stabilizer by $\eta_1$.
By similar reasoning we see that all the stabilizers $\alpha_i, \beta_i$ are removed upon measuring all the merging operators $\zeta_i$, but their products $\eta_i:= \alpha_i\beta_i$ are left as new stabilizers of the merged code.
These new $\eta_i$ stabilizers are shown in blue in Figure~\ref{fig:stabilizersMerge} (c). In what follows, we will refer to this new code as the merged code (MC).

For $P\in \{X, Z\}$, we define the MC syndrome graph $\LP$ as follows:
the vertices of $\LP$ are associated with the $P$-type stabilizers of the MC.
For the MC, we add the three $P$-type virtual stabilizers $\{v_R^{(MC,P)}, v_G^{(MC,P)}, v_B^{(MC,P)}\}$ to $\LP$.
The six total virtual stabilizers of $\mathcal{L}^{(MC,X)*}$ and $\mathcal{L}^{(MC,Z)*}$ are thus:
\begin{align}
    \left\{v_R^{(MC,X)}, v_G^{(MC,X)}, v_B^{(MC,X)}, v_R^{(MC,Z)}, v_G^{(MC,Z)}, v_B^{(MC,Z)}\right\}.
\end{align}
All of the MC virtual stabilizers {\it except for } $v_R^{(MC,X)}$ have the qubit support given by the union of the corresponding original code supports.
Specifically,
\begin{align}
&\forall (P, C)\neq (X, R), \text{support}(v_C^{(MC,P)}) =\nonumber \\
&\text{support}(v_C^{(CC,P)}) \cup \text{support}(v_C^{(SC,P)}),
\end{align}
with the convention that \begin{align*}
    \text{support}(v_B^{(SC,X)}) &=\emptyset \\ \text{support}(v_R^{(SC,Z)}) &= \emptyset\\ \text{support}(v_G^{(SC,X)}) &= \emptyset\\ \text{support}(v_G^{(SC,Z)}) &= \emptyset
\end{align*}
(as these virtual stabilizers were never defined for the SC.)
The support of the final virtual stabilizer $v_R^{(MC,X)}$ is the set of surface code qubits which are contained in the support of exactly one red surface code stabilizer and no red, green, or blue stabilizers in the MC. This corresponds to the leftmost column of qubits in Figure~\ref{fig:stabilizersMerge}.

Finally, we once again add edges to the syndrome graph $\LP$ between all pairs of stabilizers (including virtual stabilizers) sharing one or more common qubits.
The resulting syndrome graph edges are illustrated for $L=5$ in \cref{fig:stabilizersMerge}.
We also define a classification of the qubits of the MC as either {\bf face-qubits} or {\bf edge-qubits}. 
For a qubit $q$, its classification is determined by the number $t$ of stabilizers (including virtual stabilizers) which contain $q$. By inspection, we either have $t=2$ or $t=3$. If $t=2$, we say that $q$ is an edge qubit. Otherwise $t=3$ and we say that $q$ is a face qubit.
We define the {\bf 1-boundary} denoted $\partial_2 q$ of a face qubit $q$, as the set of the 3 edges in $\LP$ which connect the 3 stabilizers of $q$. Note that we must indeed have all three edges in $\LP$ as any stabilizers sharing a qubit share an edge in $\LP$.
We now extend this notion to a set of face-qubits.
For a set of face-qubits $S = \{q_1, \ldots, q_\ell\}$, we define the 1-boundary denoted $\partial_2 S$ as
\begin{equation}
    \label{eq:oneBoundaryDef}
    \partial_2 S = \bigoplus_{q \in S}\partial_2 q,
\end{equation}
in which $\oplus$ denotes the symmetric difference of sets, and $\partial_2 q$ is just the 1-boundary of the individual face qubit $q$.
The edges $e\in \LP$ are given a real-valued weight which we denote $\text{wt}(e)$, which is set to either $w_1$ or $w_2$, where $w_1, w_2\in \R$.
Intuitively, $w_1$ sets the weight of edges in the surface code, while $w_2$ sets the weight of edges in the color code.
Specifically, if the stabilizers associated with vertices $u$ and $v$ share any face qubits, we set the weight of $(u, v)$ to $w_2$, and otherwise we set the weight to $w_1$.
Optimal values for $w_1$ and $w_2$ are found numerically by computing the logical error rates of the MC code for a given noise model.

The $X$ distance of MC is $d_X = L$, and the $Z$ distance is $d_Z = 2L$.
Note that this is an advantageous configuration for biased noise models where $Z$ errors are more likely than $X$ errors. Furthermore, the orientation can be swapped such that the $X$ distance is larger in the case of $X$-biased noise.

For each $P\in \{X, Z\}$, we also define three restricted graphs $\LPRB, \LPRG, \LPBG$ as subgraphs of the full MC syndrome graph $\LP$. These graphs are defined such that:
\begin{enumerate}
\item $\LPRB\cup \LPRG\cup \LPBG = \LP$
\item $\LP_{C_1C_2}$ contains all vertices of color $C\in \{C_1, C_2\}$ and edges between these vertices.
\end{enumerate}
The restricted graphs will be used for the decoding algorithm as explained in \cref{sec:decoder}.

\begin{algorithm}[H]
\caption{Produces a minimum weight $A$-perfect matching $M\subset E$ of a weighted graph $G=(V, E)$ with $A \subset V$ and edge weights $w(u,v)$ for $(u,v)\in E$}\label{algorithm:minWeightSemiPerfectMatching}
\begin{algorithmic}[1]
    \State Set $w(u_1, u_2)= \infty \Leftrightarrow (u_1, u_2)\notin E$
    \For{$v\in A$}
        \State $q(v) \leftarrow \min_{u\in V\setminus A}w(v, u)$\label{algmwspm:setQValue}
        \State $n(v) \leftarrow \text{argmin}_{u\in V\setminus A}w(v, u)$.\label{algmwspm:setNValue}
    \EndFor
    \For{$(u_1, u_2) \in E$}
        \If{$u_1 \in A$ and $u_2 \in A$}
	        \State $L(u_1, u_2) \leftarrow 0$
	        \State $w'(u_1, u_2)\leftarrow q(u_1)+q(u_2)$
	        \If{$w'(u_1,u_2)\leq w(u_1, u_2)$}
		        \State $L(u_1, u_2) \leftarrow 1$\label{almwspm:setEdgeLabel}
		        \State $w(u_1, u_2) \leftarrow w'(u_1, u_2)$
	        \EndIf
	    \EndIf
    \EndFor
    \State $G'  = (V', E')\leftarrow G\setminus (V\setminus A)$
    \If{$|V'|$ is odd}\label{algmwspm:VOddHandle}
        \State $G'\leftarrow G'\cup\{v_0\}$
        \For{$u \in A$}
            	\State $G' \leftarrow G' + (u, v_0)$
            	\State $w(u, v_0) \leftarrow q(u)$
        \EndFor
    \EndIf
    \State Find a min. weight PM $M'\subset E'$\label{algmwspm:mwpmSubroutine}
    \State $M \leftarrow\emptyset$
    \For{$ (u_1, u_2) \in M'$}
	\If{$v_0\in \{u_1, u_2\}$}
    		\State Let $u\in  \{u_1, u_2\}\setminus\{v_0\}$
		\State $M \leftarrow M \cup(u, n(u))$
	\Else
	\If{$L(u_1, u_2) = 1$}
		\State $M \leftarrow M \cup(u_1, n(u_1))$
		\State $M \leftarrow M \cup(u_2, n(u_2))$
	\Else
		\State $M \leftarrow M \cup(u_1, u_2)$
	\EndIf\EndIf
    \EndFor
\end{algorithmic}
\end{algorithm}

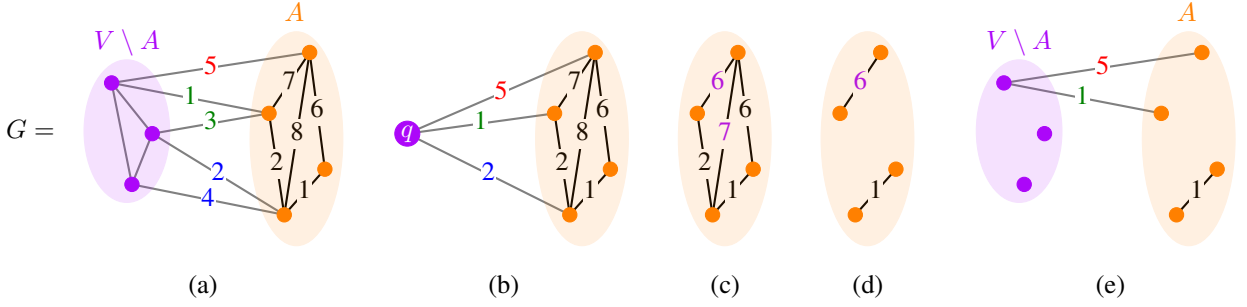
\begin{figure*}
    \centering
    \sethlcolor{white}
    \begin{tikzpicture}[scale=1.35]
        \node at (.8,-1) {(a)};

        \node (Gequals) at (-.9,.55) {$G =$};

        \draw[thick, gray] (.1,0) -- (-.1, 1);
        \draw[thick, gray] (.1,0) -- (.3, .5);
        \draw[thick, gray] (-.1, 1) -- (.3, .5);
        
        \draw[thick,gray] (-.1, 1) -- (1.85,1.3) node[midway, red]{\hl{5}};
        \draw[thick,gray] (-.1, 1) -- (1.45,.7) node[midway, VeryDarkGreen]{\hl{1}};
        \draw[thick,gray] (.3, .5) -- (1.45,.7) node[midway, VeryDarkGreen]{\hl{3}};
        \draw[thick,gray] (.3, .5) -- (1.6,-.3) node[midway, blue]{\hl{2}};
        \draw[thick,gray] (.1,0) -- (1.6,-.3) node[midway, blue]{\hl{4}};
        
        \draw[thick] (1.6,-.3) -- (1.45,.7) node[midway, black]{\hl{2}};
        \draw[thick] (1.6,-.3) -- (2,.15) node[midway, black]{\hl{1}};
        \draw[thick] (1.85,1.3) -- (1.6,-.3) node[midway, black]{\hl{8}};
        \draw[thick] (2,.15) -- (1.85,1.3) node[midway, black]{\hl{6}};
        \draw[thick] (1.45,.7) -- (1.85,1.3) node[midway, black]{\hl{7}};

        \draw[KiannaPurple, fill=KiannaPurple, opacity=0.12] (0.05,.525) ellipse (1.2em and 2em);
        \node at (0.05,1.4) {${\color{KiannaPurple}V\setminus A}$};

        \draw[KiannaPurple, fill=KiannaPurple] (.1,0) circle (0.2em);
        \draw[KiannaPurple, fill=KiannaPurple] (-.1,1) circle (0.2em);
        \draw[KiannaPurple, fill=KiannaPurple] (.3,.5) circle (0.2em);
        
        \draw[orange, fill=orange] (1.85,1.3) circle (0.2em);
        \draw[orange, fill=orange] (1.45,.7) circle (0.2em);
        \draw[orange, fill=orange] (2,.15) circle (0.2em);
        \draw[orange, fill=orange] (1.6,-.3) circle (0.2em);

        \draw[orange, fill=orange, opacity=0.12] (1.72,.45) ellipse (1.3em and 3em);
        \node at (1.7,1.7) {${\color{orange}A}$};

        \begin{scope}[xshift=80]
            \node at (.95,-1) {(b)};
        
            \draw[thick,gray] (0,.5) -- (1.85,1.3) node[midway, red]{\hl{5}};
            \draw[thick,gray] (0,.5) -- (1.45,.7) node[midway, VeryDarkGreen]{\hl{1}};
            \draw[thick,gray] (0,.5) -- (1.6,-.3) node[midway, blue]{\hl{2}};
        
            \draw[thick] (1.6,-.3) -- (1.45,.7) node[midway, black]{\hl{2}};
            \draw[thick] (1.6,-.3) -- (2,.15) node[midway, black]{\hl{1}};
            \draw[thick] (1.85,1.3) -- (1.6,-.3) node[midway, black]{\hl{8}};
            \draw[thick] (2,.15) -- (1.85,1.3) node[midway, black]{\hl{6}};
            \draw[thick] (1.45,.7) -- (1.85,1.3) node[midway, black]{\hl{7}};

            \draw[KiannaPurple, fill=KiannaPurple] (0,.5) circle (0.35em);
            \node at (0,.5) {
            ${\color{white}q}$
            };

            \draw[orange, fill=orange] (1.85,1.3) circle (0.2em);
            \draw[orange, fill=orange] (1.45,.7) circle (0.2em);
            \draw[orange, fill=orange] (2,.15) circle (0.2em);
            \draw[orange, fill=orange] (1.6,-.3) circle (0.2em);

            \draw[orange, fill=orange, opacity=0.12] (1.72,.45) ellipse (1.3em and 3em);
        \end{scope}

        \begin{scope}[xshift=120]
            \node at (1.725,-1) {(c)};

            \draw[thick] (1.6,-.3) -- (1.45,.7) node[midway, black]{\hl{2}};
            \draw[thick] (1.6,-.3) -- (2,.15) node[midway, black]{\hl{1}};
            \draw[thick] (1.85,1.3) -- (1.6,-.3) node[midway, KiannaPurple]{\hl{7}};
            \draw[thick] (2,.15) -- (1.85,1.3) node[midway, black]{\hl{6}};
            \draw[thick] (1.45,.7) -- (1.85,1.3) node[midway, KiannaPurple]{\hl{6}};

            \draw[orange, fill=orange] (1.85,1.3) circle (0.2em);
            \draw[orange, fill=orange] (1.45,.7) circle (0.2em);
            \draw[orange, fill=orange] (2,.15) circle (0.2em);
            \draw[orange, fill=orange] (1.6,-.3) circle (0.2em);

            \draw[orange, fill=orange, opacity=0.12] (1.72,.45) ellipse (1.3em and 3em);
        \end{scope}
        \begin{scope}[xshift=160]
            \node at (1.725,-1) {(d)};

            \draw[thick] (1.6,-.3) -- (2,.15) node[midway, black]{\hl{1}};
            \draw[thick] (1.45,.7) -- (1.85,1.3) node[midway, KiannaPurple]{\hl{6}};

            \draw[orange, fill=orange] (1.85,1.3) circle (0.2em);
            \draw[orange, fill=orange] (1.45,.7) circle (0.2em);
            \draw[orange, fill=orange] (2,.15) circle (0.2em);
            \draw[orange, fill=orange] (1.6,-.3) circle (0.2em);

            \draw[orange, fill=orange, opacity=0.12] (1.72,.45) ellipse (1.3em and 3em);
        \end{scope}
        \begin{scope}[xshift=250]
            \node at (.95,-1) {(e)};

        \draw[thick,gray] (-.1, 1) -- (1.85,1.3) node[midway, red]{\hl{5}};
        \draw[thick,gray] (-.1, 1) -- (1.45,.7) node[midway, VeryDarkGreen]{\hl{1}};

            \draw[thick] (1.6,-.3) -- (2,.15) node[midway, black]{\hl{1}};

            \draw[orange, fill=orange] (1.85,1.3) circle (0.2em);
            \draw[orange, fill=orange] (1.45,.7) circle (0.2em);
            \draw[orange, fill=orange] (2,.15) circle (0.2em);
            \draw[orange, fill=orange] (1.6,-.3) circle (0.2em);

            \draw[orange, fill=orange, opacity=0.12] (1.72,.45) ellipse (1.3em and 3em);
            \node at (1.7,1.7) {${\color{orange}A}$};
            
        \draw[KiannaPurple, fill=KiannaPurple, opacity=0.12] (0.05,.525) ellipse (1.2em and 2em);
        \node at (0.05,1.4) {${\color{KiannaPurple}V\setminus A}$};

        \draw[KiannaPurple, fill=KiannaPurple] (.1,0) circle (0.2em);
        \draw[KiannaPurple, fill=KiannaPurple] (-.1,1) circle (0.2em);
        \draw[KiannaPurple, fill=KiannaPurple] (.3,.5) circle (0.2em);
        \end{scope}
    \end{tikzpicture}
    \caption{Illustration of \cref{algorithm:minWeightSemiPerfectMatching} on a small graph $G$.
    (a) A graph $G = (V, E)$ with nonnegative edge weights and a vertex subset $A\subset V$. The weights on edges incident to vertices in $V\setminus A$ are omitted as these edges can be removed from any $A$-perfect matching without increasing its total weight. When applying this algorithm to the matching graph of the surface code, all vertices in $V\setminus A$ correspond to virtual boundary vertices (and thus all edges within $V\setminus A$ have zero weight). (b) A new graph is assembled by replacing all vertices in $V\setminus A$ with a single new vertex $q$. The edges between $A$ and $V\setminus A$ are replaced with some edges from $q$ to vertices in $A$, as follows. For each vertex $v\in A$ with an edge to $V\setminus A$ in $G$, we add an edge $(q, v)$, setting the weight of this new edge to the minimum of all edge weights between $v$ and $V\setminus A$ in $G$. This is analogous to the computation of $q(v)$ in \cref{algmwspm:setQValue}. (c) For each edge $(v_1, v_2)\in E$ with $v_1,v_2\in A$, such that $w(v_1, u) + w(v_2, u) < w(v_1, v_2)$, we update the edge weight to $w(v_1, u) + w(v_2, u)$ and set the label function $L(v_1,v_2)$ to 1 (\cref{almwspm:setEdgeLabel}) (indicated by the purple weight). (d) A minimum-weight perfect matching $M'$ of the graph is computed (in this example, resulting in highlighted edges of weight 6 and 1). (e) A minimum-weight $A$-perfect matching $M$ is recovered by replacing all edges $e\in M'$ such that $L(e)=1$ with the two minimum-weight edges from the endpoints of $e$ to $V\setminus A$ (corresponding to $n(v)$ as set in \cref{algmwspm:setNValue}).
    }
    \label{fig:MWSPMintuition}
\end{figure*}

We will now define a few graph-theoretic notions which are used in our MC decoding algorithm.
For a graph $G = (V, E)$ with nonnegative edge weights $E\xrightarrow{w} \mathbb{R}_{\geq 0},e\mapsto w(e)$, we recall that $M\subset E$ is a {\bf perfect matching} (PM) if each vertex $v\in V$ has exactly one edge in the set $M$.
More generally, for a subset $A\subset V$, we will say that $M\subset E$ is an {\bf $A-$perfect matching} ($A$-PM) if each vertex $v\in A$ has exactly one edge in the set $M$.
Note that this permits vertices $v\in V\setminus A$ to have any number of edges in the $A-$perfect matching. Note also that a $V-$perfect matching is simply a perfect matching.
Finally, we say that $M\subset E$ is a {\bf minimum-weight $A$-perfect matching} if for all $A-$perfect matchings $M'$, $\sum_{e\in M}{w(e)} \leq \sum_{e\in M'}{w(e)}$.
There exists a polynomial-time algorithm \cite{edmonds1965paths} for finding a minimum weight perfect matching of a weighted graph, when one exists.
Given a weighted graph $G$ and vertex subset $A$ as above, it is possible to construct a new graph $G'$ such that a minimum weight perfect matching of $G'$ can be used to recover a minimum weight $A$-perfect matching of $G$. This construction is illustrated for a small example in \cref{fig:MWSPMintuition} and exploited by Algorithm~\ref{algorithm:minWeightSemiPerfectMatching}, which uses the minimum weight perfect matching algorithm as a subroutine on \cref{algmwspm:mwpmSubroutine}.

\subsection{Merged Surface-Color Code Decoding Algorithm}\label{sec:decoder}

A decoding algorithm for triangular color code families was provided in Ref.~\cite{chamberland2020triangular}.
We now provide a modified version of this decoder to handle effects of the surface code boundary thus making it compatible with the MC code.
In this section we explain how our decoder works in detail. Our full decoding algorithm is then shown in \cref{algorithm:decoding}.
For convenience, in this section we will refer to specific numbered lines of \cref{algorithm:decoding}.

Let $\eE$ be a physical Pauli error operator on the data qubits with $X$ and $Z$ error syndromes $S_X(\eE_Z), S_Z(\eE_X)$ respectively.
That is, the set $S_P(\eE_{P'})$ is a subset of the real vertices of $\LP$.
The $X$ and $Z$ syndromes will be decoded independently to produce $Z$ and $X$ correction operators $\eE_Z', \eE_X'$, respectively.
Hence, in what follows, let $P\in \{X, Z\}$. As in \cref{algdecoding:initialiseP}, fix $P'\in \{X, Z\}$ and $P'\neq P$.
Initialize the correction operator $\eE_{P'}' = \emptyset$ as in \cref{algdecoding:emptyPartialCorrectionInitialize}.
We will refer to the stabilizers contained in 
$S_P(\eE_{P'}'\eE_{P'})$ as the {\it marked stabilizers}. Note that as the correction operator is empty at the start of the algorithm, the initial set of marked stabilizers is the same as the $P$-syndrome input to the decoder.

We decode $S_P(\eE_{P'})$ in two stages which we call the {\it color code stage} and the {\it surface code stage}.
The color code stage corresponds to \cref{algdecoding:colorCodeStageStart} through \cref{algdecoding:colorCodeStageEnd}, while the surface code stage corresponds to \cref{algdecoding:surfaceCodeStageStart} through \cref{algdecoding:surfaceCodeStageEnd}.

The color code stage produces a partial correction $\eE_{P'}'$ which only contains color code qubits. We use this partial correction to update the $P$-syndrome marked stabilizers $S_{P}(\eE_{P'}' \eE_{P'})$.
At the end of the color code stage, no color code stabilizers are marked except possibly some of the $\eta_i$.
In contrast, surface code stabilizers may still be marked.
This is because no surface code qubits are contained in the partial correction $\eE_{P'}'$ at this stage, and hence any initially-marked surface code stabilizers will remain marked at the end of the color code stage. To be clear, the partial correction $\eE_{P'}'$ at the end of the color code stage is stored in the software implementing \cref{algorithm:decoding}, and does not need to be actively applied to the physical data qubits.

After the color code stage, we run the surface code stage beginning on \cref{algdecoding:surfaceCodeStageStart}. The surface code stage is simpler than the color code stage. It makes some final modifications to the partial correction $\eE_{P'}'$ so that there are no marked MC stabilizers in the marked set $S_{P}(\eE_{P'}'\eE_{P'})$. At this point the $P'$ correction $\eE_{P'}'$ is completed.

\begin{algorithm}[H]
\caption{Produces $Z$ and $X$ corrections $\mathcal{E}_{Z}', \mathcal{E}_{X}'$ given $X$ and $Z$ syndromes $S_X(\mathcal{E}_Z), S_Z(\mathcal{E}_X)$} \label{algorithm:decoding}
\begin{algorithmic}[1]
    \For{$P \in \{X, Z\}$}
        \State Let $P'\in \{X, Z\}, P'\neq P$\label{algdecoding:initialiseP}
        \State $\eE_{P'}' \leftarrow \emptyset$ \Comment{Initialize empty partial correction}\label{algdecoding:emptyPartialCorrectionInitialize}
        \State Let $A = S_P(\eE_{P'}' \eE_{P'})$\label{algdecoding:colorCodeStageStart}
        \For{$C\in \{R, G, B\}$}\label{algdecoding:colorCodeStageColorIteration}
            \State Let $\{C_1, C_2\}= \{R, G, B\}\setminus\{C\}$
            \State $A_C\leftarrow A\cap \LP_{C_1,C_2}$
            \State $V_C\leftarrow A_C \cup \{v^{(MC,P)}_{C_1}, v^{(MC,P)}_{C_2}\}$
            \State $w_C\leftarrow \emptyset$
            \For{$(u, v)\in V_C\times V_C$}
                \State $\Gamma_{u,v,C}\leftarrow \text{MinWeightLegalPath}(u, v, \LP_{C_1,C_2})$\label{algdecoding:findMinWeightLegalPath}
                \If{$\Gamma_{u,v,C} = \perp$}
                    \State $w_C(u,v)\leftarrow \infty$
                \Else
                    \State $w_C(u,v)\leftarrow \sum_{e\in \Gamma_{u,v,C}} \text{wt}(e)$
                \EndIf
            \EndFor
            \State $G_C \leftarrow (V_C, E_C, w_C)$ \Comment{Initialize Weighted $C$-Matching Graph}
            \State $M_C\leftarrow$ min. weight $A_C$-PM of  $G_C$ using Algorithm~\ref{algorithm:minWeightSemiPerfectMatching}
            \State $\Gamma_C\leftarrow \oplus_{(u,v)\in M_C}\Gamma_{u,v,C}$
        \EndFor \label{algdecoding:colorCodeStageColorIterationEnd}
        \State $\Gamma_T \leftarrow \Gamma_R\cup\Gamma_G\cup\Gamma_B$ \label{algdecoding:colorCodeStageGammaTotalInitialize}
        \For{BCC $\theta = (\{v_i\}, \{\chi_i\})$}\label{algdecoding:BCCiterationStart}
        \State $\Gamma_\theta \leftarrow \cup_{i=1}^{\ell-1}\Gamma_{v_i, v_{i+1}, \chi_i}$\label{algdecoding:GammaThetaInitialize}
		\If{$v_\ell \in \{v_G^{(MC,P)}, v_R^{(MC,P)}\}$}\label{algdecoding:BCCColorDecide}
			\State Set $\text{color}(\theta) = B$
		\Else
			\State Set $\text{color}(\theta) = R$
		\EndIf
		\For{$u\in \Gamma_\theta|_{\text{color}(\theta)}$}
			\State $\mathcal{E}_{P'}' \leftarrow \mathcal{E}_{P'}' \oplus \text{Lift}(u, \Gamma_\theta)$\label{algdecoding:LiftAllBCCVerticesOfColor}
		\EndFor
		\If{$\text{color}(\theta)=R$}
		    \State $\mathcal{E}_{P'}' \leftarrow \mathcal{E}_{P'}'\oplus \text{SCRedLift}(\Gamma_\theta)$\label{algdecoding:SCRedLiftCall}
		\EndIf
		\State Set $\Gamma_T \leftarrow \Gamma_T \oplus \Gamma_\theta$\label{algdecoding:removeBCCPathsFromPairingPaths}
        \EndFor\label{algdecoding:BCCiterationEnd}
        \For{$u \in \Gamma_T|_{G}$}
		\State $\mathcal{E}_{P'}' \leftarrow \mathcal{E}_{P'}' \oplus \text{Lift}(u, \Gamma_T)$\label{algdecoding:LiftAllRemainingGreenVertices}
    \EndFor\label{algdecoding:colorCodeStageEnd}
	\State $A \leftarrow S_P(\eE_{P'}' \eE_{P'})$\label{algdecoding:updateMarkedforSCstage}\label{algdecoding:surfaceCodeStageStart}
	\State $\eE_{P'}' \leftarrow \text{SurfaceCodeCorrection}(A)$\label{algdecoding:SCCorrection}
    \EndFor\label{algdecoding:surfaceCodeStageEnd}
\end{algorithmic}
\end{algorithm}

\begin{figure*}
\begin{center}
\includegraphics[width=1\linewidth]{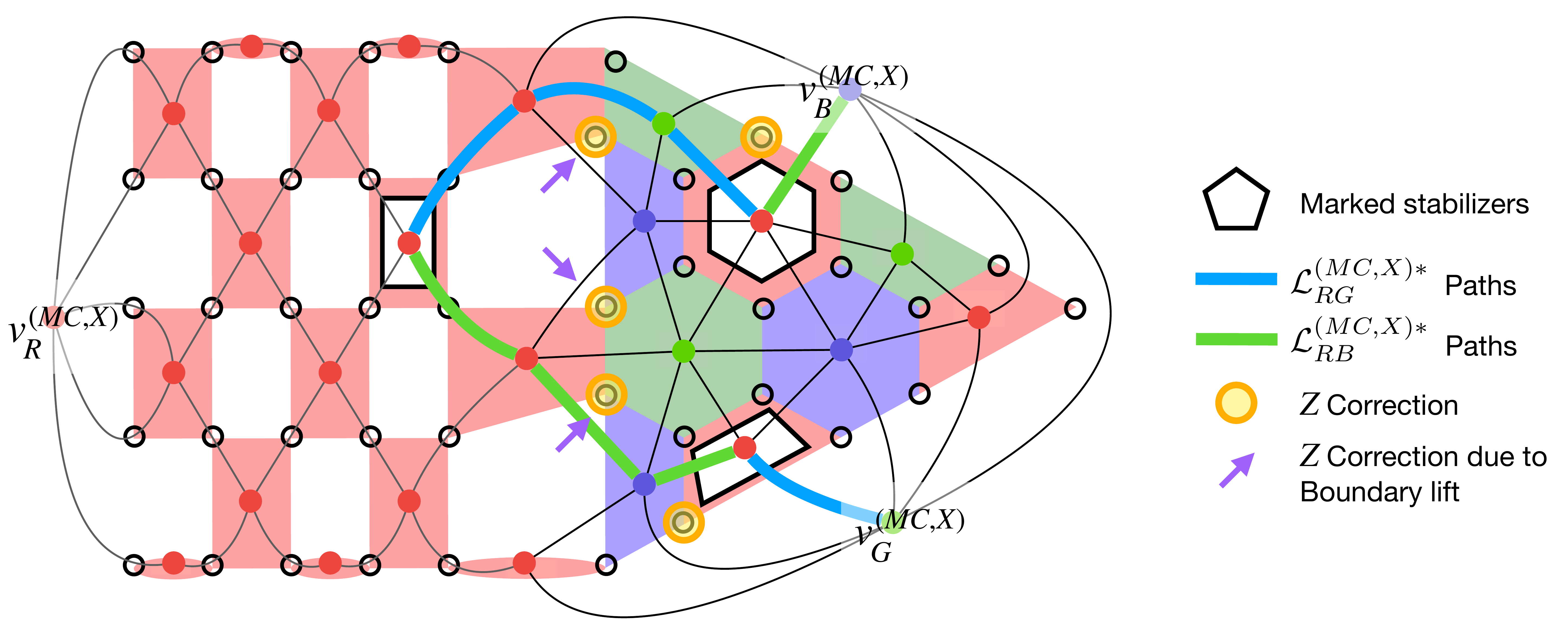}
\end{center}
\caption{Lifting at red surface code vertices in the $X$ syndrome graph is unavoidable when a component connects the blue to green virtual stabilizer and walks through such vertices along the way, as shown in the example.}\label{fig:purpleLift}
\end{figure*}

We will now explain the steps of the color code stage in detail.
The first step is to produce three {\it colored pairings} $(M_C, \{\Gamma_{u,v,C}\})$ for each of $C\in \{R, G, B\}$.
This is done in \cref{algdecoding:colorCodeStageColorIteration} through \cref{algdecoding:colorCodeStageColorIterationEnd}.
Fix any $C\in \{R, G, B\}$. Let $\{C_1, C_2\} = \{R, G, B\}\setminus \{C\}$.
The colored pairing $(M_C, \{\Gamma_{u,v,C}\})$
consists of a set $M_C = \{(u, v)\}$ of pairs of vertices of $\LP_{C_1,C_2}$, along with a path $\Gamma_{u,v,C}$ for each pair $(u, v)\in M_C$. This path $\Gamma_{u,v,C}$ joins $u$ and $v$ through the restricted graph $\LP_{C_1,C_2}$.
The path $\Gamma_{u,v,C}$ must be {\it legal}, as specified by the following conditions:
\begin{enumerate}
    \item $\Gamma_{u,v,C}$ is a path from $u$ to $v$ through $\LP_{C_1,C_2}$.
    \item Among the edges of $\Gamma_{u,v,C}$, there is at most one edge which is incident to any virtual vertex.
    \item There is at most one virtual vertex visited by $\Gamma_{u,v,C}$.
    \item If $u$ and $v$ are vertices in the color code (including all the $\eta_i$ vertices), then $\Gamma_{u,v,C}$ does not visit any vertices in the surface code.
\end{enumerate}
We choose $\Gamma_{u,v,C}$ to be the minimum weight legal path, which is found by the subroutine $\text{MinWeightLegalPath}$ in \cref{algdecoding:findMinWeightLegalPath}.
Specifically, $\text{MinWeightLegalPath}$ returns the minimum weight legal path if one exists, and returns the placeholder symbol $\perp$ if no path exists satisfying the above conditions.
The subroutine $\text{MinWeightLegalPath}$ can be easily implemented by modifying Dijkstra's pathfinding algorithm to take into account the legality conditions above. Specifically, during the Dijkstra search, the virtual vertices should be treated as having zero out-edges, and if $u$ and $v$ are color code vertices as defined above, then all edges to surface code vertices are ignored.

The colored pairings $(M_C,\{ \Gamma_{u,v,C} : (u,v)\in M_C\})$ enable us to recover a set of qubits in the color code which we add to the partial correction $\eE_{P'}'$.
Specifically, we define a subroutine ($\text{Lift}$), which is applied at some of the vertices visited by the pairing paths.
The Lift subroutine has arguments $\text{Lift}(u, \Gamma)$.
Here, $u$ is a real (non-virtual) vertex of the graph $\LP_{C_1,C_2}$ and $\Gamma$ is a path through (i.e. a subset of edges of) $\LP_{C_1,C_2}$.
The Lift subroutine returns a set of qubits, as follows.
If $u$ is a real vertex of the surface code, then the $\text{Lift}$ subroutine returns the empty set.
Otherwise $u$ is vertex of the color code and 
$\text{Lift}(u, \Gamma)$ returns a set of face-qubits to add to $\eE_{P'}$.
Specifically, let $\Gamma|_{u}$ denote the set of edges contained in $\Gamma$ which are incident to the vertex $u$. Then we have that
\begin{equation}\label{eq:liftDefinition}
    \partial_2\text{Lift}\left(u, \Gamma\right) = \Gamma|_{u}.
\end{equation}
(We remind the reader that $\partial_2$ denotes the 1-boundary as we have defined in \cref{sec:notation}.)
The implementation of the $\text{Lift}$ subroutine is easily and efficiently implemented using brute-force search or linear algebra techniques, as described in \cite{chamberland2020triangular}.
Note that above we have not defined the behavior of the $\text{Lift}$ subroutine for virtual vertices $v_{C}^{(MC,P)}$. As we will explain, we take care never to apply the $\text{Lift}$ subroutine to any such virtual vertices in our decoding algorithm, and so this is not an issue.
As a cautionary note, please note that above we have used $\Gamma$ as a placeholder for a general subset of edges through the graph $\LP$, and $\Gamma$ is {\it not to be confused} with the specific symbol $\Gamma_{u,v,C}$ which specifically denotes the minimum-weight legal path connecting $u$ and $v$ as we have just described above. In fact, it is impossible to correctly Lift a vertex $u$ with the path $\Gamma = \Gamma_{u,v,C}$, as for this path there would be no subset of face-qubits such that \cref{eq:liftDefinition} is satisfied. However in our algorithm the $\text{Lift}$ subroutine is always used in such a way that it has a valid output satisfying \cref{eq:liftDefinition}.

In our algorithm we use $\text{Lift}$ on \cref{algdecoding:LiftAllBCCVerticesOfColor} and \cref{algdecoding:LiftAllRemainingGreenVertices} as follows:
\begin{equation}
\eE_{P'}'\leftarrow \eE_{P'}' \oplus \text{Lift}\left(u, \Gamma\right)
\end{equation}
Above, the parity symbol $\oplus$ denotes the symmetric difference of the two sets $\eE_{P'}'$ and $\text{Lift}(u, \Gamma)$. That is, we update the partial correction to be the set of qubits which are contained in exactly one of $\eE_{P'}'$ or $\text{Lift}(u, \Gamma)$.

We must apply the $\text{Lift}$ subroutine carefully. As shown in \cite{chamberland2020triangular}, we must avoid lifting at virtual vertices because it decreases the effective distance of the decoder.
To avoid lifting at virtual vertices we pre-process the paths which visit the green virtual stabilizer $v_{G}^{(MC,P)}$.
This pre-processing step takes place in \cref{algdecoding:BCCiterationStart} through \cref{algdecoding:BCCiterationEnd}.
To explain this pre-processing step we will define the notion of a {\it boundary-connected component} (BCC).
These components are sequences of colored pairing paths which begin on the green virtual stabilizer $v_{G}^{(MC,P)}$.
Specifically, a BCC  $\theta = (\{v_i\}, \{\chi_i\})$ of length $\ell$ is a sequence of vertices $v_i\in \LP$ and colors $\chi_i\in \{R, G, B\}$:
\begin{align}
    &(v_1, v_2, v_3, \ldots, v_{\ell-1}, v_\ell)
    &(\chi_1, \ldots, \chi_{\ell-1})
\end{align}
for which
\begin{enumerate}
    \item $v_1 = v_G^{(MC,P)}$
    \item $v_\ell=v_C^{(MC,P)} \in \{v_R^{(MC,P)}, v_G^{(MC,P)}, v_B^{(MC,P)}\}$ (including possibly $v_C^{(MC,P)}=v_G^{(MC,P)}$)
    \item For each $i\in 1, \ldots, \ell-1$, we have $(v_{i}, v_{i+1})\in M_{\chi_i}$.
    \item All of the $\ell-2$ pairs $(v_{i}, \chi_{i-1})$, $(v_{i}, \chi_{i})$ for $i\in 2, \ldots, \ell-1$  are distinct.
\end{enumerate}
Additionally, the pairs $(v_i, \chi_i)$ are unique across the BCCs. That is if $\theta' = (\{v_i'\}, \{\chi_i\})$ is any other BCC of length $\ell'$, and $i\in 2, \ldots, \ell-1$ and $j\in 2, \ldots, \ell'-1$, then we have $\{(v_i, \chi_i), (v_i, \chi_{i-1})\}\cap\{ (v_{j}', \chi_{j}'), (v_{j}', \chi_{j-1}')\}=\emptyset$.
The BCCs are obtained by doing a depth-first search starting from $v_{G}^{(MC,P)}$ through the colored multigraph with edges given by $M_R \cup M_G \cup M_B$. In this multigraph, each edge in $M_C$ is given color $C$. During the depth-first search starting from $v_{G}^{(MC,P)}$, each edge of color $C$ is removed from the appropriate matching set $M_C$ as soon as it is crossed by the search. The depth-first search is halted when a virtual vertex $v_{C}^{(MC,P)}$ is reached. This is then set as the final vertex $v_\ell = v_{C}^{(MC,P)}$ of the BCC.
Following this implementation, the BCCs will satisfy the above properties.

We now iterate over each BCC $\theta = (\{v_i\}, \{\chi_i\})$ in \cref{algdecoding:BCCiterationStart} through \cref{algdecoding:BCCiterationEnd}.
In \cref{algdecoding:GammaThetaInitialize} we declare the path (that is, the set of edges) along the minimum weight legal paths between subsequent $v_i$. That is, we set
$$\Gamma_\theta = \cup_{i=1}^{\ell-1}\Gamma_{v_i, v_{i+1}, \chi_i}.$$
We then assign $\theta$ a color denoted by $\text{color}(\theta)\in \{R, B\}$ (\cref{algdecoding:BCCColorDecide}).
This is done so that $\Gamma_\theta$ never visits the virtual vertex $v_{\text{color}(\theta)}^{(MC,P)}$.
We apply the Lift subroutine at each vertex visited by $\Gamma_\theta$ with color equal to $\text{color}(\theta)$.
Recall that by definition, the $\text{Lift}$ subroutine returns the empty set when passed a surface code vertex.
However, if the color of the BCC is set to red and its path $\Gamma_\theta$ crosses between the color and surface code, then we must do some {\it variant of} the lift subroutine to find an appropriate correction in the vicinity of these red surface code vertices.
This is achieved by the subroutine 
$\text{SCRedLift}(\Gamma_\theta)$.
An example of applying  $\text{SCRedLift}$ is shown for illustration in \cref{fig:purpleLift}.
Formally, the $\text{SCRedLift}$ subroutine finds the ordered sequence of edges
$$e_1, e_2, \ldots, e_{2m-1}, e_{2m}\in \Gamma_\theta$$
which are incident to both surface code and color code stabilizer, and are ordered by increasing vertical position in the 2D layout. There must be an even number of such edges, since by construction the path $\Gamma_\theta$ must connect between $v_{G}^{(MC,P)}$ and $v_{B}^{(MC,P)}$, so $\Gamma_\theta$ must move from the color code  to the surface code an even number of times.
Now for two such crossing edges $(e_i, e_{i+1})$, let $\text{SandwichedQubits}(e_i, e_{i+1})$ denote the set of qubits which line between $e_i$ and $e_{i+1}$, and are contained in the intersection of the color code $X$-stabilizers and the $\zeta_i$ $X$-stabilizers of the MC code.
Then $\text{SCRedLift}(\Gamma_\theta)$ returns the union of these sandwiched qubits across all the pairs:
\begin{equation}\label{eq:SCRedLiftDef}
    \text{SCRedLift}(\Gamma_\theta) = \bigcup_{i=1}^{m} \text{SandwichedQubits}(e_{2i-1}, e_{2i})
\end{equation}
After applying the appropriate $\text{Lift}$ subroutine calls (\cref{algdecoding:LiftAllBCCVerticesOfColor}) and $\text{SCRedLift}$ subroutine calls (\cref{algdecoding:SCRedLiftCall}) as appropriate, we can remove the BCC paths $\Gamma_\theta$ from the combined pairing paths (\cref{algdecoding:removeBCCPathsFromPairingPaths}). After finishing this for all the BCCs $\theta$ we are finished with the pre-processing step.

By carefully processing the BCCs as described and updating the partial correction $\eE_{P'}'$ with lifts, we avoided lifting at $v_{G}^{(MC,P)}$ or any other virtual vertex and we and removed all of the paths to $v_{G}^{(MC,P)}$ from the pairing paths (due to \cref{algdecoding:removeBCCPathsFromPairingPaths}).
The final step of the color code stage is to iterate through the remaining pairing paths in \cref{algdecoding:LiftAllRemainingGreenVertices}, applying the aforementioned lift operation to all of the green vertices visited by the paths.
Since we remove all paths to $v_{G}^{(MC,P)}$ in the pre-processing step, we will not apply Lift at any virtual stabilizers in this final step.

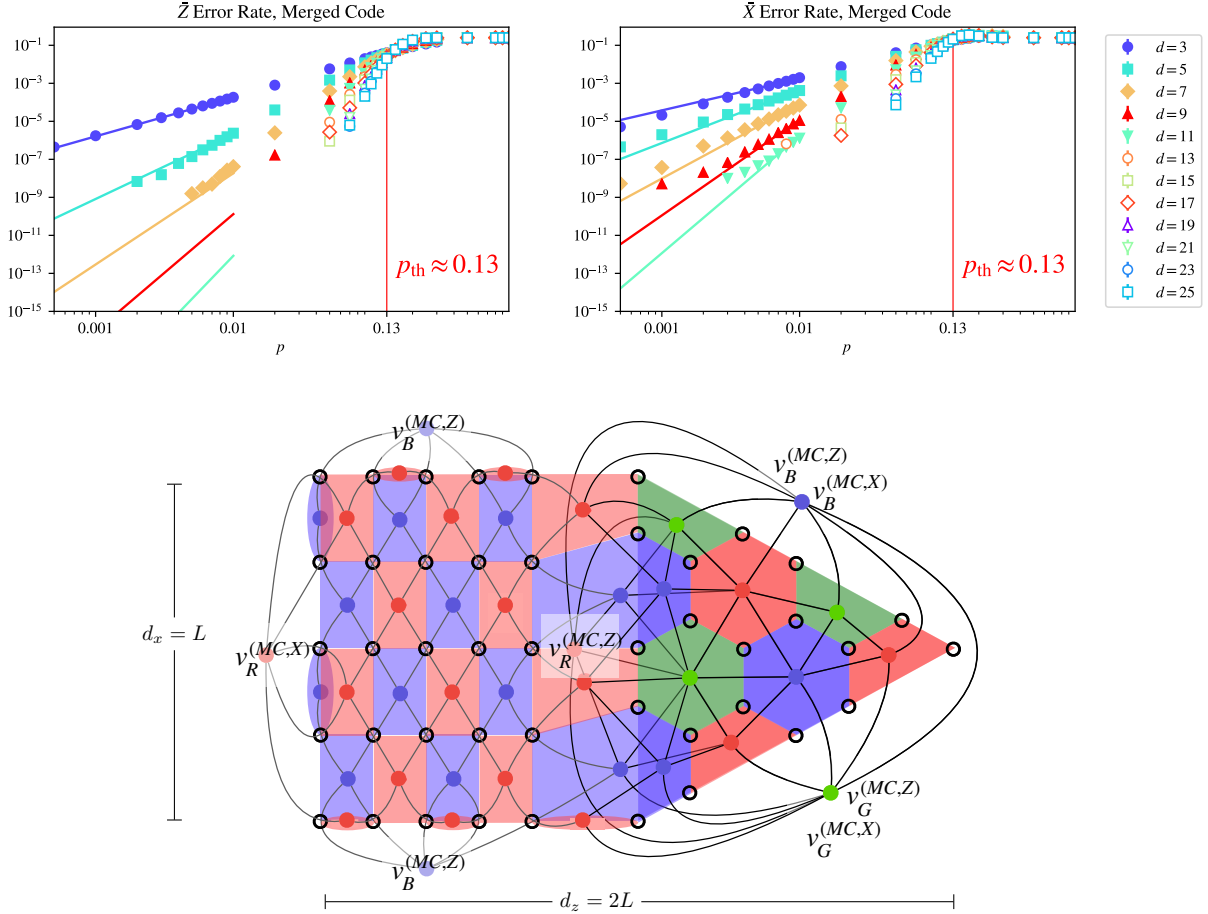
\begin{figure*}
	\begin{center}
	    \begin{tikzpicture}[scale=1.2]
            \node (xerror) at (-2,2.4) {
                 \includegraphics[height=15em]{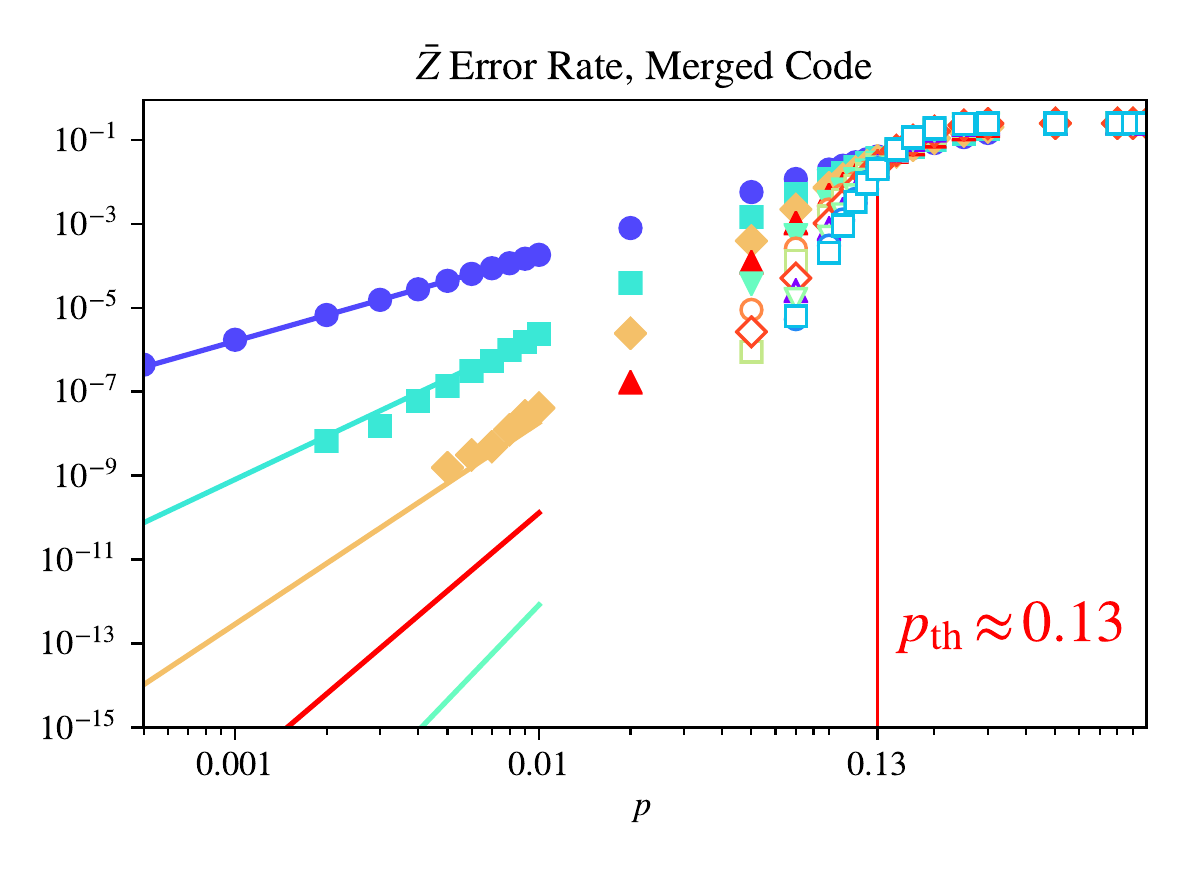}
            };
            \node (zerror) at (5,2.4) {
                 \includegraphics[height=15em]{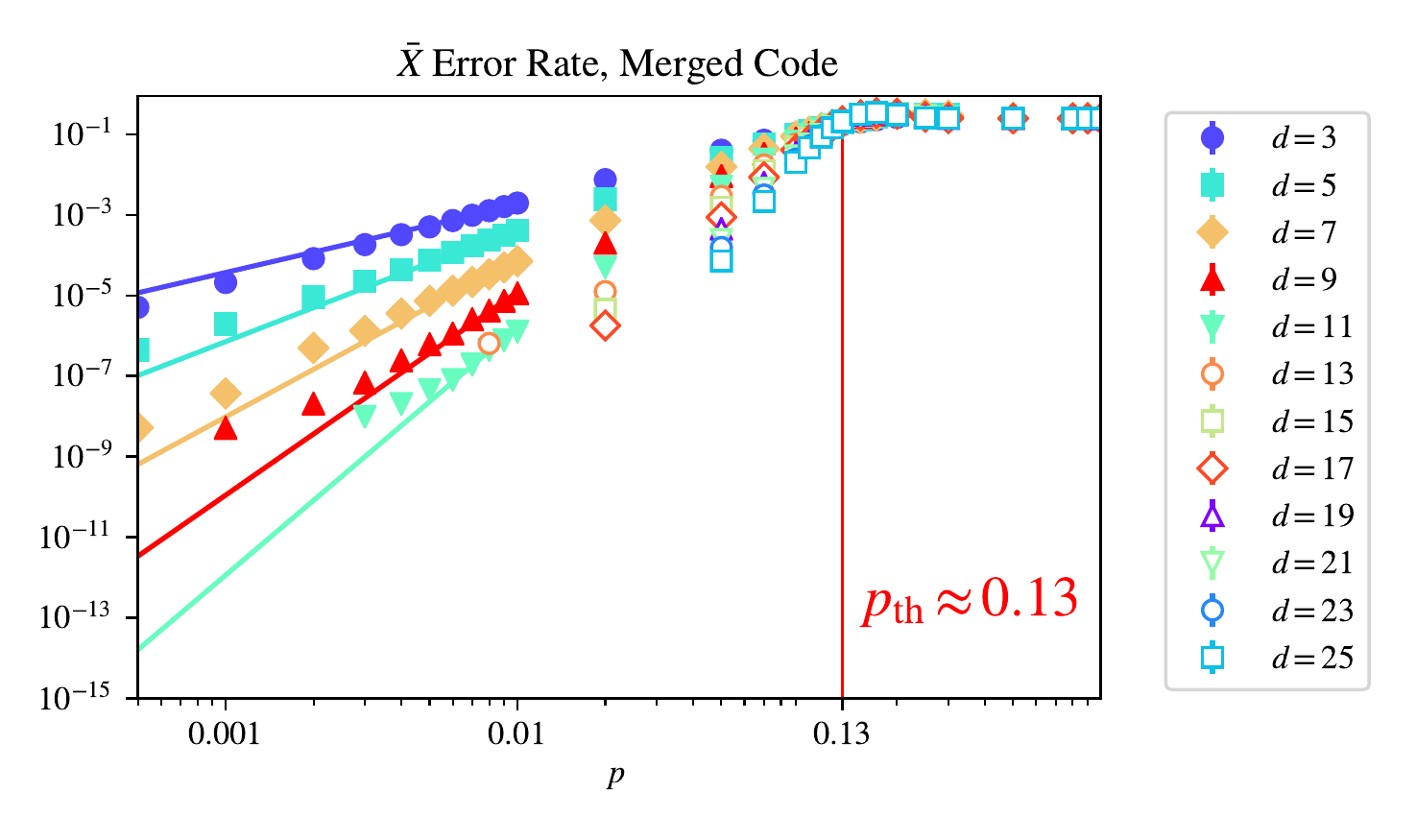}
            };
	    \begin{scope}[yshift=-10,xshift=110]
            \begin{scope}[xshift=-60,yshift=-70,scale=1.8]
	            \node (image) at (0.1,0) {
                    \includegraphics[width=0.57\linewidth]{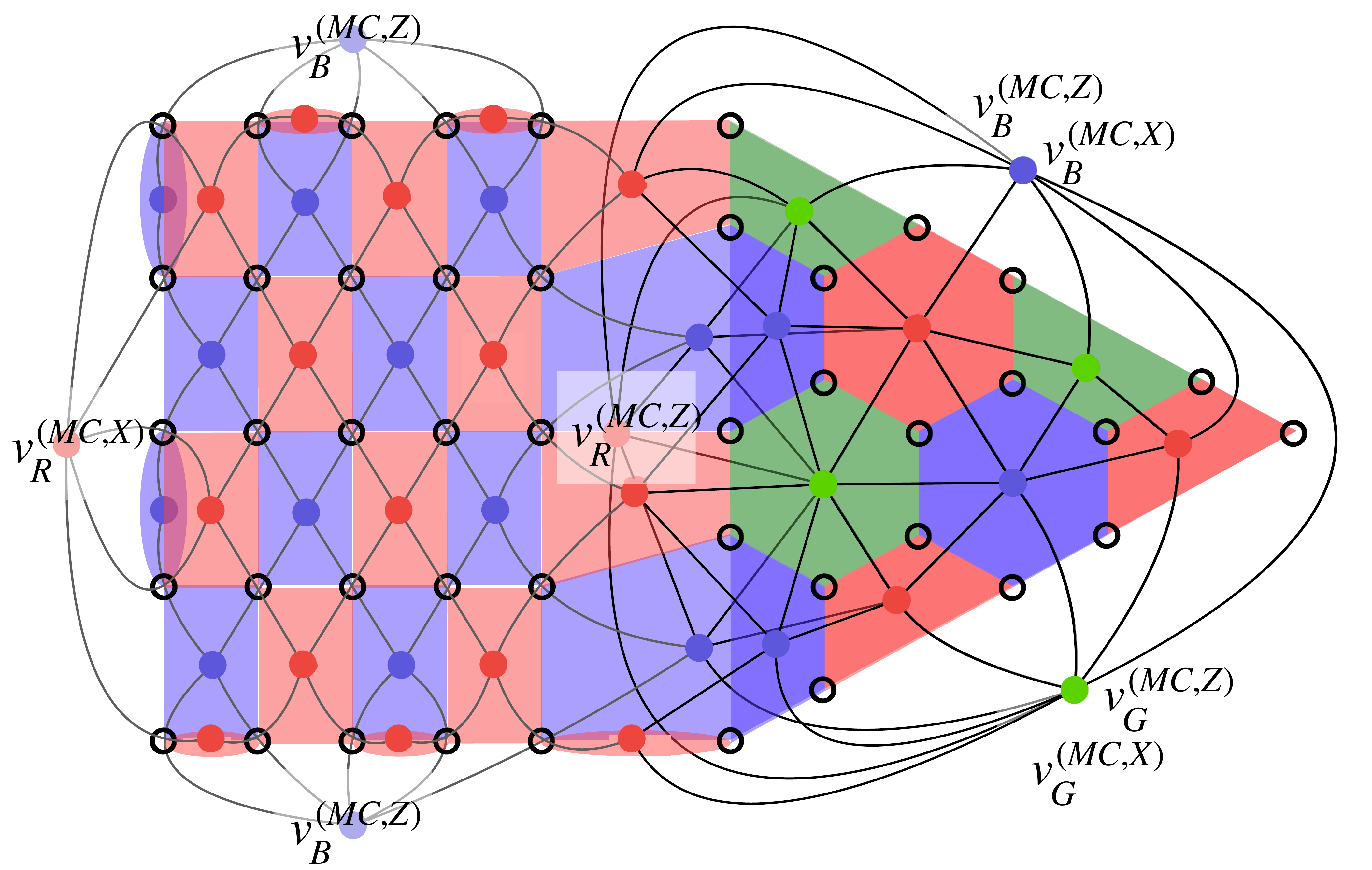}
                };
                \node[scale=0.8] (DZ) at (0, -1.55) {$d_z=2L$};
                \draw [-|] (DZ) -- (2.2,.-1.55);
                \draw [-|] (DZ) -- (-1.675,.-1.55);
                \node[scale=0.8] (DX) at (-2.6, .1) {$d_x=L$};
                \draw [-|] (DX) --(-2.6,1.025);
                \draw [-|] (DX) --(-2.6,-1.05);
            \end{scope}
        \end{scope}
    \end{tikzpicture}
	\end{center}
        \caption{(Top) $\bar{Z}$ and $\bar{X}$ logical error rates of the merged code for various code distances using the decoder described in \cref{algorithm:decoding}. In obtaining the plots, we used a code capacity depolarizing noise model. Solid lines are obtained by performing a best fit to the ansatz described in \cref{subsec:MCnumerics}. (Bottom) 
        The lattice illustrates the $d_x$ and $d_z$ code parameters as a function of the lattice size $L$.
        The effective $d_x$ and $d_z$ distances are obtained in \cref{tab:Bestfit} using the ansatz described in \cref{eq:ansatzFit}. 
        }
        \label{fig:colorcode_onestep_logicalz}
\end{figure*}

After the color code stage is concluded on \cref{algdecoding:colorCodeStageEnd}, we run the surface code stage on \cref{algdecoding:surfaceCodeStageStart} through \cref{algdecoding:surfaceCodeStageEnd}.
First we update the set of marked vertices $A = S_{P}(\eE_{P'}'\eE_{P'})$ based on the partial $P'$-type correction produced by the previous stage (\cref{algdecoding:updateMarkedforSCstage}).
We then obtain a set of qubits $\text{SurfaceCodeCorrection}(A)$ using the standard surface code decoder on \cref{algdecoding:SCCorrection}.
That is, we set
$V_{SC} = A\cup \{v_{R}^{(MC,P)},v_{B}^{(MC,P)}\}$, and for each pair $(u,v)\in V_{SC}\times V_{SC}, $ such that $u\neq v$, we set $\Gamma_{u,v}$ to be the minimum weight path which joins $u$ and $v$ through $\LP|_{SC \cup \{\eta_i\}}$ (or else $\perp$ if none exists), and set $w(u, v)=\sum_{e\in \Gamma_{u,v}} \wt (e)$. We let $G_{SC} = (V_{SC}, E_{SC}, w)$ be a weighted graph.
We then find a minimum weight $A$-perfect matching $M_{SC}$ of the graph
$G_{SC}$, and for each $(u, v) \in M_{SC}$, we set $\eE_{P'}'\leftarrow \eE_{P'}'\oplus \Gamma_{u,v}$.
This concludes the surface code stage. After this point there are no more marked vertices and $\eE_{P'}'$ is completed.

By avoiding all lifts at virtual vertices our decoder obtains good performance as shown by our numerics in \cref{subsec:MCnumerics}. Specifically, we maintain the full effective $X$ distance $d_X$ of the surface code decoder, and the combined effective $Z$ distance $\left(1+\frac{2}{3}\right)d_Z$ of the surface code decoder along with the color code decoder (\cite{chamberland2020triangular}).

\subsection{MC code capacity simulation results}
\label{subsec:MCnumerics}

\begin{table}[]
\begin{tabular}{l|l|l|}
\cline{2-3}
                        & \multirow{2}{*}{$p_L^{(X)}$} & \multirow{2}{*}{$p_L^{(Z)}$} \\
                        &                      &                      \\ \hline
\multicolumn{1}{|l|}{$a$} & .0964                & .0728                \\ \hline
\multicolumn{1}{|l|}{$b$} & .0108                & .2944                \\ \hline
\multicolumn{1}{|l|}{$c$} & .3441                & .6857                \\ \hline
\end{tabular}
\caption{Best fit parameters for the logical error rate curves $p_L^{(X)}$ and $p_L^{(Z)}$.}
\label{tab:Bestfit}
\end{table}

In this section, we describe Monte Carlo simulation results of the MC code under a code-capacity depolarizing noise model where data qubits are afflicted by $X$, $Y$ and $Z$ errors, each occurring with the same probability $p$. Data for the logical $X$ and $Z$ failure rates of the MC code for various values of $L$ are shown in \cref{fig:colorcode_onestep_logicalz}. 

By carefully analyzing the data, we find the threshold for logical $X$ and $Z$ errors to be approximately $p_{th} \approx 0.13$. However, for many noise parameter regimes, the logical $Z$ error rate is substantially lower than the logical $X$ error rate. In particular, for both logical $X$ and $Z$ error rates, we performed a best-fit analysis to obtain logical error rate curves $p_L^{(X)}$ and $p_L^{(Z)}$ using the following ansatz:
\begin{align}
    p_{L} = aL^{2}(bp)^{cL},
    \label{eq:ansatzFit}
\end{align}
where $L$ is shown in \cref{fig:colorcode_onestep_logicalz}, $a$, $b$ and $c$ are parameters obtained by the fit. As such, the parameter $c$ describes the effective $d_x$ and $d_z$ distances, i.e. the minimum-weight $X$ and $Z$ errors which can cause a logical fault. The best fit parameters are given in \cref{tab:Bestfit}. As can be seen from the parameter $c$ in \cref{tab:Bestfit}, the effective $d_x$ distance is roughly half the effective $d_z$ distance. The reason is that horizontal $Z$ error chains which can result in a logical $Z$ error must span a length of size $2L$. In contrast, vertical $X$ error chains which can cause a logical $X$ error must span a length of size $L$.

\section{Conclusion}
\label{Sec:Conclusion}

In this paper, we showed in \cref{sec:SMTsolvers} how Clifford circuits can be designed where the desired constraints (such as certain fault-tolerance proprieties, degree of connectivity between the qubits etc) can be formulated as an SMT decision problem. We provided several examples of how an SMT formula's, which evaluates to a Boolean value, can be formulated for various Clifford circuits. 

In \cref{sec:FaultTolerantHtype}, we applied our SMT formalism to derive fault-tolerant flag-based circuits to prepare $\ket{H}$-type magic states encoded in the color code. In particular, we discussed how $v$-flag circuits can be derived with the added constraint that qubits must interact via nearest neighbors with low degree connectivity constraints. Examples of such circuits for $d=3$ color codes were provided in \cref{fig:geometryToProtocol}. A clear direction of future work would be to obtain such circuits for larger code distances, and optimize iterative solving techniques described in \cref{susec:Iterative} to reduce the computation time required to find a solution. 

Lastly, in \cref{sec:DecodingMerge}, we considered converting states encoded in the color code to states encoded in the surface code. Performing such conversions was motivated by the fact that surface codes are much better suitable candidates for implementing algorithms via lattice surgery, while at the same time, color codes are particularly well suited for fault-tolerantly preparing magic states. 

To convert color codes to surface codes, we provided a decoding algorithm for a code obtained when merging the color code with the surface code via lattice surgery, an integral part of the teleportation step. We then analyzed the performance of the merged code for code capacity noise. A direction of future work would be to extend our decoder to be compatible with lattice surgery protocols, as was done for instance in Ref.~\cite{ChamberlandLatticeSurgery} and analyze the final logical error rates of the prepared magic states under a full circuit level noise model.  

We close with some further suggested directions for designing quantum circuits with SMT solvers.
We mention that there is room to optimize our encoding and solver techniques further. For example, using a variant of the Strassen algorithm \cite{strassen1969gaussian}, one could reduce the number of costly symbolic bit matrix multiplications when constructing $F_{\text{SMT}}$. SMT solvers support a wide variety of strategies which should be fine-tuned to get the best solver performance.
It is also possible that a simpler standalone algorithm could supplant the use of SMT solvers for certain design problems.
Lastly, in this work, we only considered the design of deterministic Clifford-like circuits, i.e., the class of unitary circuits that are equivalent to Clifford circuits up to conjugation by local unitaries.
It would be interesting if our techniques can be extended beyond this restricted circuit class to nondeterministic and / or non-Clifford circuits. For circuits that are dominated by Clifford gates but that contain a relatively small number of non-Clifford operations, the algorithm of Ref.~\cite{PhysRevLett.116.250501} could be used as a starting point for encoding in an SMT decision problem.

\section{Acknowledgments}
We thank Markus Kesselring for his comments on our manuscript and for pointing out the connection between our lattice surgery methods and the use of domain walls between the color code and surface code.

\bibliography{refs}

\end{document}